\documentclass[iop,apj,useAMS,usenatbib]{emulateapj-rtx4}

\usepackage{graphicx}
\usepackage{psfrag}
\usepackage{lscape}
\usepackage{amsmath}
\usepackage{longtable}
\usepackage[usenames,dvipsnames]{color}

\newcommand{\herschel}{{\it Herschel}}

\newcommand{\bootes}{Bo\"{o}tes}
\newcommand{\sersic}{S\'{e}rsic}
\newcommand{\munir}{$\mu_{\rm{NIR}}$}
\newcommand{\musubmm}{$\mu_{880}$}
\newcommand{\tnm}{\tablenotemark}
\newcommand{\tnt}{\tablenotetext}
\newcommand{\msun}{${\rm M_{\odot}}$}

\def\mnras{MNRAS}
\def\aap{A\&A}

\def\aj{AJ}
\def\araa{ARA\&A}
\def\apjl{ApJ}
\def\apj{ApJ}
\def\apjs{ApJS}
\def\pasp{PASP}
\def\nat{Nature}

\keywords{ submillimeter: galaxies gravitational lensing: strong emission galaxies: starburst}

\begin{document}

\shorttitle{Near-IR Lens Models of \herschel-selected Galaxies}
\shortauthors{Calanog et al.}

\title{Lens Models of {\it Herschel}-Selected Galaxies From High-Resolution near-IR Observations}
\author{J.A.~Calanog\altaffilmark{1},
Hai~Fu\altaffilmark{2},
A.~Cooray\altaffilmark{1,3},
J.~Wardlow\altaffilmark{4},
B.~Ma\altaffilmark{1},
S.~Amber\altaffilmark{5},
A.J.~Baker\altaffilmark{6},
M.~Baes\altaffilmark{7},
J.~Bock\altaffilmark{3,8},
N.~Bourne\altaffilmark{9},
R.~S.~Bussmann\altaffilmark{10},
C.M.~Casey\altaffilmark{1},
S.C.~Chapman\altaffilmark{11},
D.L.~Clements\altaffilmark{12},
A.~Conley\altaffilmark{13},
H.~Dannerbauer\altaffilmark{14},
G.~De Zotti\altaffilmark{15,16},
L.~Dunne\altaffilmark{17},
S.~Dye\altaffilmark{9},
S.~Eales\altaffilmark{17},
D.~Farrah\altaffilmark{18,19},
C.~Furlanetto\altaffilmark{9},
A.I.~Harris\altaffilmark{20},
R.J.~Ivison\altaffilmark{21,22},
S.~Kim\altaffilmark{1},
S.J.~Maddox\altaffilmark{9},
G.~Magdis\altaffilmark{23},
H.~Messias\altaffilmark{24,25},
M.J.~Micha{\l}owski\altaffilmark{22},
M.~Negrello\altaffilmark{15},
J.~Nightingale\altaffilmark{9},
J.M.~O'Bryan\altaffilmark{1},
S.J.~Oliver\altaffilmark{18},
D.~Riechers\altaffilmark{10},
D.~Scott\altaffilmark{26},
S.~Serjeant\altaffilmark{5},
J.~Simpson\altaffilmark{27},
M.~Smith\altaffilmark{17},
N.~Timmons\altaffilmark{1},
C.~Thacker\altaffilmark{1},
E.~Valiante\altaffilmark{26},
J.D.~Vieira\altaffilmark{3}}
\altaffiltext{1}{Dept. of Physics \& Astronomy, University of California, Irvine, CA 92697}
\altaffiltext{2}{Department of Physics and Astronomy, University of Iowa, Van Allen Hall, Iowa City, IA 52242}
\altaffiltext{3}{California Institute of Technology, 1200 E. California Blvd., Pasadena, CA 91125}
\altaffiltext{4}{Dark Cosmology Centre, Niels Bohr Institute, University of Copenhagen, Juliane Maries Vej 30, 2100 Copenhagen, Denmark}
\altaffiltext{5}{Department of Physical Sciences, The Open University, Milton Keynes MK7 6AA, UK}
\altaffiltext{6}{Department of Physics and Astronomy, Rutgers, The State University of New Jersey, 136 Frelinghuysen Rd, Piscataway, NJ 08854}
\altaffiltext{7}{1 Sterrenkundig Observatorium, Universiteit Gent, Krijgslaan 281, B-9000 Gent, Belgium}
\altaffiltext{8}{Jet Propulsion Laboratory, 4800 Oak Grove Drive, Pasadena, CA 91109}
\altaffiltext{9}{School of Physics and Astronomy, University of Nottingham, NG7 2RD, UK}
\altaffiltext{10}{Department of Astronomy, Space Science Building, Cornell University, Ithaca, NY, 14853-6801}
\altaffiltext{11}{Institute of Astronomy, University of Cambridge, Madingley Road, Cambridge CB3 0HA, UK}
\altaffiltext{12}{Astrophysics Group, Imperial College London, Blackett Laboratory, Prince Consort Road, London SW7 2AZ, UK}
\altaffiltext{13}{Center for Astrophysics and Space Astronomy 389-UCB, University of Colorado, Boulder, CO 80309}
\altaffiltext{14}{Laboratoire AIM-Paris-Saclay, CEA/DSM/Irfu - CNRS - Universit\'e Paris Diderot, CE-Saclay, pt courrier 131, F-91191 Gif-sur-Yvette, France}
\altaffiltext{15}{INAF - Osservatorio Astronomico di Padova, Vicolo dell'Osservatorio 5, I-35122 Padova, Italy.}
\altaffiltext{16}{SISSA, Via Bonomea 265, I-34136 Trieste, Italy}
\altaffiltext{17}{School of Physics and Astronomy, Cardiff University, Queens Buildings, The Parade, Cardiff CF24 3AA, UK}
\altaffiltext{18}{Astronomy Centre, Dept. of Physics \& Astronomy, University of Sussex, Brighton BN1 9QH, UK}
\altaffiltext{19}{Department of Physics, Virginia Tech, Blacksburg, VA 24061}
\altaffiltext{20}{Department of Astronomy, University of Maryland, College Park, MD 20742-2421}
\altaffiltext{21}{UK Astronomy Technology Centre, Royal Observatory, Blackford Hill, Edinburgh EH9 3HJ, UK}
\altaffiltext{22}{Institute for Astronomy, University of Edinburgh, Royal Observatory, Blackford Hill, Edinburgh EH9 3HJ, UK}
\altaffiltext{23}{Department of Astrophysics, Denys Wilkinson Building, University of Oxford, Keble Road, Oxford OX1 3RH, UK}
\altaffiltext{24}{Universidad de Concepci\'on, Barrio Universitario, Concepci\'on, Chile}
\altaffiltext{25}{Centro de Astronomia e Astrof\'isica da Universidade de Lisboa, Observat\'orio Astron\'omico de Lisboa, Tapada da Ajuda, 1349-018, Lisbon, Portugal}
\altaffiltext{26}{Department of Physics \& Astronomy, University of British Columbia, 6224 Agricultural Road, Vancouver, BC V6T~1Z1, Canada}
\altaffiltext{27}{Institute for Computational Cosmology, Durham University, South Road, Durham DH1 3LE, UK}


\label{firstpage}

\begin{abstract}
We present Keck-Adaptive Optics and {\it Hubble Space Telescope} high resolution near-infrared (IR) imaging for $500\,\mu$m-bright candidate lensing systems identified by the {\it Herschel} Multi-tiered Extragalactic Survey (HerMES) and {\it Herschel} Astrophysical Terahertz Survey (H-ATLAS). Out of 87 candidates with near-IR imaging, 15 ($\sim17\%$) display clear near-IR lensing morphologies. We present near-IR lens models to reconstruct and recover basic rest-frame optical morphological properties of the background galaxies from 12 new systems. Sources with the largest near-IR magnification factors also tend to be the most compact, consistent with the size bias predicted from simulations and previous lensing models for sub-millimeter galaxies. For four new sources that also have high-resolution sub-mm maps, we test for differential lensing between the stellar and dust components and find that the $880\,\mu$m magnification factor ($\mu_{880}$) is $\sim1.5$ times higher than the near-IR magnification factor ($\mu_{{\rm NIR}}$), on average. We also find that the stellar emission is $\sim2$ times more extended in size than dust. The rest-frame optical properties of our sample of \herschel-selected lensed SMGs are consistent with those of unlensed SMGs, which suggests that the two populations are similar. 
\end{abstract}

\section{Introduction}
Dusty star-forming galaxies~\citep[DSFGs; For a recent review, see ][]{Casey14}, selected for being bright in the infrared or sub-mm regimes, are responsible for the bulk of cosmic star-formation in the early Universe~\citep[e.g.][]{LeFloch05, Takeuchi05}. Sub-millimeter galaxies (SMGs, \citealt{Smail97, Hughes98, Barger98} and see \citealt{Blain02} for a review), an $850-880\,\mu$m-bright subset of the DSFG population, present an appealing opportunity to study an important phase in galaxy evolution at the peak of cosmic star-formation. The negative $K$-correction in the Rayleigh-Jeans tail of thermal dust emission at the (sub-)mm regime forms an approximately constant infrared (IR) luminosity limit across a wide range in redshift ($z=1-8$). This effectively allows SMGs to be readily detected in sub-mm surveys. Since their discovery 17 years ago, we have learned that SMGs are massive ($M_{\ast}\sim10^{11}$~{\msun}; \citealt{Michalowski10,Hainline11,Bussmann12,Targett13}), gas-rich ($M_{\rm{gas}}\sim10^{10-11}$~{\msun}; \citealt{Greve05, Tacconi08, Ivison11, Bothwell13}), and metal-rich ($Z \sim Z_\odot$; \citealt{Swinbank04}) galaxies at a median redshift of $z\sim2.5$ \citep{Chapman05} that could be undergoing a short burst of star-formation ($t\sim50-100$ Myr; \citealt{Tacconi08,Narayanan10,Lapi11,Hickox12,Simpson13}). They have the most extreme star-formation rates, which can be as high as $10^{3}~{\rm M}_{\odot}~{\rm yr}^{-1}$ and compose $20-30\%$ of the total comoving star-formation rate density ($\rho_{\rm{SFR}}$) at $z\sim2.5$ \citep{Chapman05,Wardlow11,Casey13}. This is comparable to the total contribution of mid-IR selected galaxies at the same epoch, although SMGs are fewer in number but have larger IR luminosities~\citep[e.g.][]{Farrah08,Hernan09,Calanog13}. 
 
From an evolutionary standpoint, it has long been proposed that ultra-luminous infrared galaxies (ULIRGs, $L_{\rm IR}\ge10^{12}~{\rm L}_{\odot}$), which include SMGs, is an intense star-forming phase that precedes the growth of the AGN hosted by massive elliptical galaxies \citep{Sanders88}. Multiple lines of evidence  suggest that SMGs are the likely progenitors of massive elliptical galaxies \citep{Lilly99, Swinbank06, Tacconi08, Michalowski10, Lapi11, Hickox12, Toft14}. For instance, $\le30\%$ of SMGs are known to harbor AGN, supporting formation scenarios in which massive elliptical galaxies evolve from a quasar-dominated phase \citep{Alexander03,Pope08,Coppin10}. Furthermore, clustering analyses indicate that SMGs are hosted by $10^{13}${\msun} dark matter halos and have space densities of $\sim10^{-5}~{\rm Mpc}^{-3}$, consistent with optically-selected quasars at $z\sim2$ and $2-3~L^{\ast}$ elliptical galaxies at $z\sim0$ \citep[e.g.][]{Blain04b,Farrah06,Hickox12}. 

While our knowledge of SMGs have definitely advanced, their dominant formation mechanism is still unclear. One picture proposes that SMGs are a result of gas-rich major-mergers~\citep{Tacconi06,Schinnerer08,Tacconi08, Engel10} while another favors them as being extreme analogues of normal star-forming galaxies, fed with gas through minor mergers and smooth infall \citep{Finlator06, Dekel09, Dave10}. Observational studies that focus on SMG morphologies can help clarify this issue, and would require analysis in wavelength regimes that trace the constituent gas, dust, and stars. However, SMG morphologies are difficult to study with current instruments because of poor spatial resolution, insufficient sensitivity, or both. Here, we circumvent these difficulties by studying SMGs that are strongly gravitationally lensed. The lensed background source receives a boost in apparent flux by a factor of $\mu$, where $\mu$ is the magnification factor, enabling the study of emission that would otherwise be too faint to detect. In addition, the apparent size of the background source is increased by a factor of $\sim\sqrt{\mu}$  (Schneider 1992) -- allowing high-spatial resolution studies of the lensed galaxies, even if they are at high redshift. 

The obvious benefits of studying SMGs via gravitational lensing sparked interest in producing an efficient and straight-forward method to identify strong-lensing events. Efficient strong lensing event identification through bright source selection in wide-area extragalactic sub-mm/mm surveys has been long proposed \citep{Blain96, Perrotta02, Negrello07, Paciga09}.  The idea behind this selection method exploits the fact that sources that are intrinsically sub-mm bright are also very rare~\citep[e.g. see ][]{Weiss09}. This implies that a significant fraction of the sub-mm bright population could be lensed and flatten the observed declining number counts at large flux densities. This flattening however, could also be caused by contaminants such as local late-type spiral galaxies and flat spectrum radio quasars \citep{Negrello07} which can be removed trivially through optical and radio surveys (e.g. SDSS, \citealt{Abazajian03}; NVSS, \citealt{Condon98}). Thus, after removing such contaminants, a large fraction of the brightest sub-mm sources are expected to be strongly lensed and lie at $z\ge1$. 

The launch of the {\it Herschel Space Observatory}\footnote{Herschel is an ESA space observatory with science instruments provided by European-led Principal Investigator consortia and with important participation from NASA.} \citep{Pilbratt10} ushered in the possibility of confirming these theoretical predictions. Indeed, the two largest wide-area sub-mm surveys, the {\it Herschel} Multi-Tiered  Extragalactic Survey  (HerMES, \citealt{Oliver12}) and the {\it Herschel} Astrophysical Terahertz Large Area Survey (H-ATLAS, \citealt{Eales10}) have provided the first samples of candidate lensing systems by selecting $500\,\mu$m-bright sources. Since then, high-resolution, spatially-resolved multi-wavelength follow-up observations have confirmed that a large fraction ($70-100\%$) of these candidates are undoubtedly lensed \citep{Negrello10,Gavazzi11,Bussmann12,Wardlow13,Bussmann13}. 

This paper focuses on studying the background lensed galaxies with new high-resolution near-IR data for 87 $500\,\mu$m-bright candidate lensing systems discovered by H-ATLAS and HerMES. A comprehensive analysis of the properties of the foreground lenses is deferred to a future publication (Amber et al., in prep.). Near-IR observations of \herschel-selected $500\,\mu$m-bright lensed SMGs allow one to characterize the stellar distribution at spatial resolutions that are unachievable with the current facilities. Furthermore, since classically-selected SMGs are $850-880\,\mu$m-bright, we can directly compare their rest-frame optical properties, such as their luminosities, against the $500\,\mu$m-bright population. This comparison can help clarify any differences between these two SMG populations, which can potentially arise from their sub-mm selections. Aside from their rest-frame optical luminosities, the morphological information recovered from reconstructing the background galaxy can also be used to compare against previous studies of unlensed SMGs~\citep{Swinbank10,Targett11,Targett13,Aguirre13}.  In this context, the morphological  study of lensed SMGs at an unprecedented spatial resolution can provide observational evidence to determine the formation mechanisms that are present. Finally, these high-resolution near-IR observations compliments previous studies done on lensed SMGs using high-resolution sub-mm facilities \citep{Bussmann13,Weiss13,Hezaveh13,Vieira13}. Any sources that overlap between the near-IR and the sub-mm can be used to study the morphologies, spatial distribution, and the effects of differential magnification between the older stellar population and the dust-emitting star-forming regions of the same galaxy. 

 All of the candidate lensing systems in this paper have been observed using either the {\it Hubble Space Telescope's} ({\it HST}) Wide Field Camera 3 (WFC3) in the $J$ band (F110W, $\lambda$=$1.15\,\mu$m) or Keck II Near-Infrared Camera 2 (NIRC2) with laser guide star adaptive optics system (LGS-AO, \citealt{Wizinowich06}) in the $K$ ($\lambda$=$2.2\,\mu$m) band. We model the lensing in 12 galaxy-scale lensing systems with new near-IR data that have high-significance lensing morphology detections and sufficiently constrained configurations. From our lens models, we determine the magnification in the near-IR and the source-plane emission regions. Of these 12, six of the systems were also studied in the sub-mm by~\citet{Bussmann13}. By comparing the lensing in the sub-mm and near-IR, we quantify the effects of differential lensing and measure the size difference of stellar and dust components. Using our near-IR data and lens models, we measure the intrinsic photometry for lensed SMGs and estimate their rest-frame absolute $B$-band magnitudes.

\begin{figure}
\begin{minipage}[b]{0.50\textwidth}
\includegraphics[width=\textwidth,keepaspectratio]{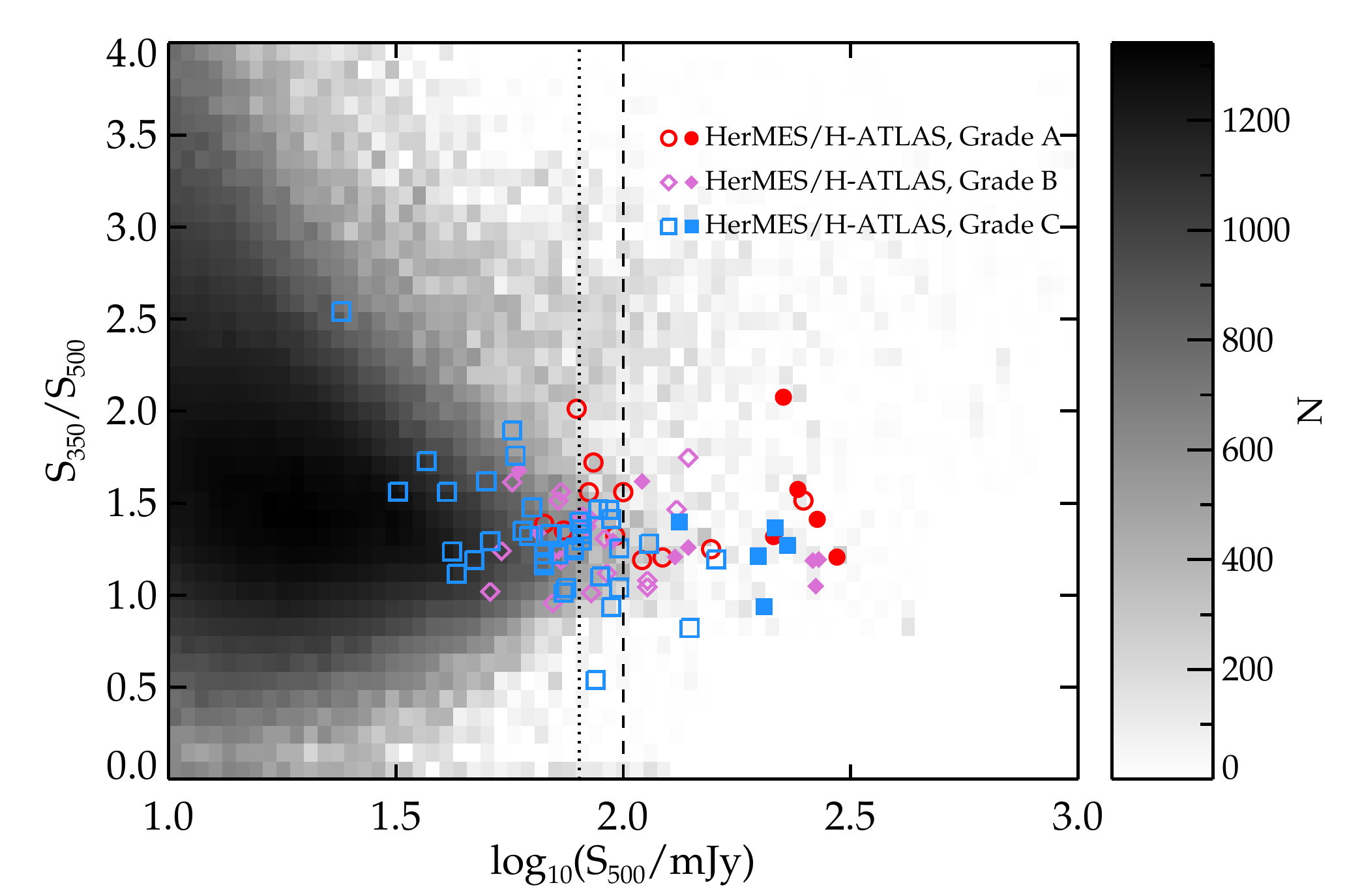}
\caption{$S_{350}/S_{500}$ as a function of $S_{500}$ for SPIRE galaxies in HerMES and H-ATLAS.
Open and filled symbols correspond to HerMES and H-ATLAS candidate lensing systems with high-resolution near-IR imaging, respectively. Red circles, violet diamonds and blue squares are assigned the Grade A, B, and C, respectively, on the basis of their near-IR lensing morphologies, as discussed in Section~\ref{sec:cl}. The vertical dotted and dashed lines correspond to $S_{500}=80$ and 100 mJy. HerMES lensing systems with $S_{500}\le80\,$mJy were selected from an initial source catalog and here we show the most updated $S_{500}$ value. The majority of the targeted candidate lensing systems are biased towards larger $500\,\mu$m flux densities, but have $S_{350}/S_{500}$ ratios similar to the fainter population.}
\label{fig:fircolorvsflux}
\end{minipage}
\end{figure}

This paper is organized as follows. In Section~\ref{sec:data}, we summarize the sub-mm lensed candidate selection and describe our high-resolution near-IR observations and data reduction process. Our classification of candidate lensing systems is presented in Section~\ref{sec:cl}. Section~\ref{sec:lm} describes our lens modeling methodology and individual notes on each strong lensing system. We then discuss our results and compare them with previous studies of both lensed and unlensed SMGs in Section~\ref{sec:res}. Finally, we summarize our findings and conclusions in Section~\ref{sec:conc}. 

We adopt a $\Lambda$CDM cosmology, with $H_{o}=70\,$km s$^{-1}$ Mpc$^{-1}$, $\Omega_{\rm{m}}=0.3$, and $\Omega_{\Lambda}=0.7$. Unless otherwise stated, all magnitudes reported are in the AB system~\citep{Fukugita96}.

\label{sec:intro}

\section{Lensed Candidate Selection and near-IR Observations}

\label{sec:data}
In this section we summarize the selection criteria used to define our sample and describe the data acquisition and reduction of our high-resolution near-IR imaging of the galaxies. A summary of all the targets observed, along with their integration times and observation dates are found on Table~\ref{tab:nirdata}. Of the 87 near-IR targets, 49 (56\%) HerMES/H-ATLAS sources are observed with Keck/NIRC2-LGS-AO, 42 (48\%) HerMES sources with {\it HST}/WFC3 F110W (with 15 (17\%) HerMES sources observed using both instruments).

\subsection{Selection of Candidate Lensing Systems}
The targets of this study are selected from the Spectral and Photometric Imaging REceiver (SPIRE, \citealt{Griffin10}) maps in the HerMES~\citep{Oliver12} and H-ATLAS~\citep{Eales10} fields. Targets are identified in the same way in both surveys, using the SPIRE $500\,\mu$m channel to minimize the number of contaminants \citep{Negrello07,Negrello10}. The {\it Herschel}-SPIRE data reduction and photometry procedures differ slightly for each survey, with the main difference being that HerMES accounts for blending from positional priors that can result in detecting fainter objects while H-ATLAS only retains sources above $5\sigma$. Even with this difference, the $500\,\mu$m number counts appear consistent \citep{Oliver10,Clements10}. Full details of the H-ATLAS map-making data reduction and source extraction are presented in \citet{Pascale11} and \citet{Rigby11}. For HerMES, see \citet{Levenson10}, \citet{Roseboom10}, and \citet{Smith12}, with updates in \citet{Viero13} and \citet{Wang13}. Both procedures are summarized below. 

For HerMES, SPIRE maps were generated using the SPIRE-HerMES Iterative Mapper (\texttt{SHIM}) algorithm \citep{Levenson10}. The most updated point-source catalogues use an iterative source-detection scheme of \texttt{STARFINDER} \citep{Diolaiti00} and De-blend SPIRE Photometry (\texttt{DESPHOT}) algorithm \citep{Roseboom10,Roseboom12,Wang13}. \texttt{STARFINDER} is used to detect and find the optimal positions of point sources in SPIRE maps by assuming that the observed images can be modeled as a superposition of point-response functions (PRF). These source positions are then used as inputs for \texttt{DESPHOT} to perform map segmentation (de-blending), source photometry, background estimation and noise (instrumental and confusion) estimation. 

For sources identified by H-ATLAS fields, source extraction is performed using the Multi-band Algorithm for Source eXtraction (\texttt{MADX}; Maddox et al. in prep) on {\it Herschel} Interactive Processing Environment (HIPE) generated SPIRE maps \citep{Pascale11}. \texttt{MADX} iteratively performs PSF fitting and subtraction to measure flux densities and positions for each band. Sources that are detected at $\ge5\sigma$ (including confusion noise of $\approx6$ mJy at all bands, \citealt{Nguyen10}) in any of the bands are retained in the final catalogues. 

\LongTables
\begin{deluxetable*}{lccc}
\small
\tablecolumns{4}
\tablewidth{0pt}
\tablecaption{Summary of High Resolution Data} 
\tablehead{
\colhead{IAU Name} & 
\colhead{Short Name} & 
\colhead{Exp. Time} & 
\colhead{Depth\tnm{d}} 
\\
\colhead{} & 
\colhead{} & 
\colhead{Filter\tnm{a}  = $t_{\rm{int}}$\tnm{b} $\times N_{\rm{frames}}$ \tnm{c}} &
\colhead{AB  mag}  
}
\startdata
1HerMES S250 J002854.0-420457 & HELAISS04 & $J = 62 \times 4$ & $J = 25.8$ \\ 
1HerMES S250 J002906.3-421420 & HELAISS01 & $J = 62 \times 4$ & $J = 25.4$ \\ 
1HerMES S250 J003823.7-433705 & HELAISS02 & $J = 125 \times 4$ & $J = 25.7$ \\ 
1HerMES S250 J021620.0-032520 & HXMM26 & $K_{\rm p} = 60 \times 30$ & $K_{\rm p} = 25.6$\tnm{e}\\ 
1HerMES S250 J021632.1-053422 & HXMM14 & $J = 125 \times 4$ & $J = 25.6$ \\ 
1HerMES S250 J021830.6-053125 & HXMM02 & $J = 177 \times 4$, $K_{\rm p} = 60 \times 18$ & $J = 26.3$, $K_{\rm p} = 25.6$\tnm{e}\\ 
1HerMES S250 J021836.7-035316 & HXMM13 & $J = 62 \times 4$ & $J = 25.6$ \\ 
1HerMES S250 J021942.9-052433 & HXMM20 & $J = 125 \times 4$ & $J = 25.6$ \\ 
1HerMES S250 J022016.6-060144 & HXMM01 & $J = 62 \times 4$, $K_{\rm s} = 80 \times 35$ & $J = 25.5$, $K_{\rm s} = 25.6$ \\ 
1HerMES S250 J022021.8-015329 & HXMM04 & $J = 62 \times 4$ & $J = 25.6$ \\ 
1HerMES S250 J022029.2-064846 & HXMM09 & $J = 62 \times 4$, $H = 120 \times 12$, $K = 80 \times 15$ & $J = 25.2$, $H = 24.8$, $K = 24.5$ \\ 
1HerMES S250 J022135.2-062618 & HXMM03 & $J = 62 \times 4$ & $J = 25.4$ \\ 
1HerMES S250 J022201.7-033340 & HXMM11 & $K_{\rm s} = 100 \times 18$ & $K_{\rm s} = 25.6$\tnm{e}\\ 
1HerMES S250 J022205.5-070727 & HXMM23 & $J = 62 \times 4$ & $J = 25.2$ \\ 
1HerMES S250 J022212.9-070224 & HXMM28 & $J = 125 \times 4$ & $J = 25.6$ \\ 
1HerMES S250 J022250.8-032414 & HXMM22 & $J = 62 \times 4$ & $J = 25.4$ \\ 
1HerMES S250 J022515.3-024707 & HXMM19 & $J = 62 \times 4$ & $J = 25.3$ \\ 
1HerMES S250 J022517.5-044610 & HXMM27 & $J = 62 \times 4$ & $J = 25.6$ \\ 
1HerMES S250 J022547.9-041750 & HXMM05 & $J = 62 \times 4$ & $J = 25.8$\\ 
1HerMES S250 J023006.0-034153 & HXMM12 & $J = 62 \times 4$ & $J = 25.2$ \\ 
1HerMES S250 J032434.4-292646 & HECDFS08 & $J = 62 \times 4$ & $J = 25.4$ \\ 
1HerMES S250 J032443.1-282134 & HECDFS03 & $J = 125 \times 4$ & $J = 25.4$ \\ 
1HerMES S250 J032636.4-270045 & HECDFS05 & $J = 62 \times 4$ & $J = 25.6$ \\ 
1HerMES S250 J032712.7-285106 & HECDFS09 & $J = 62 \times 4$ & $J = 25.5$ \\ 
1HerMES S250 J033118.0-272015 & HECDFS11 & $J = 62 \times 4$ & $J = 25.3$ \\ 
1HerMES S250 J033210.8-270536 & HECDFS04 & $J = 62 \times 4$ & $J = 26.0$ \\ 
1HerMES S250 J033732.5-295353 & HECDFS02 & $J = 177 \times 4$ & $J = 26.8$ \\ 
1HerMES S250 J043340.5-540338 & HADFS04 & $J = 62 \times 4$ & $J = 25.6$ \\ 
1HerMES S250 J043829.8-541832 & HADFS02 & $J = 62 \times 4$ & $J = 25.7$ \\ 
1HerMES S250 J044154.0-540351 & HADFS01 & $J = 62 \times 4$ & $J = 25.5$ \\ 
1HerMES S250 J044946.6-525427 & HADFS09 & $J = 125 \times 4$ & $J = 25.3$ \\ 
1HerMES S250 J045027.1-524126 & HADFS08 & $J = 62 \times 4$ & $J = 25.1$ \\ 
1HerMES S250 J045057.6-531654 & HADFS03 & $J = 62 \times 4$ & $J = 25.3$ \\ 
HATLASJ083051.0+013224 & G09v1.97 & $K_{\rm s} = 80 \times 41$ & $K_{\rm s} = 25.5$ \\ 
HATLASJ084933.4+021443 & G09v1.124 & $K = 80 \times 17$ & $K = 24.5$ \\ 
HATLASJ084957.6+010712 & G09v1.1259 & $K_{\rm s} = 80 \times 30$ & $K_{\rm s} = 25.7$ \\ 
HATLASJ085358.9+015537 & G09v1.40 & $K_{\rm s} = 80 \times 45$ & $K_{\rm s} = 26.2$ \\ 
HATLASJ090319.6+015636 & SDP.301 & $K_{\rm s} = 80 \times 26$ & $K_{\rm s} = 25.7$ \\ 
HATLASJ090542.1+020734 & SDP.127 & $K_{\rm s} = 80 \times 24$ & $K_{\rm s} = 25.4$ \\ 
HATLASJ091840.8+023047 & G09v1.326 & $K_{\rm s} = 80 \times 41$ & $K_{\rm s} = 25.9$ \\ 
1HerMES S250 J100030.6+024142 & HCOSMOS03 & $K_{\rm s} = 80 \times 45$ & $K_{\rm s} = 25.6$\tnm{e}\\ 
1HerMES S250 J100057.1+022010 & HCOSMOS02 & $J = 177 \times 4$, $K_{\rm s} = 80 \times 45$ & $J = 26.3$, $K_{\rm s} = 25.6$\tnm{e}\\ 
1HerMES S250 J100144.2+025712 & HCOSMOS01 & $J = 62 \times 4$, $K_{\rm s} = 80 \times 23$ & $J = 25.4$, $K_{\rm s} = 25.6$\tnm{e}\\ 
1HerMES S250 J103330.0+563315 & HLock15 & $J = 125 \times 4$ & $J = 25.5$ \\ 
1HerMES S250 J103618.5+585456 & HLock05 & $J = 62 \times 4$, $K_{\rm s} = 80 \times 44$ & $J = 26.0$, $K_{\rm s} = 25.6$\tnm{e}\\ 
1HerMES S250 J103826.6+581543 & HLock04 & $J = 62 \times 4$, $H = 120 \times 30$, $K_{\rm s} = 80 \times 33$ & $J = 25.6$, $H = 25.5$, $K_{\rm s} = 25.2$ \\ 
1HerMES S250 J103957.8+563120 & HLock17 & $J = 62 \times 4$ & $J = 25.6$ \\ 
1HerMES S250 J104050.6+560653 & HLock02 & $J = 62 \times 4$ & $J = 25.9$ \\ 
1HerMES S250 J104140.3+570858 & HLock11 & $J = 177 \times 4$, $K_{\rm s} = 80 \times 40$ & $J = 26.4$, $K_{\rm s} = 26.1$ \\ 
1HerMES S250 J104549.2+574512 & HLock06 & $J = 62 \times 4$, $K_{\rm s} = 80 \times 34$ & $J = 25.5$, $K_{\rm s} = 25.6$ \\ 
1HerMES S250 J105551.4+592845 & HLock08 & $J = 62 \times 4$ & $J = 25.7$ \\ 
1HerMES S250 J105712.2+565458 & HLock03 & $J = 62 \times 4$, $K_{\rm s} = 80 \times 41$ & $J = 26.2$, $K_{\rm s} = 25.8$ \\ 
1HerMES S250 J105750.9+573026 & HLock01 & $J = 62 \times 4$, $K_{\rm p} = 64 \times 15$, $K_{\rm s} = 80 \times 12$ & $J = 25.5$, $K_{\rm p} = 25.4$, $K_{\rm s} = 25.6$\tnm{e}\\ 
1HerMES S250 J110016.3+571736 & HLock12 & $J = 62 \times 4$ & $J = 25.9$ \\ 
HATLASJ113526.4-014606 & G12v2.43 & $K_{\rm s} = 80 \times 26$ & $K_{\rm s} = 26.0$ \\ 
HATLASJ114638.0-001132 & G12v2.30 & $K_{\rm s} = 80 \times 42$ & $K_{\rm s} = 25.3$ \\ 
HATLASJ115101.8-020024 & G12v2.105 & $K_{\rm s} = 80 \times 26$ & $K_{\rm s} = 25.$7 \\ 
HATLASJ132426.9+284452 & NB.v1.43 & $H = 120 \times 14$, $K_{\rm s} = 80 \times 48$ & $H = 25.6$, $K_{\rm s} = 26.0$ \\ 
HATLASJ132630.1+334410 & NA.v1.195 & $K_{\rm s} = 80 \times 35$ & $K_{\rm s} = 25.9$ \\ 
HATLASJ132859.3+292327 & NA.v1.177 & $K_{\rm s} = 80 \times 28$ & $K_{\rm s} = 25.9$ \\ 
HATLASJ133008.3+245900 & NB.v1.78 & $H = 120 \times 9$, $K_{\rm s} = 80 \times 20$ & $H = 25.5$, $K_{\rm s} = 25.7$ \\ 
HATLASJ133255.8+342209 & NA.v1.267 & $K_{\rm s} = 80 \times 42$ & $K_{\rm s} = 26.4$ \\ 
HATLASJ141351.9-000026 & G15v2.235 & $K_{\rm s} = 80 \times 16$ & $K_{\rm s} = 25.3$ \\ 
1HerMES S250 J142201.4+533214 & HEGS01 & $J = 125 \times 4$ & $J = 26.1$ \\ 
HATLASJ142413.9+022303 & G15v2.779 & $K_{\rm s} = 80 \times 27$ & $K_{\rm s} = 25.4$ \\ 
1HerMES S250 J142557.6+332547 & H{\bootes}09 & $J = 62 \times 4$ & $J = 25.5$ \\ 
1HerMES S250 J142650.6+332943 & H{\bootes}04 & $K_{\rm s} = 80 \times 36$ & $K_{\rm s} = 25.8$ \\ 
1HerMES S250 J142748.7+324729 & H{\bootes}11 & $K_{\rm s} = 80 \times 35$ & $K_{\rm s} = 25.4$ \\ 
1HerMES S250 J142824.0+352620 & H{\bootes}03 & $J = 62 \times 4$ & $J = 25.6$ \\ 
1HerMES S250 J142825.7+345547 & H{\bootes}02 & $J = 62 \times 4$, $H = 120 \times 28$, $K_{\rm s} = 80 \times 27$ & $J = 25.6$, $H = 25.9$, $K_{\rm s} = 25.2$ \\ 
HATLASJ142935.3-002836 & G15v2.19 & $H = 120 \times 10$, $K_{\rm s} = 80 \times 15$ & $H = 25.6$, $K_{\rm s} = 25.2$ \\ 
1HerMES S250 J143204.9+325908 & H{\bootes}10 & $K_{\rm s} = 80 \times 46$ & $K_{\rm s} = 25.3$ \\ 
1HerMES S250 J143330.7+345439 & H{\bootes}01 & $J = 62 \times 4$ & $J = 25.5$ \\ 
1HerMES S250 J143543.5+344743 & H{\bootes}12 & $J = 62 \times 4$, $K_{\rm s} = 80 \times 36$ & $J = 25.5$, $K_{\rm s} = 25.9$ \\ 
1HerMES S250 J143702.0+344635 & H{\bootes}08 & $K_{\rm s} = 80 \times 36$ & $K_{\rm s} = 25.8$ \\ 
1HerMES S250 J144015.7+333055 & H{\bootes}13 & $K_{\rm s} = 80 \times 37$ & $K_{\rm s} = 25.9$ \\ 
1HerMES S250 J144029.8+333845 & H{\bootes}07 & $K_{\rm s} = 80 \times 36$ & $K_{\rm s} = 25.9$ \\ 
HATLASJ144556.1-004853 & G15v2.481 & $K_{\rm s} = 80 \times 34$ & $K_{\rm s} = 26.0$ \\ 
1HerMES S250 J161331.4+544359 & HELAISN01 & $J = 125 \times 4$ & $J = 25.4$ \\ 
1HerMES S250 J161334.4+545046 & HELAISN04 & $K_{\rm s} = 80 \times 45$ & $K_{\rm s} = 25.6$ \\ 
1HerMES S250 J170507.6+594056 & HFLS07 & $J = 62 \times 4$ & $J = 25.5$ \\ 
1HerMES S250 J170607.7+590922 & HFLS03 & $J = 62 \times 4$ & $J = 26.7$ \\ 
1HerMES S250 J170817.6+582845 & HFLS05 & $J = 125 \times 4$ & $J = 24.5$ \\ 
1HerMES S250 J171450.9+592634 & HFLS02 & $J = 62 \times 4$ & $J = 25.3$ \\ 
1HerMES S250 J171544.9+601239 & HFLS08 & $J = 62 \times 4$ & $J = 25.5$ \\ 
1HerMES S250 J172222.3+582609 & HFLS10 & $J = 355 \times 4$, $K_{\rm s} = 80 \times 18$ & $J = 26.5$, $K_{\rm s} = 25.1$ \\ 
1HerMES S250 J172612.0+583743 & HFLS01 & $J = 177 \times 4$ & $J = 25.2$
\enddata
\tnt{a}{Filters are $J$ = HST F110W, $H$ = Keck H-band, $K_{s}$ = Keck Ks band, $K$ = Keck K-band, and $K_{p}$ = Keck Kp-band.}
\tnt{b}{$t_{\rm{int}}$ is exposure time per frame}
\tnt{c}{$N_{\rm{frames}}$ is number of independent frames}
\tnt{d}{$5\sigma$ point-source depth calculated using the specifications outlined in Section~\ref{sec:nirc2} and~\ref{sec:wfc3}.}
\tnt{e}{Depth calculated using average zero point ($\Delta m_{\rm{zpt}}=0.4$) due to absence of a suitable point source in the frame.}
\label{tab:nirdata} 
\end{deluxetable*}


In both surveys lensing candidates are selected by applying a high flux cut at $500\,\mu$m, which for H-ATLAS is $S_{500} \ge 100$ mJy~\citep{Negrello10}, where $S_{500}$ is the $500\,\mu$m flux density, and for HerMES is $S_{500}\ge80$ mJy~\citep{Wardlow13}. Sources that are not associated with local late-type galaxies or flat-spectrum radio galaxies are retained as lensing candidates. The targeted sources are presented in Table~\ref{tab:obsprops}, along with their SPIRE 250, 350 and 500$\,\mu$m flux densities and redshifts. 

We should also clarify that our selection in HerMES at $S_{500}\ge80$ mJy was applied on an initial source catalog, extracted from blind detections using \texttt{SUSSEXtractor} \citep{Savage07,Smith12}, but subsequent iterations of HerMES data products resulted in better deblending of $500\,\mu$m flux densities with $250\,\mu$m positions as a prior \citep{Wang13}. This results in some of the sources initially categorized as candidate lensing systems (having $S_{500}\ge80$ mJy), with a final lower probability of being lensed at $\le40\%$, based on the statistical models of \citet{Wardlow13} that uses the foreground lensing matter distribution, unlensed SMG number counts, and an assumed SMG redshift distribution. As a result, some are confirmed as bonafide lenses and we keep them in our sample, as they have been followed-up but we exclude them for statistics involving lensed SMGs at the bright $500\,\mu$m flux densities. 

Figure~\ref{fig:fircolorvsflux} shows $S_{500}$ as a function of the flux density ratio $S_{350}/S_{500}$ for the targeted candidate lensing systems with high-resolution near-IR imaging. By design, our targeted sources are biased towards those that are brightest at 500$\,\mu$m, although they have similar 350/500$\,\mu$m colors (with $S_{350}/S_{500}\ge1$ for most systems) to the bulk of the SPIRE population. This indicates that \herschel-selected lensed galaxies and the SPIRE population have similar far-IR SED shapes, dust temperatures, and redshift distribution but will have larger apparent IR luminosities due to flux boosting effects from lensing~\citep{Wardlow13,Bussmann13}. 

\subsection{Keck NIRC2/LGS-AO}
\label{sec:nirc2}
We have obtained Keck/NIRC2 LGS-AO imaging for {\it Herschel}-candidate lensing systems. Conditions were typically good, characterized by clear skies and seeing values of $\sim0.8''$ from our successful observing runs from 2011 to 2013. We observe our targets primarily using the $K_{\rm s}$~filter ($\lambda_{\rm c} = 2.2\,\mu$m), mainly because Keck-AO performs the best at longer wavelengths and $K_{\rm s}$ gives the optimal sensitivity because the background is minimal at this wavelength \citep{Simons02}. Typical integration times for each source are $\sim45$ minutes to acquire a $5\sigma$ point source depth of 25.7 AB using a $0.1''$ aperture radius. We use the wide camera that has a $40''\times40''$ field of view and sub-arcsecond dithering steps. The spatial resolution with AO correction reaches $0.1''$ in the best conditions and the estimated Strehl ratios were $\sim15-25\%$. Some of the targets showing clear signs of lensing, are also observed in the $H$ ($\lambda_{\rm c}=1.6\,\mu$m) band. However, we do our lens modeling (Section~\ref{sec:lm}) only in the $K$ band where the signal to noise is at its highest. We used custom \texttt{IDL} scripts to reduce the images, following standard procedures \citep{Fu12,Fu13}. Briefly, after bad pixel masking, background subtraction, and flat-fielding, sky background and object masks were updated iteratively. For each frame, after subtracting a scaled median sky, the residual background was removed with 2-dimensional B-spline models. In the last iteration, we discard frames of the poorest image quality and correct the camera distortion using the on-sky distortion solution from observations of the globular cluster M92\footnote{http://www2.keck.hawaii.edu/inst/nirc2/dewarp.html}.
Since image distortion has been removed in previous steps, astrometry is calibrated against four to five non-saturated SDSS sources in the final mosaicked field of view with a linear offset. The mean offset is weighted by the S/N of the sources, so that offsets computed from brighter sources carry more weight. 

The NIRC2 images are flux calibrated against UKIDSS $K_{\rm s}$-band photometry, when available. Each frame is PSF matched and corrected for airmass and we use the UKIDSS aperture radius of $1''$ to perform our calibration. Photometric zero points are derived by calculating the magnitude difference for overlapping sources. For NIRC2 frames that do not overlap with UKIDSS footprints, we use the night-averaged zero point and its standard deviation to account for the associated systematic error. 

For the PSF used in our lens modeling analysis (Section~\ref{sec:lm}), we use a nearby unsaturated point source, whenever available. Otherwise, point sources from other images observed on the same day are used, while keeping the airmass difference within 0.2 and applying the appropriate rotation. 

\begin{figure}
\begin{minipage}[b]{0.5\textwidth}
\begin{center}
\includegraphics[width=\textwidth,height=\textheight,keepaspectratio]{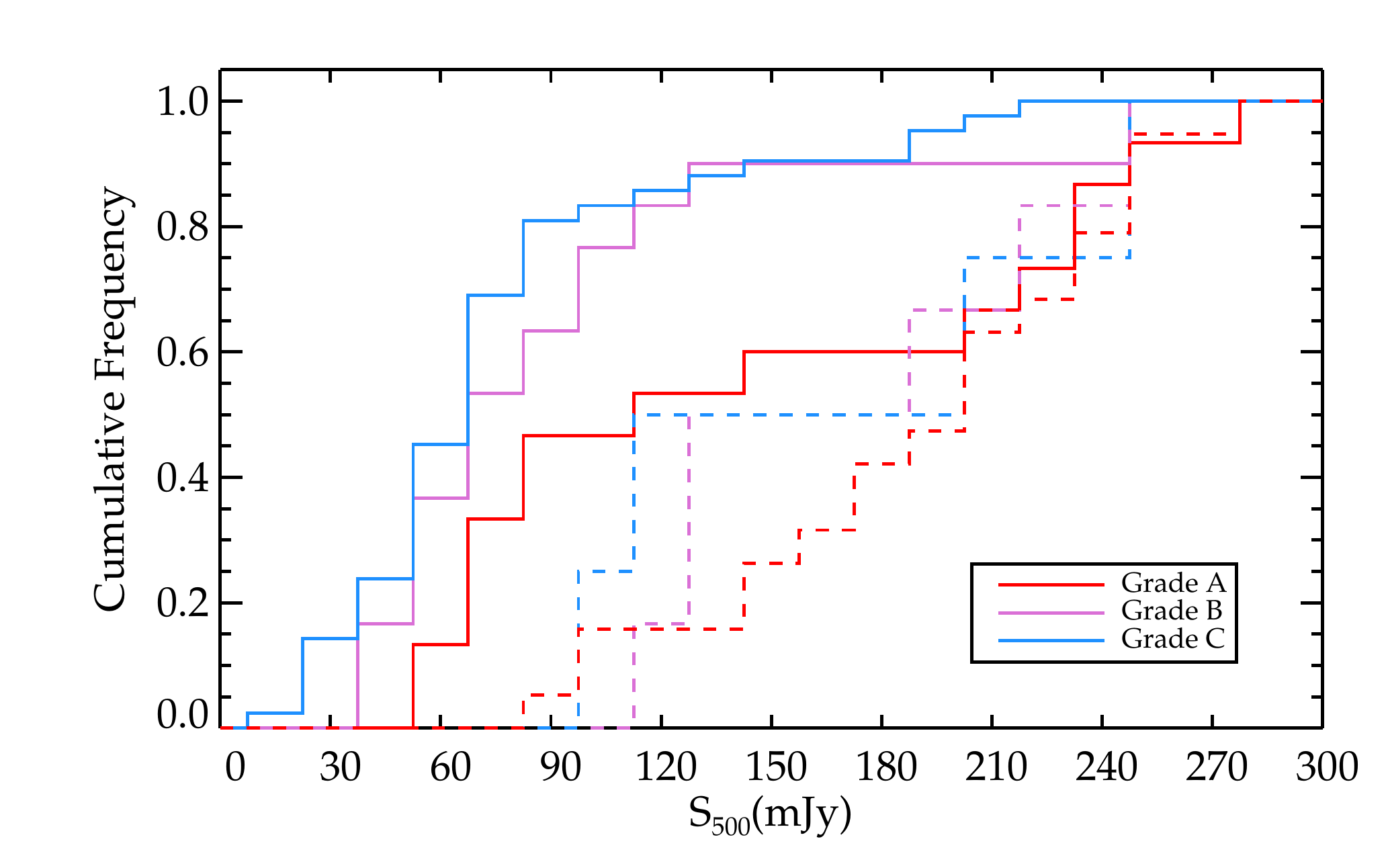}
\caption{Cumulative frequency distribution of $S_{500}$ flux densities for the near-IR subset of SMG candidate lensing systems. The solid red, purple, and blue lines represent Grade A, B, and C sources respectively. For comparison, the dashed lines are from the sub-mm sample from \citet{Bussmann13}, converted to the same near-IR grading scheme. Grade A sources tend to have smaller cumulative fractions than Grade B and C sources with increasing values of $S_{500}$, which supports the idea that $500\,\mu$m-bright sources have a higher probability of being lensed.}
\label{fig:s500hist}
\end{center}
\end{minipage}
\end{figure}

\subsection{{\it HST}/WFC3} 
\label{sec:wfc3}
{\it Herschel}-lensing candidates in the HerMES fields have also been observed as part of the {\it HST} WFC3 Cycle 19 snapshot program (P.I. M. Negrello). All are observed with the F110W filter ($\lambda_{\rm c}=1.15\,\mu$m), using a 4-point parallelogram dither pattern with point and line spacings of $0.57''$ and $0.36''$, respectively. Most of the images have a total integration time of 4 minutes per target, while a few sources that have red SPIRE colors ($S_{500} \ge S_{350}$) have doubled integration times, because these sources could be at higher redshifts and thus likely fainter the in near-IR \citep{Dowell14}. 

The~\texttt{calwfc3}~processed flat-fielded data from the {\it HST}/WFC3 pipeline are used as inputs for \texttt{multidrizzle} \citep{Koekemoer03}, producing an output image with a pixel scale of 0.04$''$ to allow adequate sampling of the PSF and to match the pixel scale of the Keck images. Due to some fields being crowded by bright sources, we turn off sky subtraction on all WFC3 frames and set the drop size parameter, ``pixfrac" = 1, in order to minimize additional noise due to sky variations. We set the ``bits'' parameter to the value of 4608 to include pipeline-rejected pixels and dust motes, since our dithering pattern is not large enough to fill in these regions with good data. To account for the uncertainty in each pixel value, an error map is generated to account for the RMS value of the sky and the Poisson error each pixel. The resulting output images have a spatial resolution of $\sim0.2''$ and an average $5\sigma$ point source depth of 25.4 and 26.2 AB mag for integration times of 4 and 8 minutes, using a $0.2''$ aperture radius. 

We use a different PSF extraction method for {\it HST}/WFC3 images. Since {\it HST}/WFC3 covers a field of view of $~2'\times2'$, we use \texttt{starfinder} to stack on unsaturated point sources within the image to generate the PSF used for our lens modeling analysis. 

\section{Classification of Lensing Candidates}
\label{sec:cl}
For our 87 lensing candidates with high-resolution near-IR data, we implement a two-step grading rubric to identify sources for which we could perform our lens modeling analysis to derive magnification factors and recover the intrinsic properties of the SMG. In this section, we describe our rubric that prioritizes bonafide lensing morphologies and available redshifts for the background source. The resulting grade for each candidate lensing system is listed in Table~\ref{tab:obsprops} and our grading rubric is summarized in Table~\ref{tab:lg}.  

\subsection{Visual Identification of Lensing Morphologies}
\begin{figure*}
 \includegraphics[width=\textwidth,keepaspectratio]{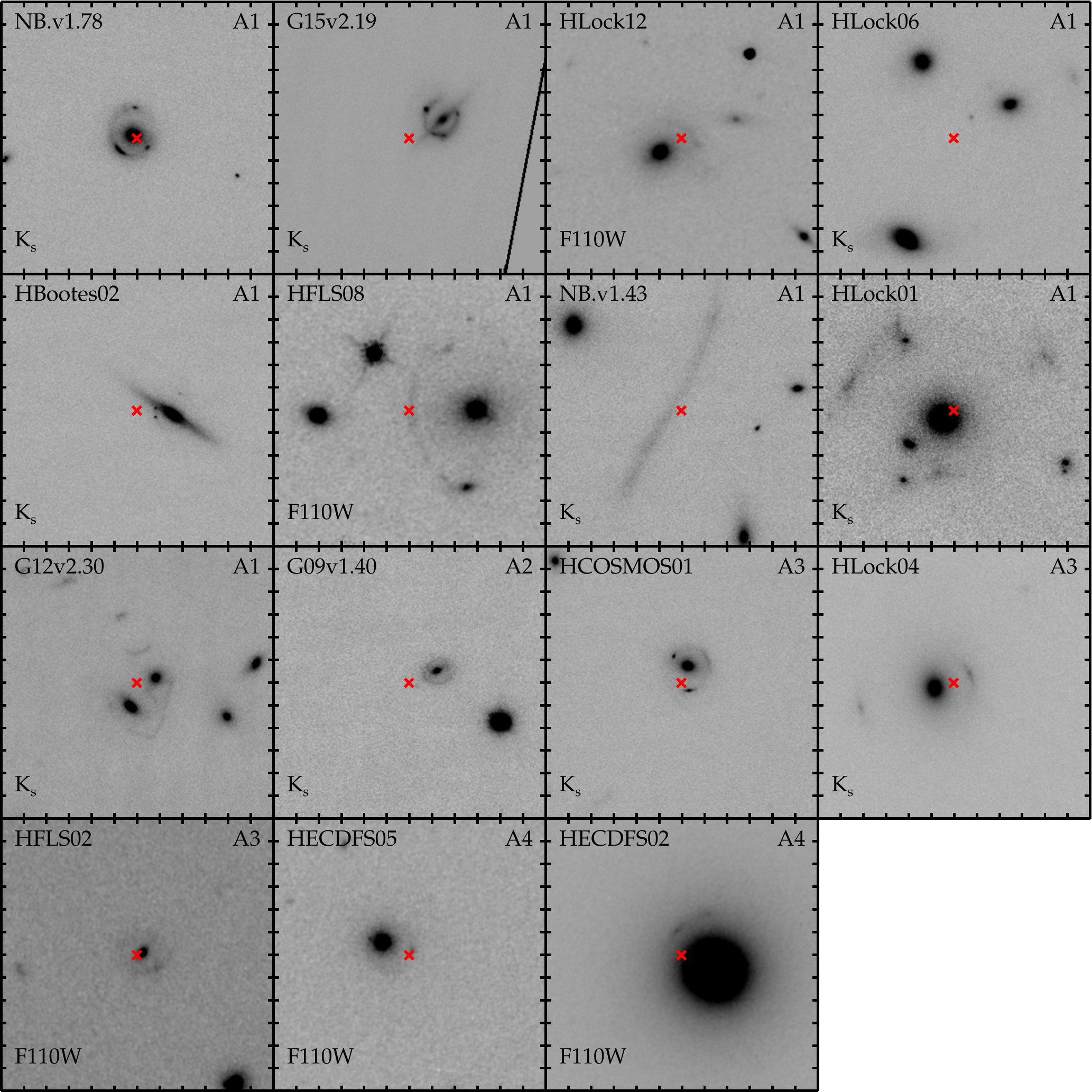} 
\caption{\label{first}12$''$ cutouts of all lens Grade A lensed SMGs, with each tick mark corresponding to 1$''$ and oriented with north is up and east is left. All have either been observed using Keck NIRC2-LGS-AO $K_{\rm s}$ or {\it HST}/WFC3 F110W. The red cross marks the measured {\it Herschel} position. Contrast levels are varied in each image to highlight the observed lensing morphology.}
\label{fig:lg}
\end{figure*}

For each target we assign a letter grade based on the existence and quality of any lensing features that are present in the near-IR data. Candidates that are classified as Grade A are of high-priority and are what we assume to be confirmed lensing systems. To the zeroth order, these are typically sources that show obvious lensing morphology such as rings, arcs, and counter-images, detected at high-significance. Some candidates that are more ambiguous (e.g. HLock12, HFLS08, and HECDFS05) are also classified as Grade A when a possible counter-image after subtracting the foreground galaxy is revealed and the observed lensing configuration can be successfully modeled. As an additional check to boost our confidence, we also confirm if the suspected near-IR lensing morphologies trace the observed configuration from existing high-resolution sub-mm data \citep{Bussmann13} or be located within the beam ($3-4''$) of radio observations for blind spectroscopy (Riechers et al., in prep.). Grade B sources can usually be described as systems with ambiguous low signal-to-noise features surrounding a relatively brighter galaxy which could either be due to lensing or be part of the galaxy itself. Deeper high-resolution data or observations in different wavelength regimes are needed to confirm the lensing status of these systems. These sources may also be intrinsically unlensed~\citep{Dowell14} or only moderately lensed, such is the case with HXMM01 \citep{Fu13}. Grade C sources are assigned to candidates of lowest priority for our study. The near-IR images for these targets typically show no detections within $15''$ of the measured $250\,\mu$m SPIRE position or sources with compact irregular morphologies that do not resemble any lensing morphologies. Like Grade B systems, we also interpret that our sample of Grade C sources could also include sources that are intrinsically bright in the far-IR. The near-IR lens models presented in this paper focuses on Grade A sources, which are shown in Fig.~\ref{fig:lg}. 

\begin{deluxetable*}{ c c c c c }[H]
\tablecolumns{5}
\tablewidth{7.0in}
\tablecaption{Grading Rubric Summary for Lensed SMGs}
\tablehead{
\colhead{NIR Lens Morphology} &
\colhead{SMG and Lens Redshift} &
\colhead{SMG Only Redshift} &
\colhead{Lens Only Redshift} &
\colhead{Neither} 
}
\startdata
Obvious & A1 & A2 & A3 & A4 \\
Marginal & B1 & B2 & B3 & B4 \\
None & C1 & C2 & C3 & C4 
\enddata
\label{tab:lg}
\end{deluxetable*}


\subsection{Redshift Availability}

Redshifts are needed to convert observed parameters into physical quantities. Spectroscopic followup programs at (sub)mm and optical/near-IR wavelengths are still ongoing \cite[e.g.,][Riechers et al\,in prep.]{Harris12, Bussmann13}. The existing redshifts are presented in Table~\ref{tab:obsprops}, and we use these data to assign a secondary letter grade from $1$ through $4$: $1$ -- redshifts available for both foreground lens and background SMG; $2$ -- redshift only available for the background SMG; $3$ -- redshift only available for foreground lens; $4$ -- no foreground lens or background SMG redshift. Note that our focus is to study the lensed SMG, we assign a higher grade for systems with background source redshifts. 

For Grade A3 and A4 systems, we estimate the lensed SMG redshifts by fitting a modified blackbody using fixed parameters of $T = 35$K and dust-emissivity parameter $\beta$ = 1.5 to the {\herschel}-SPIRE photometry, which is the typical average dust temperature for SMGs and dust emissivity parameter used for dusty galaxies at high-redshift \citep[e.g.][]{Chapman03,Kovacs06,Wardlow11}. These far-IR photometric redshifts have a large systematic uncertainty because of redshift-temperature degeneracy effects in the far-IR SED \citep{Blain04} and should therefore be used with caution. This results to a minimum uncertainty of approximately $\Delta z\pm\sim0.5$ for dust temperature variation of $\pm10$K. Due to the inherent uncertainties associated with far-IR derived photometric redshifts, we do not use them in our analysis of the intrinsic properties of lensed SMGs (Section~\ref{sec:mb}).

\subsection{Near-IR Strong Lensing Identification Efficiency}

\citet{Negrello07} predicted that, in the regime where $S_{500}\ge100$ mJy, the surface density of unlensed SMGs is extremely low, defining a flux density cut in which a large fraction of the observed source counts are strongly lensed. Out of our 87 targets, 28 satisfy $S_{500}\ge100$ mJy and 9 of these are confirmed strong lensing events (Grade A). This yields an efficiency of $\ge32\%$ at the average depth of our near-IR data (Sec.~\ref{sec:data}). The remaining $72\%$ could be unlensed or have faint lensing morphologies that fall below our near-IR detection limits. In addition, our near-IR sample of candidate lensing systems with $S_{500}\ge100$ mJy is incomplete and does not include SMGs from other studies observed at different depths and wavelengths (e.g.,\,Lensed SMGs from the H-ATLAS SDP sample, \citealt{Negrello14,Dye13}). For these reasons, we conclude that $32\%$ is a lower limit for the near-IR lensing efficiency rate. If we also treat the 11 Grade B candidates with $S_{500}\ge100$ mJy as confirmed lensing events to determine an upper limit, the near-IR lensing efficiency rate increases to $71\%$. These limits are consistent with the predicted $32-74\%$ strong lensed fraction at $S_{500}\ge100$ mJy from the statistical models of~\citet{Wardlow13}. To get an idea how this efficiency can improve as a function of near-IR depth, the H-ATLAS SDP sample~\citep{Negrello14,Dye13}, also observed using {\it HST}/WFC3 F160W with $5\sigma$ point source depths of $>$26.8 mag using $>60$ min. integration times, confirmed lensing to be present for all 5 candidate lensing systems with $S_{500}\ge100$ mJy. For comparison, the \citet{Bussmann13}'s sample of lensed SMGs with $S_{500}\ge100$ mJy observed with the Sub-Millimeter Array (SMA), 25 out of 30 candidates ($83\%$) with a depth of $5\sigma\sim15$ mJy showing evidence of moderate to strong lensing in the sub-mm maps. Of the 12 sources with high-resolution near-IR data that are confirmed to be lensed ($\mu_{880}\ge2$) in \citet{Bussmann13}, six are Grade A (NB.v1.78, H{\bootes}02, NB.v1.43, G09v1.40, HLock01, HLock04), four are Grade B (HXMM02, G09v1.97, NA.v1.195, H{\bootes}03), and the two remaining are Grade C (G09v1.124, G15v2.779). 

The lower near-IR efficiency for identifying strong-lensing events relative to sub-mm confirmations is not surprising. If a source is detected in both the sub-mm and the near-IR has two different spectroscopic redshifts, one can use small but significant offsets between the two images as evidence for lensing. This is useful in cases for which the observed sub-mm emission does not resemble convincing lensing morphologies (e.g. HXMM02, H{\bootes}03). There are also different possibilities to explain the lower efficiency associated with near-IR lensing identifications, which include the background SMGs suffering from heavy dust-obscuration, being intrinsically faint in the rest-frame optical, or lying at a high redshift. A geometric argument could also be made for the cause of non-detections, in which the near-IR emission is significantly offset from the sub-mm emission and the central caustic, thus lying in regions of low magnifications on the source-plane. In all alternative cases, this could lead to the observed near-IR emission from the background SMG to fall below our detection limits despite showing a bonafide lensing morphology in the sub-mm (e.g., G15v2.779, \citealt{Bussmann12}).

Figure~\ref{fig:s500hist} shows the cumulative frequency distribution of $S_{500}$ for all the targeted sources with high-resolution near-IR data labeled with their associated grades. \LongTables
\begin{deluxetable*}{lccccccccc}[h!]
\tablecolumns{10}
\tablewidth{0pt}
\tablecaption{Observed Properties of SMG Lens Candidates}
\tablehead{
\colhead{Name} &
\colhead{$S_{250}$}\tnm{a} &
\colhead{$S_{350}$}\tnm{a} &
\colhead{$S_{500}$}\tnm{a} &
\colhead{$S_{880}$}\tnm{b} &
\colhead{$z_{\rm{source}}$} &
\colhead{Ref.} &
\colhead{$z_{\rm{lens}}$} &
\colhead{Ref.} &
\colhead{Lens Grade} 
\\
\colhead{} &
\colhead{(mJy)} &
\colhead{(mJy)} &
\colhead{(mJy)} &
\colhead{(mJy)} &
\colhead{} &
\colhead{} &
\colhead{} &
\colhead{} &
\colhead{}
}
\startdata
HELAISS04 & 131 & 102 &  58 & \nodata & \nodata & \nodata & \nodata & \nodata & C4 \\
HELAISS01 & 129 & 116 &  81 & \nodata & \nodata & \nodata & \nodata & \nodata & B4 \\
HELAISS02 & 114 & 101 &  76 & \nodata & \nodata & \nodata & \nodata & \nodata & B4 \\
HXMM26 &  45 &  56 &  47 & \nodata & \nodata & \nodata & \nodata & \nodata & C4 \\
HXMM14 &  98 &  98 &  78 & \nodata & \nodata & \nodata & \nodata & \nodata & C4 \\
HXMM02 &  91 & 122 & 113 & 51.9 &    3.390 & R14 & 1.350 & W13 & B1 \\
HXMM13 &  55 &  88 &  94 & \nodata & 4.45\tnm{c} & R14 & \nodata & \nodata & C2 \\
HXMM20 &  85 &  79 &  67 & \nodata & \nodata & \nodata & \nodata & \nodata & C4 \\
HXMM01 & 180 & 192 & 131 & 25.1 &    2.307 & F13,W13 & 0.654 & F13,W13 & B1 \\
HXMM04 & 143 & 136 &  93 & \nodata & \nodata & \nodata & 0.210 & W13 & C3 \\
HXMM09 & 127 & 114 &  83 & \nodata & \nodata & \nodata & 0.210 & W13 & B3 \\
HXMM03 & 120 & 131 & 110 & \nodata & 2.72\tnm{c} & R14 & 0.359 & O08 & B1 \\
HXMM11 & 106 & 108 &  81 & \nodata &    2.179 & W13 & \nodata & \nodata & C2 \\
HXMM23 & 137 & 108 &  57 & \nodata & \nodata & \nodata & \nodata & \nodata & C4 \\
HXMM28 &  27 &  47 &  87 & \nodata & \nodata & \nodata & \nodata & \nodata & C4 \\
HXMM22 &  97 &  82 &  62 & \nodata & \nodata & \nodata & \nodata & \nodata & C4 \\
HXMM19 &  43 &  67 &  70 & \nodata & \nodata & \nodata & \nodata & \nodata & B4 \\
HXMM27 &   0 &  48 &  43 & \nodata & \nodata & \nodata & \nodata & \nodata & C4 \\
HXMM05 & 105 & 119 &  91 & \nodata &    2.985 & R14 & \nodata & \nodata & B2 \\
HXMM12 & 102 & 110 &  81 & \nodata & \nodata & \nodata & \nodata & \nodata & C4 \\
HECDFS08 & 104 &  67 &  54 & \nodata & \nodata & \nodata & \nodata & \nodata & B4 \\
HECDFS03 &  83 & 118 & 113 & \nodata & \nodata & \nodata & \nodata & \nodata & B4 \\
HECDFS05 & 155 & 131 &  84 & \nodata & \nodata & \nodata & \nodata & \nodata & A4 \\
HECDFS09 &  77 &  66 &  51 & \nodata & \nodata & \nodata & \nodata & \nodata & C4 \\
HECDFS11 &  45 &  52 &  42 & \nodata & \nodata & \nodata & \nodata & \nodata & C4 \\
HECDFS04 &  73 &  86 &  85 & \nodata & \nodata & \nodata & \nodata & \nodata & B4 \\
HECDFS02 & 133 & 147 & 122 & \nodata & \nodata & \nodata & \nodata & \nodata & A4 \\
HADFS04 &  76 &  90 &  72 & \nodata & \nodata & \nodata & \nodata & \nodata & B4 \\
HADFS02 &  57 &  78 &  75 & \nodata & \nodata & \nodata & \nodata & \nodata & C4 \\
HADFS01 &  79 & 103 &  92 & \nodata & \nodata & \nodata & \nodata & \nodata & B4 \\
HADFS09 & 115 &  61 &  24 & \nodata & \nodata & \nodata & \nodata & \nodata & C4 \\
HADFS08 &  88 &  81 &  50 & \nodata & \nodata & \nodata & \nodata & \nodata & B4 \\
HADFS03 & 138 & 114 &  73 & \nodata & \nodata & \nodata & \nodata & \nodata & B4 \\
G09v1.97 & 260 & 321 & 269 & 86.8 &    3.634 & R14 & 0.626 & B13 & B1 \\
G09v1.124 & 241 & 292 & 230 & 50.0 &    2.410 & H12 & 0.348 & I13 & C1 \\
G09v1.1259 &  90 & 123 &  95 & \nodata & \nodata & \nodata & \nodata & \nodata & B4 \\
G09v1.40 & 388 & 381 & 242 & 62.2 &    2.091 & L14 & \nodata & \nodata & A2 \\
SDP.301 &  83 &  87 &  65 & \nodata & \nodata & \nodata & \nodata & \nodata & B4 \\
SDP.127 & 119 &  99 &  59 & \nodata & \nodata & \nodata & \nodata & \nodata & B4 \\
G09v1.326 & 141 & 175 & 139 & 18.6 &    2.581 & H12 & \nodata & \nodata & B2 \\
HCOSMOS03 &  82 &  64 &  37 & \nodata & 3.25\tnm{c} & R14 & \nodata & \nodata & C2 \\
HCOSMOS02 &  71 &  64 &  41 & \nodata & 2.497\tnm{c} & R14 & \nodata & \nodata & C2 \\
HCOSMOS01 &  91 & 100 &  74 & \nodata & \nodata & \nodata & 0.608 & new\tnm{d} & A3 \\
HLock15 & 102 &  87 &  73 & \nodata & \nodata & \nodata & \nodata & \nodata & B4 \\
HLock05 &  71 & 102 &  98 & \nodata & 3.42\tnm{c} & R14 & 0.490 & W13 & C1 \\
HLock04 & 190 & 156 & 100 & 32.1 & \nodata & \nodata & 0.610 & W13 & A3 \\
HLock17 &  62 &  82 &  67 & \nodata & 3.039\tnm{c} & R14 & \nodata & \nodata & C2 \\
HLock02 &  53 & 115 & 140 & \nodata & \nodata & \nodata & \nodata & \nodata & C4 \\
HLock11 &  97 & 112 &  80 & \nodata & \nodata & \nodata & \nodata & \nodata & C4 \\
HLock06 & 136 & 127 &  96 & \nodata &    2.991 & R14 & 0.200 & W13 & A1 \\
HLock08 & 142 & 119 &  84 & \nodata & 1.699\tnm{c} & R14 & \nodata & \nodata & B2 \\
HLock03 & 113 & 146 & 114 & 47.0 & 2.771\tnm{c} & R14 & \nodata & \nodata & C2 \\
HLock01 & 402 & 377 & 249 & 52.8 &    2.956 & R11, S11 & 0.600 & O08 & A1 \\
HLock12 & 224 & 159 &  79 & \nodata & 1.651\tnm{c} & R14 & 0.630 & O08 & A1 \\
G12v2.43 & 289 & 295 & 216 & \nodata &    3.127 & H12 & \nodata & \nodata & C2 \\
G12v2.30 & 289 & 356 & 295 & \nodata &    3.259 & H14 & 1.225 & B13 & A1 \\
G12v2.105 & 197 & 178 & 110 & \nodata & \nodata & \nodata & \nodata & \nodata & B4 \\
NB.v1.43 & 347 & 377 & 267 & 27.0 &    1.680 & G13 & 0.997 & \nodata & A1 \\
NA.v1.195 & 179 & 278 & 265 & 57.6 &    2.951 & H14 & 0.786 & B13 & B1 \\
NA.v1.177 & 264 & 310 & 261 & 51.8 &    2.778 & K13 & \nodata & \nodata & B2 \\
NB.v1.78 & 273 & 282 & 214 & 46.0 &    3.111 & R14 & 0.428 & R14 & A1 \\
NA.v1.267 & 164 & 186 & 133 & \nodata & \nodata & \nodata & \nodata & \nodata & C4 \\
G15v2.235 & 189 & 240 & 198 & 33.5 &    2.478 & H12 & \nodata & \nodata & C2 \\
HEGS01 &  74 &  98 &  89 & \nodata & \nodata & \nodata & 0.530 & W13 & C3 \\
G15v2.779 & 115 & 191 & 204 & 90.0 &    4.243 & O13, C11 & \nodata & \nodata & C2 \\
H{\bootes}09 &  69 &  81 &  60 & \nodata & 2.895\tnm{c} & R14 & \nodata & \nodata & C2 \\
H{\bootes}04 & 141 & 133 &  94 & \nodata & \nodata & \nodata & \nodata & \nodata & C4 \\
H{\bootes}11 & 103 &  93 &  63 & \nodata & \nodata & \nodata & \nodata & \nodata & C4 \\
H{\bootes}03 & 323 & 243 & 139 & 18.4 &    1.034 & B06 & 1.034 & B06 & B1 \\
H{\bootes}02 & 159 & 195 & 156 & 35.5 &    2.804 & R14 & 0.414 & W13 & A1 \\
G15v2.19 & 778 & 467 & 225 & \nodata &    1.026 & M13 & 0.218 & M13 & A1 \\
H{\bootes}10 & 113 &  92 &  57 & \nodata & \nodata & \nodata & \nodata & \nodata & B4 \\
H{\bootes}01 & 158 & 191 & 160 & 61.0 &    3.274 & R14 & 0.590 & W13 & C1 \\
H{\bootes}12 &  11 &  52 &  51 & \nodata & \nodata & \nodata & \nodata & \nodata & B4 \\
H{\bootes}08 &  65 &  78 &  67 & \nodata & \nodata & \nodata & \nodata & \nodata & C4 \\
H{\bootes}13 & 112 & 109 &  72 & \nodata & \nodata & \nodata & \nodata & \nodata & B4 \\
H{\bootes}07 &  86 &  88 &  72 & \nodata & 4.167\tnm{c} & R14 & \nodata & \nodata & C2 \\
G15v2.481 & 141 & 157 & 130 & \nodata & \nodata & \nodata & \nodata & \nodata & B4 \\
HELAISN01 & 123 & 129 &  88 & \nodata & \nodata & \nodata & \nodata & \nodata & C4 \\
HELAISN04 &  80 &  97 &  78 & \nodata & \nodata & \nodata & \nodata & \nodata & C4 \\
HFLS07 & 115 &  92 &  69 & \nodata & \nodata & \nodata & \nodata & \nodata & C4 \\
HFLS03 &  98 & 105 &  81 & \nodata & \nodata & \nodata & 0.160 & W13 & C3 \\
HFLS05 &  40 &  75 &  74 & \nodata &    4.286 & R14 & \nodata & \nodata & C2 \\
HFLS02 & 164 & 148 &  86 & \nodata & \nodata & \nodata & 0.560 & W13 & A3 \\
HFLS08 &  86 &  93 &  67 & \nodata &    2.264 & R14 & 0.330 & O08 & A1 \\
HFLS10 &  52 &  50 &  32 & \nodata & \nodata & \nodata & \nodata & \nodata & C4 \\
HFLS01 & 107 & 123 &  98 & \nodata & \nodata & \nodata & \nodata & \nodata & C4 \\
\enddata
\tablecomments{
The following lists the reference key for redshifts: W13 = \citet{Wardlow13}; B13 = \citet{Bussmann13}; R14 = Riechers et al. (in prep.), M14 = Messias et al. (in prep.); O13 = \citet{Omont13}; C11 = \citet{Cox11}; H12 = \citet{Harris12}; H14 = Harris et al. (in prep.); I13 = \citet{Ivison13}; R11 = \citet{Riechers11}; S11 = \citet{Scot11}; O08 = \citet{Oyaizu08}; K14 = Krips et al. (in prep.); G13 = \citet{George13}; L14 = Lupu et al. (in prep.); and B06 = \citet{Borys06}.\\
The $S_{250}$, $S_{350}$, and $S_{500}$ are flux densities measured from SPIRE photometry. $S_{880}$ corresponds to the $880\,\mu$m flux density measured from SMA. $z_{\rm{source}}$ and $z_{\rm{lens}}$ refer to the redshifts of the background source and foreground lens, respectively. Lens Grade is the priority value assigned to the lensed candidate, discussed in Section~\ref{sec:cl}.}
\tnt{a}{Typical errors which include confusion and instrumental noise on SPIRE photometry are $7-10$ mJy \citep{Smith12}, which includes both statistical and confusion noise.}
\tnt{b}{$S_{880}$ is only available for sources that overlap with the sample from \citet{Bussmann13}. Typical errors for SMA photometry are $\sim15\%$ of the measured $S_{880}$ value.}
\tnt{c}{Single line redshift measurement, using CO observations.}
\tnt{d}{Based on Keck/LRIS observations, Fu et al. (in prep.)}
\label{tab:obsprops}
\end{deluxetable*}
 For comparison, we also show the  SMA sample from \citet{Bussmann13}, where we convert the sub-mm grade to an equivalent near-IR grade \footnote{The following describes the grading scheme conversion from this paper to \citet{Bussmann13}: A1 = A, A2 + A3 = B, B1 = C, A4 + B2 + B3 + B4 + C1 + C2 + C3 + C4 = X.}. In both studies, Grade A sources tend to have smaller cumulative fractions than Grade B and C sources with increasing $S_{500}$. Despite the lower efficiency of identifying lenses relative to the sub-mm, our near-IR candidate lensing system classification is consistent with the prediction that confirmed strong lensing events tend to be the brightest in $S_{500}$, having a median $S_{500}\sim122$ mJy and 9 out of the 16 (56\%) Grade A sources have $S_{500}\ge100$ mJy. Grade B sources have a median $S_{500}\sim85$ mJy (11/30 with $S_{500}\ge100$ mJy, 37\%) while Grade C sources have a median $S_{500}\sim78$ mJy (8/42 with $S_{500}\ge 100$ mJy, 19\%). The sub-mm sample from \citet{Bussmann13} shows a contrasting result and have median $S_{500}$ values of 214, 200, and 216 mJy for  Grade A, B, and C sources (using the near-IR scheme), respectively. However, we note that this could be due to the smaller sample size (30 sources total, 20 Grade A, 6 Grade B, and 4 Grade C), and the larger applied flux cut ($S_{500}\ge80$ mJy) to select the sub-mm candidate lensing systems. 

\section{Lens Models}
\label{sec:lm}
 \subsection{General Methodology} 
 \label{sec:lmgm}
  
For each lensing system we use \texttt{galfit} \citep{Peng02} to model the surface brightness profile of the foreground lens and subtract it from the image. We use {\sersic} profiles on foreground galaxies that resemble an elliptical morphology and edge-disk profiles for edge-on disks (G15v2.19 and H{\bootes}02).  Foreground lens subtraction can also reveal close counter-images required to constrain the lens model~\citep{Cooray11,Hopwood11,Negrello14,Dye13}. Any observed lensing features and nearby sources that are not associated with the lensing galaxy are masked out. The foreground lens subtracted image is then used as the input image for our lens modeling. 
   
In cases where the emission from the foreground lens and background source are blended, we implement an iterative process in order to obtain an optimal lens model \citep{Cooray11}. Using the \texttt{galfit} residual as the initial input, we derive a preliminary lens model. After achieving an acceptable fit ($\chi^{2}_{\nu}$ on the order of unity), we then subtract the lensed image of the model source from the original image. For the second iteration, we then use \texttt{galfit} on this ``lensing morphology-subtracted" image, effectively isolating the surface brightness profile of the foreground lens and eliminating the need to mask out the lensing morphology. The updated foreground lens surface brightness profile from \texttt{galfit} is subtracted from the original data, which will then serve as the new input for our lens modeling. This iterative method to obtain an optimal foreground lens subtracted image yields a $\Delta\chi^{2}_{\nu}\sim0.2-0.3$ difference from the preliminary lens model, which corresponds to a $3-5\sigma$ improvement. The best-fit model for these blended lensing systems typically converges after 1 or 2 iterations.

For gravitational lensing, the condition for strong lensing to occur is when the normalized surface mass density of the foreground lens, $\kappa$ is greater than unity. In this paper, we assume a singular isothermal ellipsoid \citep[SIE; ][]{Kormann94} for $\kappa$, with the convergence at a point ($x,y$) in the image plane defined as:
\begin{equation}
\kappa(x,y) = \frac{\Sigma}{\Sigma_{\rm{crit}}} = \frac{\sqrt{1+q^2}}{2}\frac{b}{2q\sqrt{x^{2} + y^{2}/q^{2}}}
\end{equation} 
where $\Sigma$ is the surface mass density, $\Sigma_{\rm{crit}}$ is the critical surface mass density, $b$ is the critical or Einstein radius and $q$ is the axis ratio. The SIE profile has been found to reproduce observed configurations of galaxy-galaxy strong lensing events (see \citealt{Treu10} for a recent review) and has been successfully used in modeling lensed SMGs \citep{Fu12,Bussmann12,Bussmann13, Hezaveh13}. The fitting parameters we use to describe the foreground SIE profile are the Einstein radius ($b$), distance from the measured \texttt{galfit} centroid ($\delta{x}, \delta{y}$) in RA and DEC, ellipticity ($\epsilon_{{\rm lens}}=1-q$), and the position angle ($\theta_{\rm{lens}}$, east of north). A parameter for the external shearing amplitude was also initially included in our analysis, but provided marginal to no improvement in the fit. In addition, our current data does not allow accurate redshift determination of any nearby foreground sources (with the exception of G12v2.30, which the effects of shear were accounted for by additional lensing profiles in \citealt{Fu12}). For these reasons, we do not include shearing amplitude in our models and note that additional constraints are needed in order to properly quantify its effect on the lens models. The components of the background galaxy in the source plane are assumed to have {\sersic} profiles \citep{Sersic68}. While the use of {\sersic} profiles may oversimplify the morphology of the high redshift star-forming population, previous studies have shown that this approach provide useful information about their morphologies, such as intrinsic size, shapes and orientations for both lensed and unlensed SMGs~\citep{Swinbank10,Gavazzi11,Targett11,Targett13,Aguirre13}. The fitting parameters of the background {\sersic} profile are the flux ($F$), position ($\delta u,\delta v$) from the measured foreground lens center of mass, ellipticity ($\epsilon_{\rm{source}}$), position angle ($\theta_{\rm{source}}$, defined east of north), effective semi-major axis ($a_{\rm{eff}}$), and the {\sersic} index ($n$). For all systems, we start with the simplest model for the background galaxy (1 source) and increase the components to check if this provides a significantly better fit ($\Delta\chi^{2}_{\nu}\ge0.3$). 

These model parameters are all varied consistently for each lensing system. In order to take advantage of the high-resolution data, we adopt informative priors about the foreground lens, mostly given from the \texttt{galfit} subtraction. For the background source, we adopt uniform priors for every case. The Einstein radius is typically allowed to vary within $\pm0.5''$ from a circular radius that encloses the observed lensing morphology. The lensing mass is centered on the measured \texttt{galfit} position of the foreground lens, which is varied within an area defined by the FWHM of the PSF. The ellipticities are allowed to vary from 0.0 to 0.8, and the position angles from $-90^{\circ}$ to $90^{\circ}$, with the initial values of both set to the midpoints of these ranges. The background galaxies are initially placed in perfect alignment with the foreground lens and are allowed to explore the position space within $\pm 0.75$ times the Einstein radius, which is a valid assumption, since the detection of multiple counter-images is an indication that these sources are within the vicinity of the source-plane caustics. Indeed, the maximum observed offset from direct alignment between the foreground and background galaxy is 40$\%$ of the Einstein radius (HECDFS02). The effective semi-major axis length has an initial value of $0.3''$ with a minimum value of $0.01''$ and a maximum value of $1.00''$, based on half-light radii measurements of  unlensed SMGs at $z\sim2.5$~\citep{Chapman03,Swinbank10,Targett11,Targett13,Aguirre13}. {\sersic} indices are allowed to vary from 0.10 to 4.00. The integrated flux in the lens model and the input image are normalized consistently before being compared and where there are multiple background components flux ratios are computed. For each lensing system, the total number of parameters is equal to $5\times N_{\rm{L}} + 7\times N_{\rm{S}} - 1$, where $N_{\rm{L}}$ and $N_{\rm{S}}$ represent the number of lens and source components, respectively. 

With a given set of initial parameters for the image and source plane, we use \texttt{gravlens} \citep{Keeton01} to generate a model of the lensed image. The model is convolved with the PSF to generate the expected observed image for each parameter set. This PSF-convolved model is then compared with the foreground lens subtracted image within the fitting region, shown as the green contours on Fig.~\ref{fig:lm}. These fitting regions are initially hand-drawn to enclose all the suspected lensing morphologies in the data. After a preliminary lens model is derived, the fitting region is regenerated to enclose all pixels with values $\ge 1\sigma$, measured from the data (no noise is present from the model). Defining the fitting region through this process serves three main purposes: Firstly, it helps prevent the lens model from including pixels from the background which can make the fit insensitive and degenerate from varying the input parameters. This effectively makes the model fit for shot-noise dominated pixels. Secondly, it minimizes the under or over-subtracted regions from imperfect galfit subtractions that can cause the lens model to be fixated on these unwanted features. Thirdly, it accounts for any counter-images predicted by the model but not accounted for by the data, reducing the bias in our fit. 

The process of comparing the lens model to the data is iterated using the \texttt{IDL} routine \texttt{amoeba$\_$sa}, which performs multidimensional minimization using the downhill simplex method with simulated annealing \citep{Press92} on the $\chi^{2}$ function, defined as:   
\begin{equation}
\chi^{2} = \sum\limits_{x,y}^{N} \frac{(I_{{\rm obs}}(x,y) - I_{{\rm mod}}(x,y))^{2}}{\sigma(x,y)^{2}}, 
\end{equation}
where $I_{{\rm obs}}$ and $I_{{\rm mod}}$ is the surface brightness map of the observed and the model image, respectively, $\sigma$ is the 1$\sigma$ uncertainty map for the observed image that accounts for background and shot noise, $x$ and $y$ are the pixel coordinates, and $N$ represents the number of pixels enclosed in the fitting region. Typically, $N\sim200$ for the least constrained systems (e.g., double) and $N\sim1000$ for the most constrained systems (Einstein rings or giant arcs). Depending on how well constrained the lensing system is, the correct configuration for the observed lensing morphology is usually obtained after the first few iterations of \texttt{amoeba\_sa} and the probability of accepting worse solutions decreases for each iteration due to the simulated annealing. The rest of the calls are then spent on performing an extensive search around the optimal solution with the given configuration. All parameters and calculated quantities are saved in each iteration and the $1\sigma$ confidence interval for the best fit model parameters are calculated from $\chi^{2} - \chi_{\rm{min}}^{2} \le 1$.  We note that $\chi^{2}$ is renormalized to minimize correlated noise between pixels. This is done by dividing the total number of pixels of the original unbinned $\chi^{2}$ values from the original images by the square area of the PSF \citep{Fu12}.

The near-IR magnification factor $\mu_{\rm{NIR}}$ is calculated in the same manner as in ~\citet{Bussmann13}. Briefly, we integrate the model flux ($F_{\rm{SP}}$) within elliptical apertures with the same orientations and ellipticities as the model but with double the semi-major axis length. Then, these source plane elliptical apertures are mapped on to the image plane using the foreground lens model and the image plane flux is integrated ($F_{\rm{IP}}$). The magnification factor is then simply a ratio of the two integrated fluxes, $\mu_{\rm{NIR}} = F_{\rm{IP}}/F_{\rm{SP}}$, and is representative of total from all background source components. We note that since our near-IR data is at a much higher resolution than in the sub-mm, changing the aperture size to equal the semi-major axis compared to double its value had little effect on the magnification value(within 10\%).  

To measure near-IR photometry, we use our fitting region to define the aperture and our results are listed in Table~\ref{tab:nirphot}. The same aperture is also applied when measuring available multi-wavelength high-resolution near-IR data (Fig.~\ref{fig:mw}). Photometric statistical errors are measured by calculating the standard deviation of the total counts from non-overlapping background-dominated fields on the data, using the same sized aperture. A simple aperture correction is calculated by measuring the ratio of total counts from the lens model with and without the aperture. We divide the integrated flux densities by $\mu_{\rm{NIR}}$ for each background source to obtain a magnification-corrected value.

\LongTables
\begin{deluxetable*}{ l c c c c c c c c}[ht!]
\tablecolumns{7}
\tablewidth{0pt}
\tablecaption{Properties of the Foreground Lenses of Grade 1 Systems.}
\tablehead{
\colhead{Name} &
\colhead{RA$_{\rm NIR}$} & 
\colhead{DEC$_{\rm NIR}$} &
\colhead{$b$} &
\colhead{$\delta x$} &
\colhead{$\delta y$} &
\colhead{$\epsilon$} &
\colhead{$\theta$} &
\colhead{$\chi^{2}/N_{\rm{DOF}}$} 
\\
\colhead{} &
\colhead{} &
\colhead{} &
\colhead{$('')$} &
\colhead{$('')$} &
\colhead{$('')$} &
\colhead{} &
\colhead{(deg)} &
\colhead{} 
}
\startdata
NB.v1.78& 13:30:08.513 & +24:58:59.13 & $     0.944^{+     0.002}_{    -0.001} $ & $     0.018^{+     0.001}_{    -0.003} $ & $    -0.042^{+     0.003}_{    -0.001} $ & $     0.419^{+     0.002}_{    -0.007} $ & $    80.9^{+     0.2}_{    -0.2} $ &  1455/1897 \\
G15v2.19& 14:29:35.212 & -00:28:35.94 & $     0.738^{+     0.002}_{    -0.001} $ & $     0.027^{+     0.002}_{    -0.002} $ & $     0.044^{+     0.002}_{    -0.003} $ & $     0.208^{+     0.005}_{    -0.003} $ & $   -51.0^{+     0.5}_{    -0.4} $ &  5452/2097 \\
HLock12& 11:00:16.457 & +57:17:34.96 & $     1.14^{+     0.04}_{    -0.07} $ & $    -0.15^{+     0.06}_{    -0.04} $ & $    -0.05^{+     0.03}_{    -0.05} $ & $     0.41^{+     0.05}_{    -0.08} $ & $       132^{+         7}_{       -12} $ &  2641/2871 \\
HLock06& 10:45:48.892 & +57:45:12.99  & $     2.46^{+     0.01}_{    -0.01} $ & $    -0.14^{+     0.03}_{    -0.01} $ & $    -0.14^{+     0.01}_{    -0.03} $ & $     0.067^{+     0.02}_{    -0.005} $ & $       94^{+        2}_{       -4} $ &  415/ 656\\
H{\bootes}02& 14:28:25.474 & +34:55:46.84 & $     0.56^{+     0.01}_{    -0.01} $ & $    -0.201^{+     0.005}_{    -0.01} $ & $     0.084^{+     0.005}_{    -0.005} $ & $     0.68^{+     0.01}_{    -0.01} $ & $    50.7^{+     0.3}_{    -0.5} $ &  227/172 \\
HFLS08& 17:15:44.502 & +60:12:39.02  &$     1.95^{+     0.05}_{    -0.04} $ & $    -0.42^{+     0.05}_{    -0.07} $ & $    -0.38^{+     0.07}_{    -0.05} $ & $     0.46^{+     0.04}_{    -0.04} $ & $     -110^{+        2}_{       -1} $ &  1630/1364 \\
G09v1.40& 08:53:58.864 & +01:55:37.72 &$     0.56^{+     0.01}_{    -0.02} $ & $     0.0034^{+     0.01}_{    -0.001} $ & $    -0.01^{+     0.01}_{    -0.02} $ & $     0.0^{+     0.1}_{    -0.2} $ & $      -57^{+        4}_{       -1} $ &  544/874\\
HCOSMOS01& 10:01:44.183 & +02:57:12.74 &$     0.91^{+     0.01}_{    -0.01} $ & $    -0.00^{+     0.01}_{    -0.02} $ & $    -0.01^{+     0.02}_{    -0.02} $ & $     0.26^{+     0.04}_{    -0.03} $ & $       67^{+        2}_{       -1} $ &  1182/633\\
HLock04& 10:38:26.742 & +58:15:42.61 &$     2.403^{+     0.01}_{    -0.005} $ & $     0.080^{+     0.001}_{    -0.02} $ & $    -0.092^{+     0.013}_{    -0.003} $ & $     0.22^{+     0.01}_{    -0.02} $ & $    14^{+     1}_{    -1} $ & 1268/2013\\
HFLS02 & 17:14:50.848 & +59:26:33.83 &$     0.87^{+     0.020}_{    -0.05} $ & $     0.21^{+     0.06}_{    -0.01} $ & $    -0.01^{+     0.04}_{    -0.04} $ & $     0.46^{+     0.04}_{    -0.04} $ & $      -23^{+        4}_{       -3} $ & 1644/1981 \\
HECDFS05 & 03:26:36.449 & -27:00:44.44 & $     0.96^{+     0.02}_{    -0.03} $ & $    -0.11^{+     0.02}_{    -0.02} $ & $    -0.10^{+     0.02}_{    -0.02} $ & $     0.12^{+     0.01}_{    -0.01} $ & $       -38^{+        11}_{       -11} $ &  305/369\\
HECDFS02\tnm{a}& 03:37:32.359 & -29:53:53.50 &$     1.6477^{+     0.03}_{    -0.05} $ & $     0.09^{+     0.01}_{    -0.02} $ & $    -0.10^{+     0.01}_{    -0.03} $ & [0.0] & [0.0] &  860/983 
\enddata
\tablecomments{The following parameters discussed in Section~\ref{sec:lmgm} are used to describe the foreground lens: (RA$_{\rm NIR}$, DEC$_{\rm NIR}$) = centroid of light from galfit subtraction, $b=$ Einstein radius, $(\delta x,\delta y)=$ centroid position of mass relative to light, $\epsilon=$ elongation, $\theta=$ orientation of mass profile (east of north), $\chi^{2}$/$N_{\rm{DOF}}$ = $\chi^{2}$ value and the number of degrees of freedom. }
\tnt{a}{The ellipticity and position angle is fixed to assume a circular shape, since the best fit for the foreground lens converges to this lower limit if left as free parameters.}
\label{tab:lensprop}
\end{deluxetable*}

 \subsection{Notes on Individual Lens Models}
 \label{sec:lmnotes}
\begin{figure*}
\vspace{-1 cm}
\includegraphics[width=\textwidth,keepaspectratio]{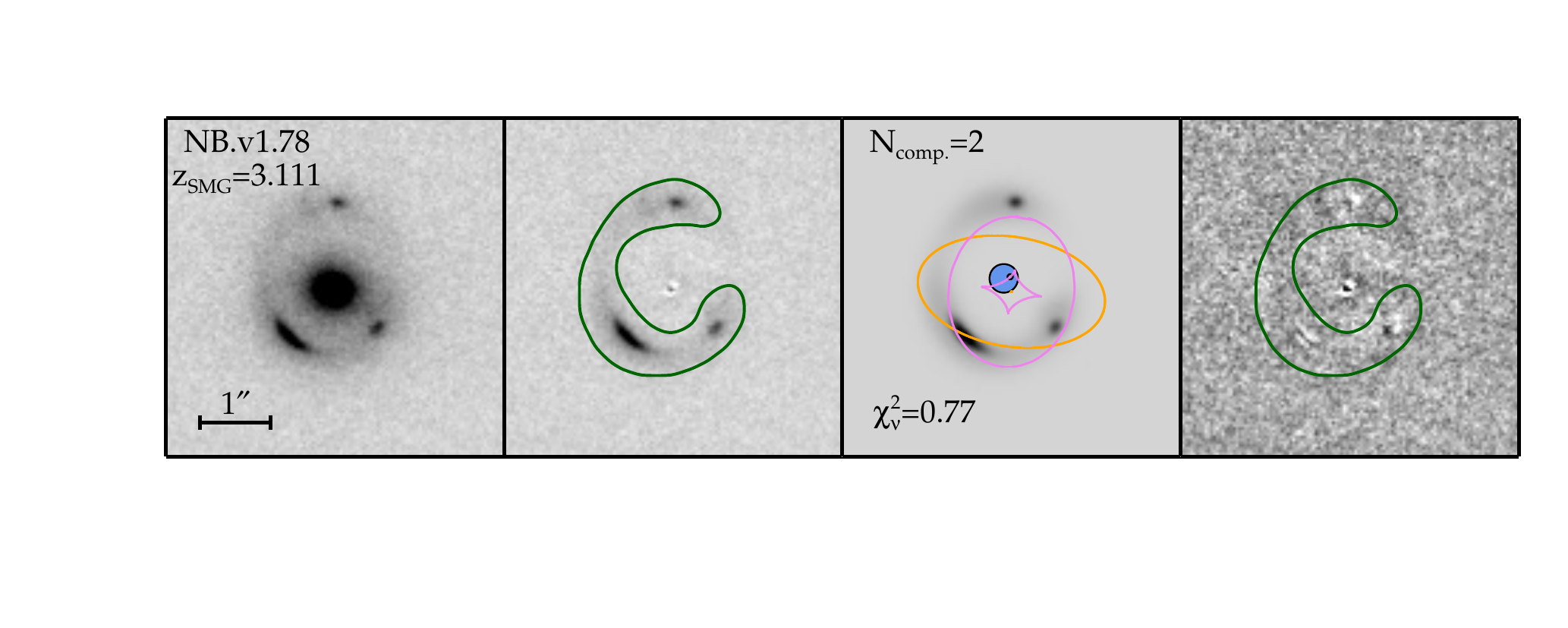}

\vspace{-3.35 cm}
\includegraphics[width=\textwidth,keepaspectratio]{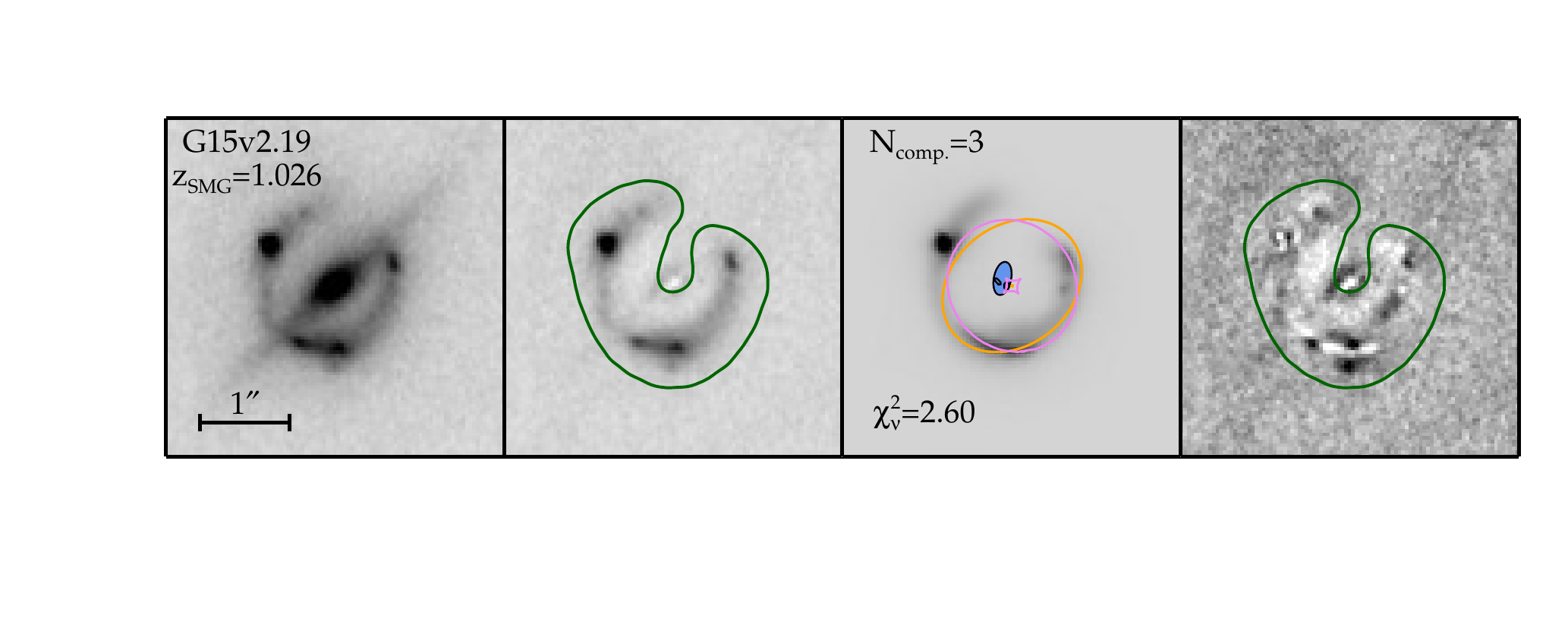}

\vspace{-3.35cm}
\includegraphics[width=\textwidth,keepaspectratio]{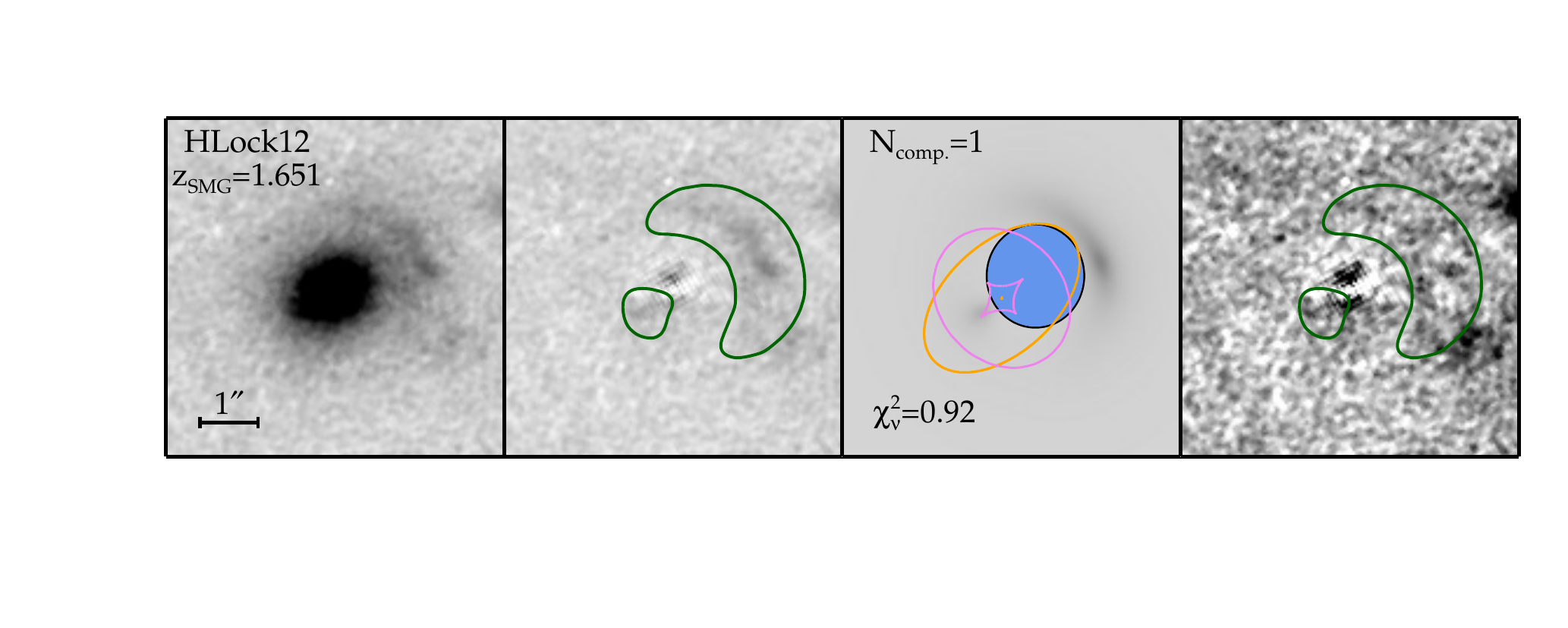} 

\vspace{-3.35cm}
\includegraphics[width=\textwidth,keepaspectratio]{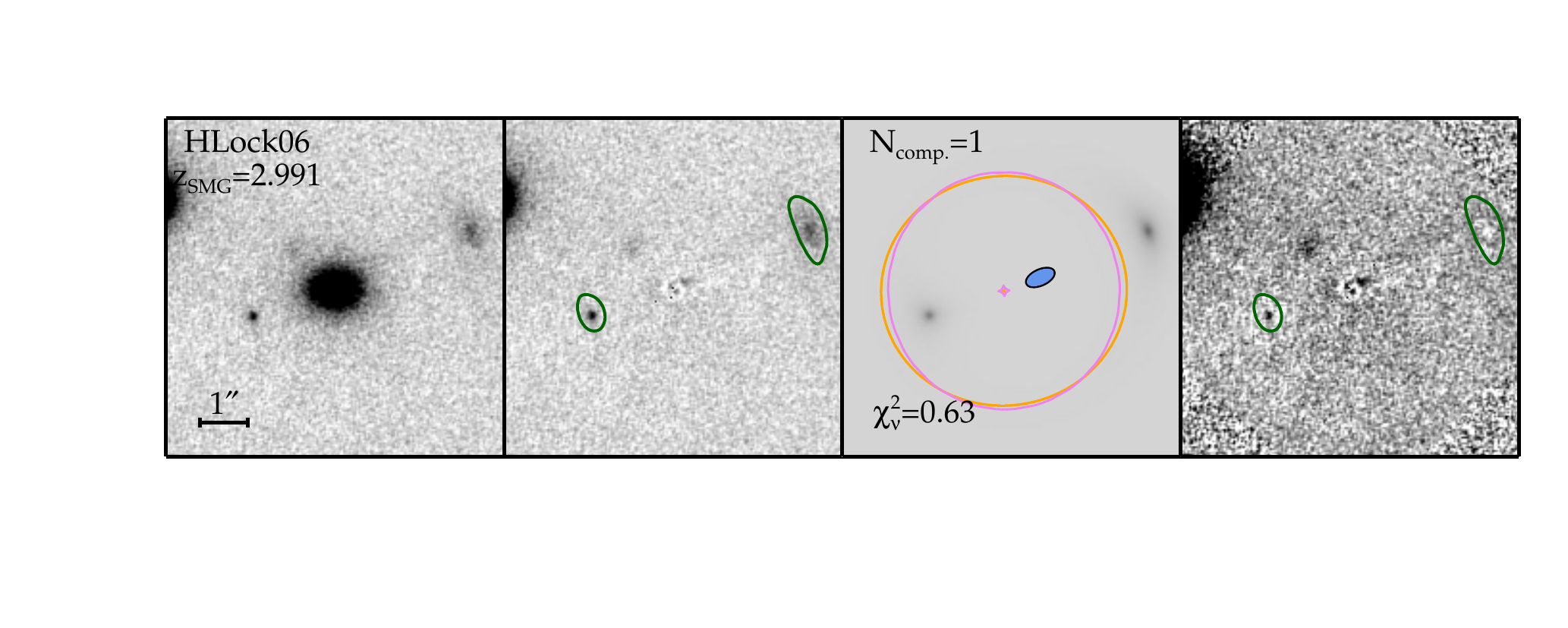} 

\vspace{-1.6cm}
\caption{Near-IR lens modeling results of selected Grade A sources, oriented with north up and east to the left for all images. From left to right: postage stamp of observed image; foreground lens subtracted image; best-fit lens model; and the residual image. Green apertures enclose  the final fitting region used. The orange and pink outlines trace the critical and caustic curves, respectively. Blue ellipses are the source plane models, displayed with the best-fit half-light semi-major axis, ellipticity, and position angle. Redshifts labeled with square brackets are photometric redshifts estimated from far-IR to sub-mm photometry and those without are spectroscopic. The third panel also lists the number of background components used in the best fit, denoted as $N_{\rm{comp}}$ and the reduced $\chi^{2}$, defined as $\chi^{2}_{\nu}=\chi^{2}/N_{DOF}$. The residual image is shown at a narrower greyscale, which is 0.2 times the minimum and maximum pixel value of the original image in order to highlight under/over-subtracted regions. }
\label{fig:lm}
\end{figure*}

 In this section, we provide notes on the basic characteristics for each lensing system with available lens models. We do not provide lens models for HLock01 and G12v2.30, as they have already been subjects of detailed studies from previous works \citep{Gavazzi11, Fu12} and are also included in the sub-mm sample from \citet{Bussmann13}. The SMGs with lens models derived here are shown in Fig.~\ref{fig:lm}. The best-fit parameters along with the $1\sigma$ errors describing the foreground lens and the background source are presented in Tables~\ref{tab:lensprop} and~\ref{tab:sourceprop}. As a test for differential lensing and size comparison analysis in Section~\ref{sec:dl}, we also generate lens models for the four new sources (NB.v1.78, H{\bootes}02, G09v1.40, and HLock04) that overlap with \citet{Bussmann13}, using the same foreground lens parameters reported in their paper, allowing the foreground lens position to vary within $0.3''$ to account for any astrometric offset between the near-IR and sub-mm data. The use of sub-mm derived foreground lens parameters generally yields poorer fits but is able to reproduce the observed near-IR lensing configuration. The lens models for this near-IR/sub-mm subsample are discussed on an object-by-object basis and shown in the Appendix. \\

\textbf{NB.v1.78 (Grade A1):} The $K_{\rm s}$-band image shows a classic configuration observed when the background source lies on top of the caustic fold, the same configuration shown by the lensing system SDSS J0737+3216 \citep{Marshall07}. The $H$-band image (Fig.~\ref{fig:mw}) shows a consistent configuration, but the lensing morphology is fainter. The multiple, well-separated arcs, in addition to the incomplete Einstein ring strongly constrains the lens model. The best-fit lens model requires two background {\sersic} profiles to account for a compact, brighter and extended, fainter, component. The best fit model shows a compact source located off-center within an extended component, indicating an asymmetric morphology. Using a single component model yields a significantly worse fit ($\chi^{2}_{\nu}$=1.50) and fails to reproduce the extended Einstein ring. This source was also discussed in \citet{Bussmann13}, in which the SMA image reveal a similar configuration to the compact component in the $K_{\rm s}$-band image. We measure a marginally lower magnification factor of $\mu_{\rm{NIR}} = 10.8^{+0.3}_{-0.2}$, compared to $\mu_{880} = 13.0\pm1.5$ for the SMA data. \\

\textbf{G15v2.19 (Grade A1):}  The observed lensing morphology features a quad-like configuration accompanied by an incomplete Einstein ring, observed in both $H$-band and $K_{\rm s}$-band images. The background source is being lensed by an edge-on disk and has the most complicated background galaxy model in our whole sample, requiring three components. It has the poorest fit, $\chi^{2}_{\nu} = 2.6$, with both over- and under-subtracted regions that can be $\ge5\sigma$. Using less than 3 components resulted in $\chi^{2}_{\nu}>5$. This system serves as an example in which substructure in the background source dominates, such that our assumed {\sersic} profile is an inadequate description of the source.  Furthermore, if all counter-images are resolved in the Keck data (as indicated by their angular sizes being larger than the Keck PSF), and if the observed emission from the individual knots are from the same source, then their surface brightnesses should be somewhat comparable, which is a property of the counter-images in the image plane \citep{Kochanek89}. Instead, we observe the surface brightness to be significantly inconsistent relative to each other, which supports our hypothesis that the morphology of the background source is highly complex and the observed emission is due to multiple background components.
  
\begin{figure*}
\vspace{-1cm}
\includegraphics[width=\textwidth,keepaspectratio]{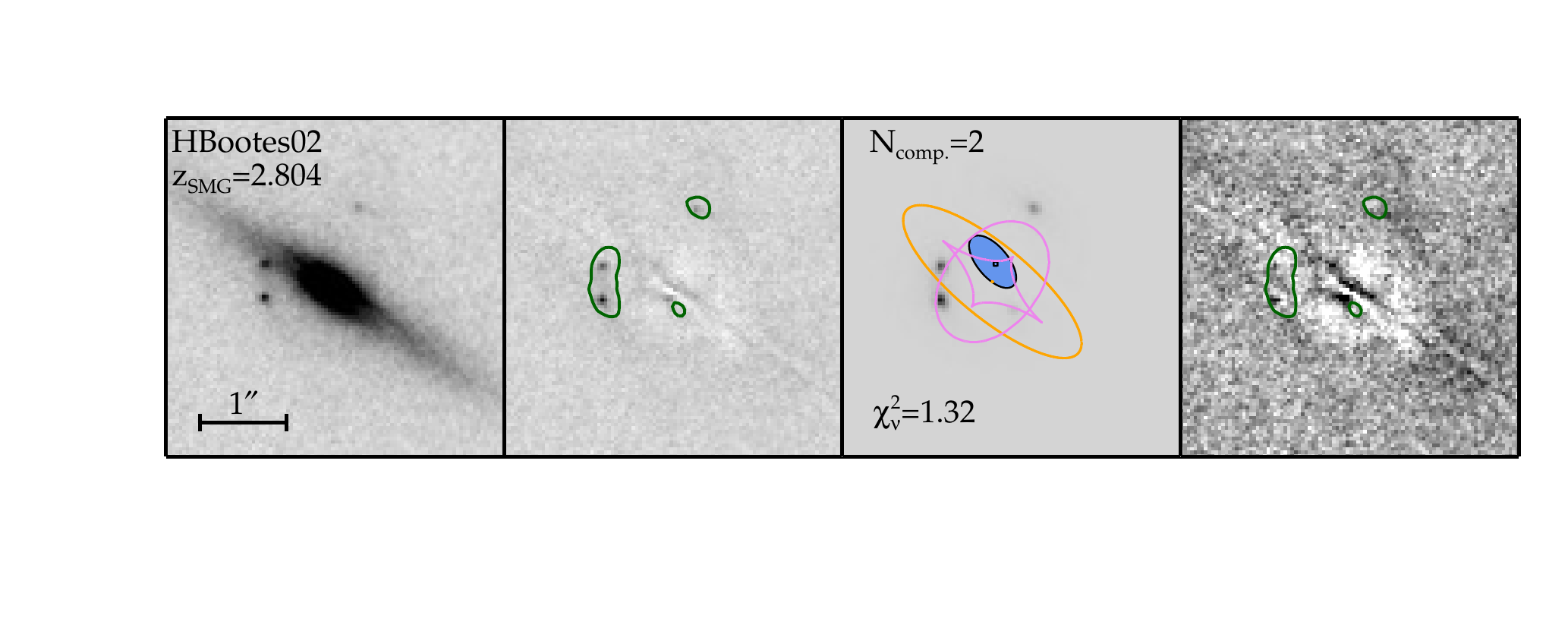} 

\vspace{-3.35cm}
\includegraphics[width=\textwidth,keepaspectratio]{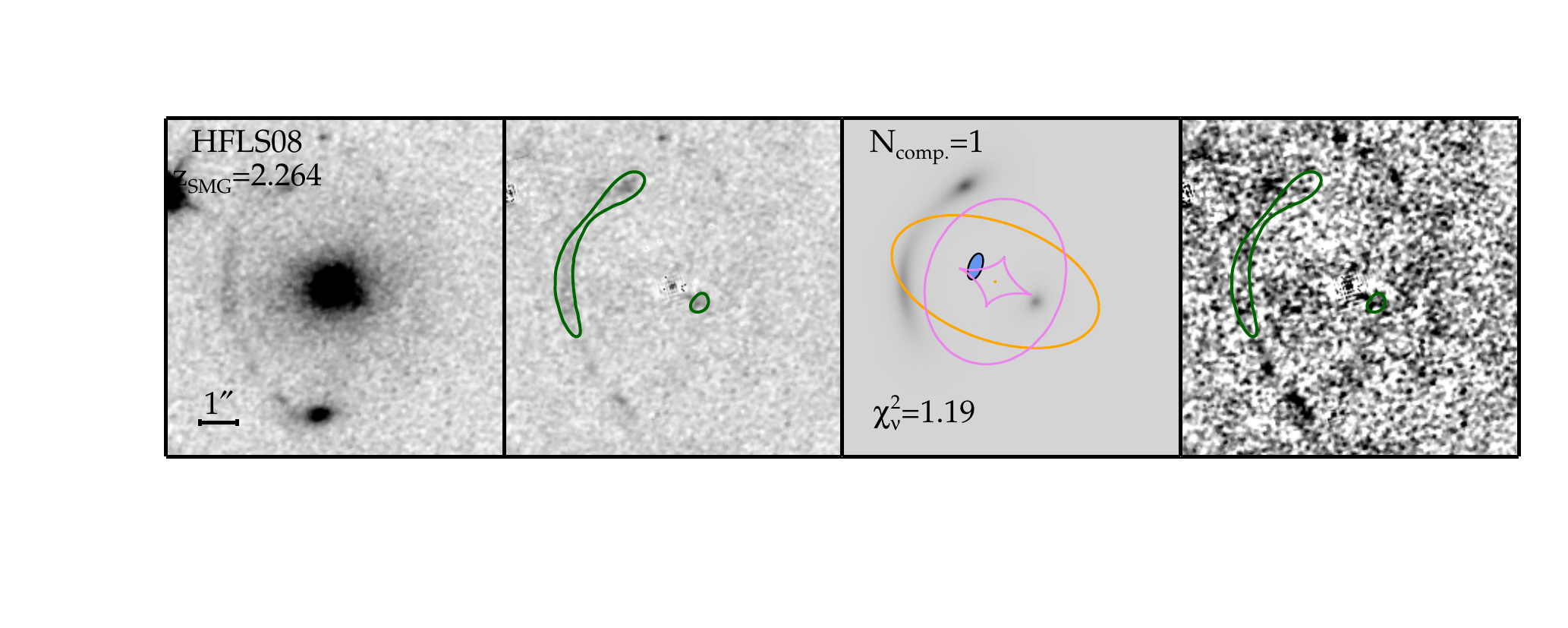} 

\vspace{-3.35cm}
\includegraphics[width=\textwidth,keepaspectratio]{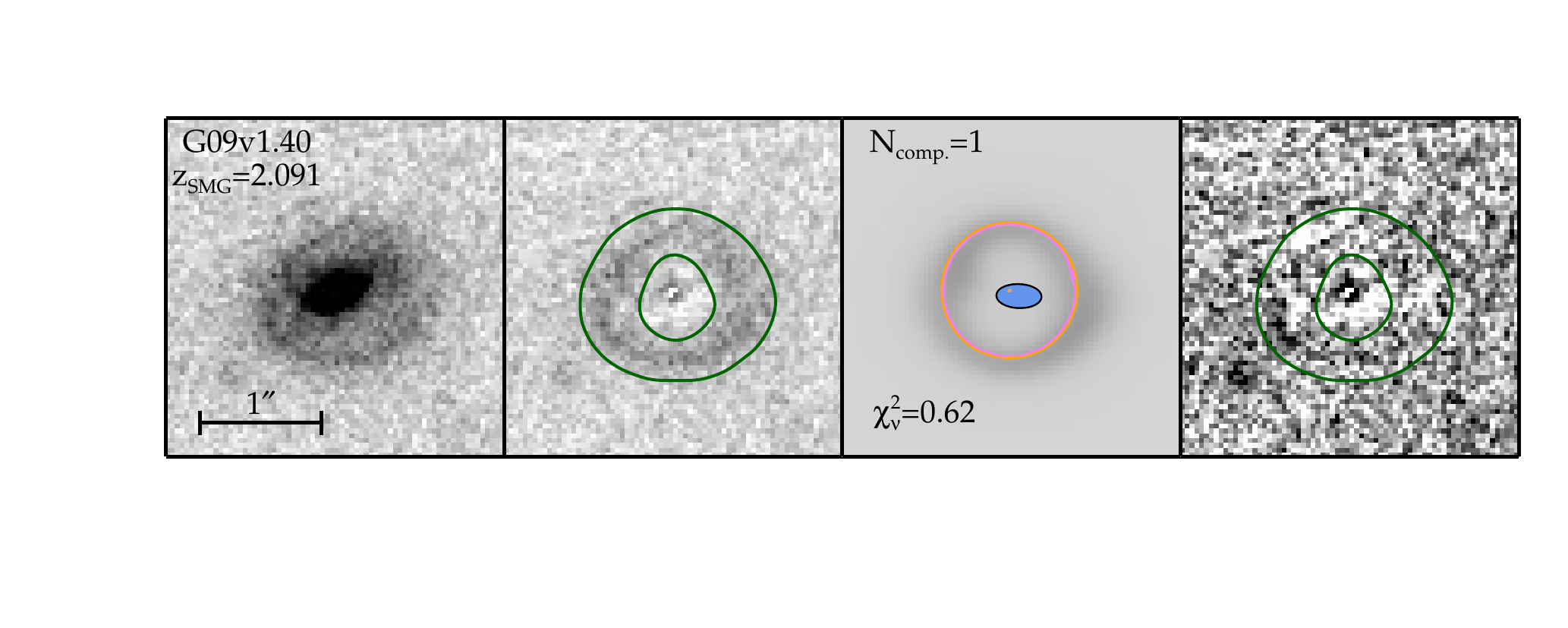} 

\vspace{-3.35cm}
\includegraphics[width=\textwidth,keepaspectratio]{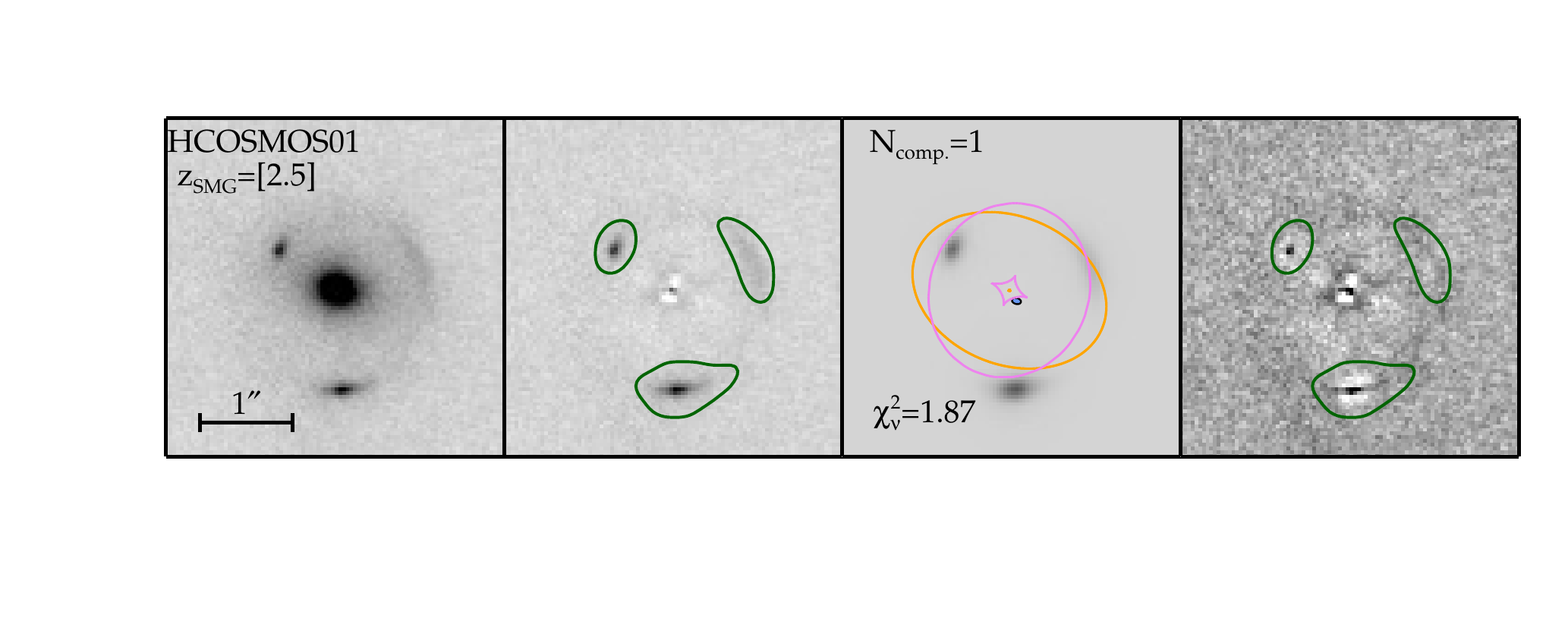} 

\vspace{-1.6cm}
\centerline{{\small \textbf{Figure~\ref{fig:lm}} --- continued.}}
\end{figure*}
 
We regard our lens model as a simple solution that can serve as a basis for future analysis on this object. Our source-plane reconstruction consists of two compact objects separated by $\sim0.1''$ within a third extended elongated source. The positions of the two compact objects forms quads and double images in the observatations, in which one of the counter-images from each component converge at roughly the same position in the image plane to produce the brightest knot located in the northeast. The extended component straddles the caustic, causing the incomplete Einstein ring. Due to the poor fit and under-subtracted regions in the residual image, the error bars in the magnification factor we report, $\mu_{\rm{NIR}} = 9.6^{+0.8}_{-0.3}$, are most likely underestimated, since the contribution for the complexity of the system is not included. For comparison,  a more extensive analysis for this system is discussed in \citet{Messias14}, which features a semi-linear inversion (SLI) approach \citep{Warren03, Dye08, Dye13} in lens modeling multi-wavelength data simultaneously\footnote{In \citet{Messias14}, G15v2.19 is identified as H1429$-$0028. For consistency with the other sources, we use the G15v2.19, as identified by H-ATLAS.}. Between the two independent analyses, a qualitative comparison of the complex background source morphologies are fairly consistent and the differences in some of the resulting parameters are mainly due to differential lensing and foreground obscuration (shown in Fig. 1 and Fig. 8 of \citealt{Messias14}). In addition, our derived magnification factor of $9.6^{+1.0}_{-0.3}$ is consistent with their result of $8.9\pm0.7$. \\
 
 \textbf{HLock12 (Grade A1):} The subtraction of the bright early-type galaxy reveals a counter-image detected at $5\sigma$ located $1''$ east of the foreground lens. This constrains the lens model, which features a classic cusp configuration. The background SMG is extended with a half-light radius comparable to the foreground lens ($\sim1''$). At $z = 1.7$, 1$''$ is $\sim$7 kpc, so this source is larger than the average for $z\sim2.5$ SMGs \citep{Aguirre13,Targett13,Targett11,Swinbank04,Chapman03}, although it is still consistent with other near-IR observations of SMGs at $z=0.5-1.5$ \citep{Mosleh11}. The {\it HST} image has multiple peaks in the arc, causing the residual image to contain under-subtracted regions. This could indicate the presence of substructure in the background source or the foreground lens. It is unlikely that the most prominent under-subtracted region, $\sim2''$ south-west from the centroid of the arc emission, is associated with the background, since all variations of the lens model fail to reproduce any emission in this area, even when it is included in the fitting region and multiple components are allowed. \\
 
\begin{figure*}
\vspace{-1.0 cm}
\includegraphics[width=\textwidth,keepaspectratio]{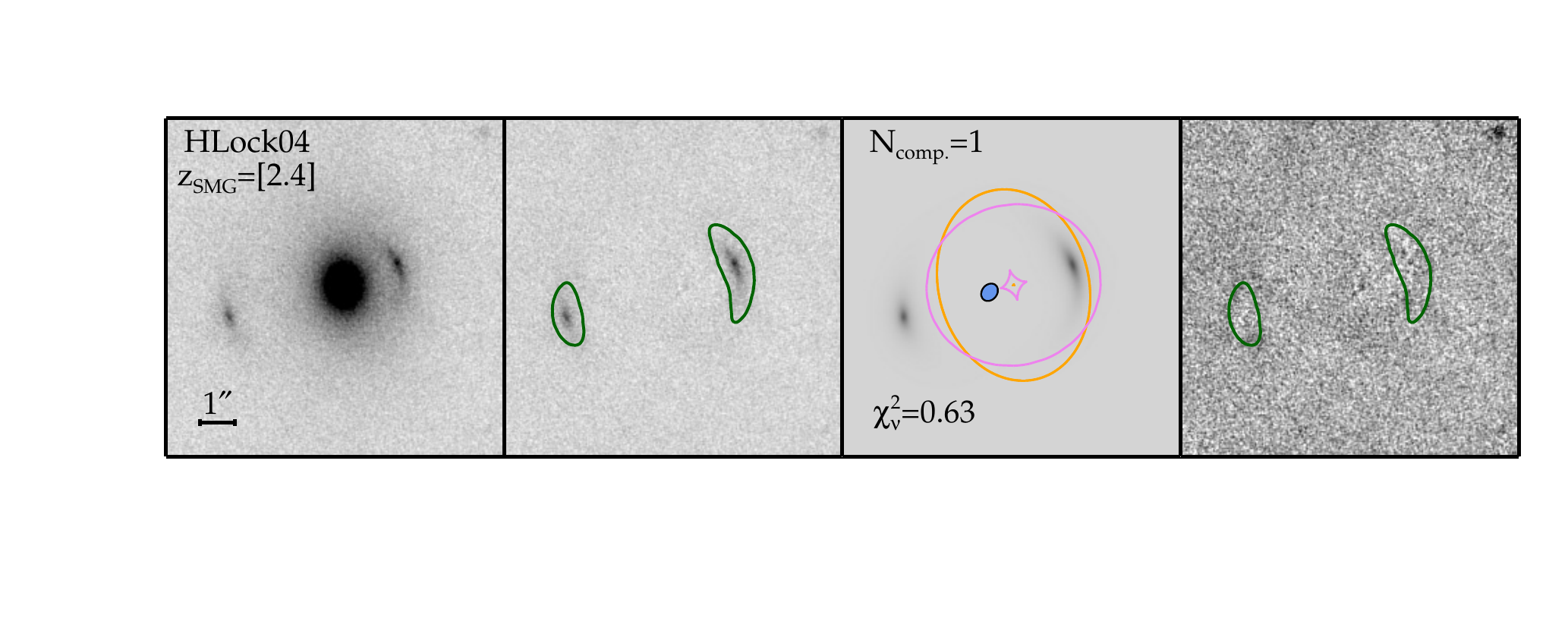} 

\vspace{-3.35cm}
\includegraphics[width=\textwidth,keepaspectratio]{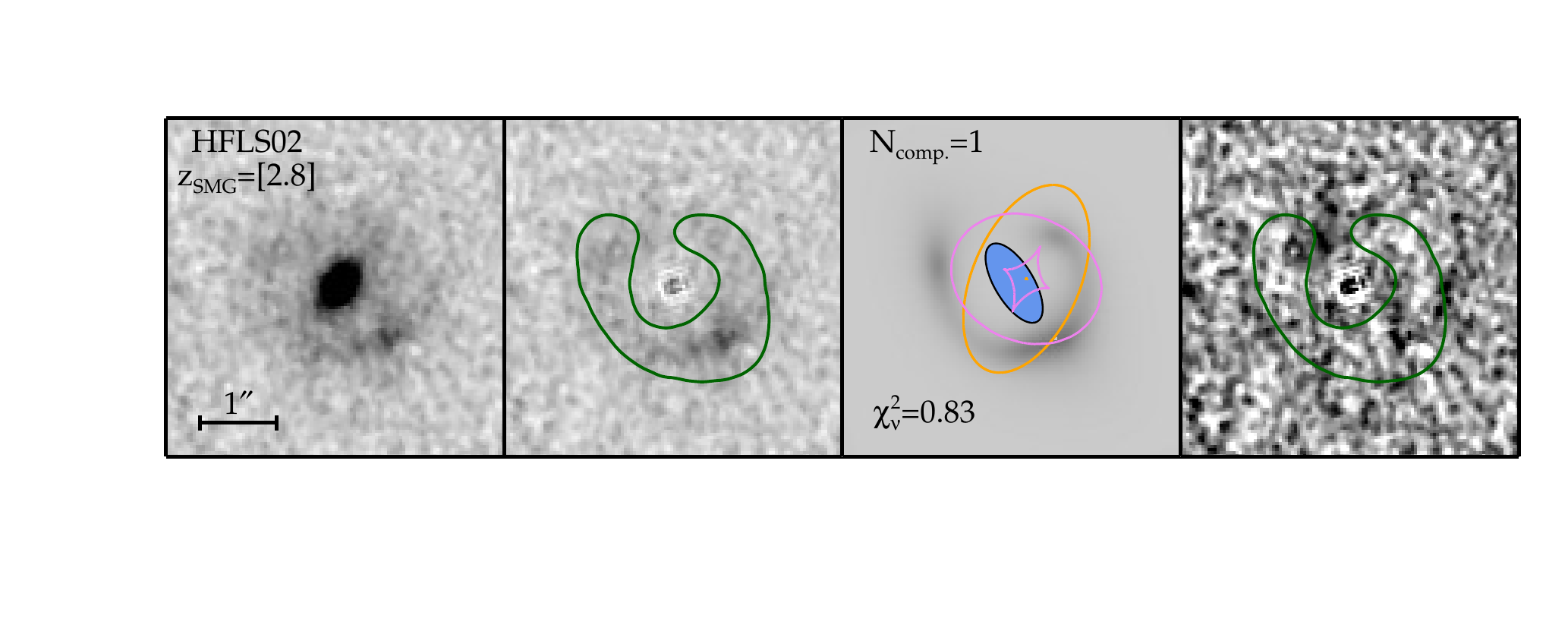} 

\vspace{-3.35cm}
\includegraphics[width=\textwidth,keepaspectratio]{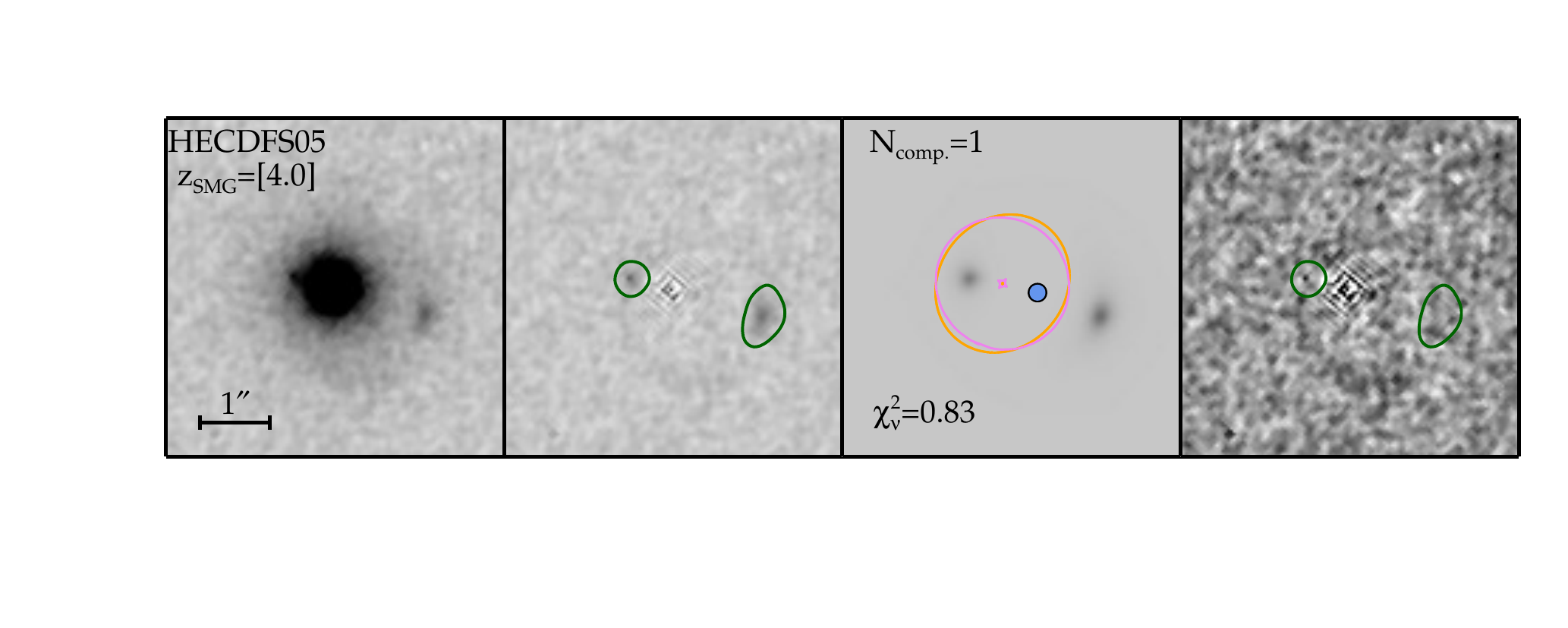} 

\vspace{-3.35cm}
\includegraphics[width=\textwidth,keepaspectratio]{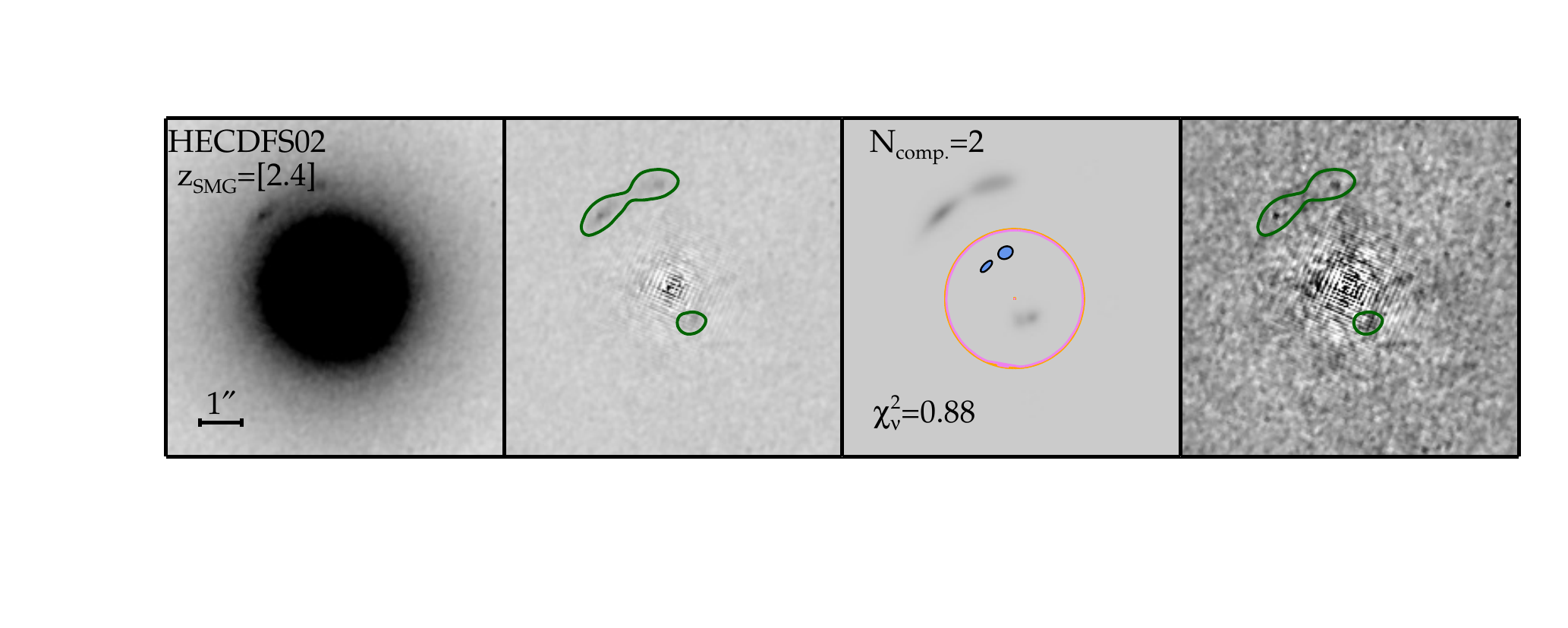} 

\vspace{-1.6cm}
\centerline{{\small \textbf{Figure~\ref{fig:lm} }--- continued.}}
\end{figure*}
 
\textbf{HLock06 (Grade A1):} The lensing morphology of this source shows an arc to the west and a counter image to the east of the foreground lensing galaxy. The same features are also detected in the {\it HST} image (Fig.~\ref{fig:mw}). The lens model shows that the Einstein radius of the foreground lens is very extended compared to the observed emission, which could be due to overlapping mass profiles from the neighboring galaxies. However, additional mass profiles or adding an external shearing amplitude has little effect on the derived source morphology so here we present the simplest best-fit model using a single mass component. There is significant under-subtraction in the eastern counter-image, which is not reproduced even when multiple components are used. This could primarily be due to systematic effects in the data. It is also unlikely that the residual emission northeast of the foreground lens is associated with the background galaxy since the lens model also fails to reproduce any counter-images in this region. \\

\textbf{H{\bootes}02 (Grade A1):} The lens model for the sub-mm emission, which shows an incomplete Einstein ring, was discussed in \citet{Bussmann13}. A multi-wavelength analysis for this object will be featured in Wardlow et al. 2014 (in prep.). The $K_{\rm s}$-band image shows an edge-on disk galaxy with an incomplete quad configuration, accompanied by faint, extended emission between the counter-images. The WFC3 F110W image shows no detections of the background source, while the detection in the NIRC2 $H$-band is marginal. 

To model the background source, we consider both a one component point-source (circular Gaussian profile) and a two component model with a point-source and an extended {\sersic} profile. The one component fit yields a $\chi_{\nu}^2 = 1.42$ and reproduces all the observed features. However, the converged solution predicts the fourth undetected counter-image in the data to be detected at $10\sigma$ in the model. One possible explanation favoring this model would be severe obscuration from the edge-on disk. However, there is also EVLA radio observations of this system \citep{Wardlow13}, which will not be affected by dust obscuration from the foreground lensing galaxy. In the EVLA data only the three near-IR luminous sources are detected, despite the sensitivity being high enough to detect the fourth image predicted by the single component model, if the flux ratios are as predicted. Therefore we consider it unlikely that the single component model is correct. 
 
Furthermore, the two component model (shown in Fig.~\ref{fig:lm}) has a marginally improved fit, with $\chi_{\nu}^2 = 1.19$ and has a configuration in which the fourth faint counter-image is faint and expected to be undetected ($<5\sigma$). This model also has some physically motivation, since the sub-mm data \citep{Bussmann13} shows an extended component, interpreted as star-forming regions, while the radio data \citep{Wardlow13} show a point source, indicative of an AGN. Both AGN and star-formation can be bright in the near-IR, which is supported by the faint extended emission in the observed frame $K_{\rm s}$-band data.

\LongTables
\begin{deluxetable*}{ l c c c c c c c c }[h!]
\tabletypesize{\scriptsize} 
\tablecolumns{9}
\tablewidth{7.5in}
\tablecaption{Properties of the Background Lensed Galaxy for Grade 1 Systems.}
\tablehead{
\colhead{Name} &
\colhead{Flux Ratio} &
\colhead{$\delta u$} &
\colhead{$\delta v$} &
\colhead{$\epsilon_{\rm{s}}$} &
\colhead{$\theta_{\rm{s}}$} &
\colhead{$a_{\rm{eff}}$} &
\colhead{$n$} &
\colhead{$\mu_{\rm{NIR}}$} 
\\
\colhead{} &
\colhead{} &
\colhead{$('')$} &
\colhead{$('')$} &
\colhead{} &
\colhead{(deg)} &
\colhead{$('')$} &
\colhead{} &
\colhead{}
}
\startdata
NB.v1.78&\nodata& $     0.11^{+     0.01}_{    -0.01} $ & $     0.19^{+     0.01}_{    -0.01} $ & $     0.01^{+     0.04}_{    -0.02} $ & $       -13^{+        14}_{       -13} $ & $     0.188^{+     0.01}_{    -0.002} $ & $     0.37^{+     0.07}_{    -0.03} $ & $    10.8^{+     0.3}_{    -0.2}$ \\
\nodata& $     0.22^{+     0.01}_{    -0.02} $ & $     0.017^{+     0.002}_{    -0.004} $ & $     0.211^{+     0.004}_{    -0.002} $ & $     0.015^{+     0.036}_{    -0.003} $ & $        24^{+        24}_{       -22} $ & $     0.0220^{+     0.0019}_{    -0.0006} $ & $     0.99^{+     0.11}_{    -0.06} $ &  \nodata \\
HLock12&\nodata& $     0.6^{+     0.1}_{    -0.1} $ & $     0.31^{+     0.04}_{    -0.1} $ & $     0.06^{+     0.1}_{    -0.02} $ & $    -0.1^{+    50}_{    -10} $ & $     0.9^{+     0.2}_{    -0.1} $ & $     2.6^{+     0.4}_{    -0.4} $ & $     4.0^{+     0.4}_{    -0.4}$ \\
HLock06&\nodata& $     0.75^{+     0.03}_{    -0.02} $ & $     0.78^{+     0.02}_{    -0.04} $ & $     0.50^{+     0.03}_{    -0.1} $ & $      114^{+        4}_{       -1} $ & $     0.30^{+     0.01}_{    -0.02} $ & $     2.5^{+     0.3}_{    -0.2} $ & $     6.9^{+     0.4}_{    -0.3}$ \\
G15v2.19&\nodata& $     0.161^{+     0.003}_{    -0.003} $ & $     0.013^{+     0.003}_{    -0.004} $ & $     0.80^{+0.01}_{    -0.02} $ & $  -136^{+     2}_{    -1} $ & $     0.031^{+     0.001}_{    -0.002} $ & $     0.34^{+     0.06}_{    -0.03} $ & $     9.6^{+     1}_{    -0.3}$ \\
\nodata& $     0.24^{+     0.05}_{    -0.02} $ & $     0.062^{+     0.003}_{    -0.004} $ & $     0.025^{+     0.003}_{    -0.01} $ & $     0.4^{+0.04}_{    -0.1} $ & $        1^{+       17}_{       -7} $ & $     0.028^{+     0.002}_{    -0.002} $ & $     0.15^{+     0.1}_{    -0.01} $ &  \nodata \\
\nodata& $     1.8^{+     0.2}_{    -0.1} $ & $     0.108^{+     0.01}_{    -0.01} $ & $     0.037^{+     0.01}_{    -0.004} $ & $     0.51^{+0.03}_{    -0.02} $ & $      -11^{+        2}_{       -1} $ & $     0.18^{+     0.01}_{    -0.01} $ & $     0.34^{+     0.1}_{    -0.02} $ &  \nodata \\
H{\bootes}02\tnm{a}&\nodata& $     0.04^{+     0.01}_{    -0.01} $ & $     0.20^{+     0.01}_{    -0.01} $ & [$0.0$] & [$0.0$] & $     0.013^{+     0.001}_{    -0.001} $ & $[0.5] $ & $     5.3^{+     1.4}_{    -0.4}$ \\
\nodata& $     1.7^{+     0.4}_{    -0.3} $ & $     0.00^{+     0.01}_{    -0.01} $ & $     0.23^{+     0.01}_{    -0.02} $ & $     0.5^{+     0.1}_{    -0.1} $ & $       40^{+    3}_{  -1} $ & $     0.35^{+     0.03}_{    -0.03} $ & $     2.0^{+     0.4}_{    -0.4} $ &  \nodata \\
HFLS08&\nodata& $     0.5^{+     0.1}_{    -0.1} $ & $     0.6^{+     0.1}_{    -0.1} $ & $     0.6^{+     0.1}_{    -0.2} $ & $       -19^{+        30}_{       -19} $ & $     0.34^{+     0.01}_{    -0.05} $ & $     2.6^{+     0.4}_{ -1} $ & $     7.7^{+     1.6}_{    -0.7}$ \\
G09v1.40&\nodata& $     0.08^{+     0.01}_{    -0.01} $ & $     0.05^{+     0.01}_{    -0.03} $ & $     0.49^{+     0.02}_{    -0.06} $ & $       87^{+        6}_{       -4} $ & $     0.18^{+     0.01}_{    -0.01} $ & $     0.51^{+     0.02}_{    -0.04} $ & $11.4^{+0.9}_{ -1}$ \\
HCOSMOS01&\nodata& $     0.08^{+     0.02}_{    -0.02} $ & $     0.12^{+     0.02}_{    -0.02} $ & $     0.4^{+     0.1}_{    -0.1} $ & $        76^{+        25}_{       -24} $ & $     0.037^{+     0.005}_{    -0.005} $ & $     1.0^{+     0.7}_{    -0.2} $ & $        9^{+        5}_{       -2}$ \\
HLock04&\nodata& $     0.69^{+     0.01}_{    -0.02} $ & $     0.714^{+     0.02}_{    -0.003} $ & $     0.22^{+     0.02}_{    -0.01} $ & $   -40.0^{+     0.1}_{    -0.1} $ & $     0.24^{+     0.01}_{    -0.01} $ & $     2.0^{+     0.1}_{    -0.2} $ & $     8.1^{+     0.2}_{    -0.3}$ \\
HFLS02&\nodata& $     0.16^{+     0.1}_{    -0.02} $ & $     0.04^{+     0.04}_{    -0.05} $ & $     0.58^{+     0.03}_{    -0.1} $ & $     -148^{+        9}_{       -6} $ & $     0.57^{+     0.01}_{    -0.1} $ & $     1.7^{+     0.2}_{    -0.3} $ & $     7.4^{+     0.5}_{    -0.6}$ \\
HECDFS05&\nodata& $     0.50^{+     0.03}_{    -0.03} $ & $     0.47^{+     0.03}_{    -0.03} $ & $     0.0018^{+     0.0003}_{    -0.0003} $ & $     -168^{+        4}_{       -4} $ & $     0.11^{+     0.01}_{    -0.01} $ & $     3.9^{+     1.1}_{    -0.5} $ & $     4.0^{+     0.8}_{    -0.7}$ \\
HECDFS02& \nodata& $     0.67^{+     0.02}_{    -0.03} $ & $     0.02^{+     0.01}_{    -0.03} $ & $     0.7^{+     0.0}_{    -0.2} $ & $       -44^{+        21}_{       -12} $ & $     0.15^{+     0.02}_{    -0.02} $ & $     0.9^{+     0.6}_{    -0.2} $ & $     3.1^{+     0.1}_{    -0.1}$ \\
\nodata& $     0.9^{+     0.1}_{    -0.1} $ & $     0.22^{+     0.03}_{    -0.03} $ & $     0.58^{+     0.02}_{    -0.04} $ & $     0.25^{+     0.12}_{    -0.07} $ & $       -61^{+        28}_{       -21} $ & $     0.16^{+     0.03}_{    -0.01} $ & $     0.17^{+     0.54}_{    -0.02} $ &  \nodata 
\tablecomments{The following parameters discussed in Section~\ref{sec:lmgm} are used to describe the background source: Flux Ratio = ratio of integrated flux, relative to the first listed component (fixed in the case of single components), $(\delta u, \delta v)=$ background source position, relative to the the centroid of the mass profile, $\epsilon_{\rm{s}}=$ elongation of the background source, $\theta_{\rm{s}}=$ orientation of the background source (east of north), $a_{\rm{eff}}=$ effective semi-major axis, $n=$ {\sersic} index, $\mu_{\rm{NIR}}$ = near-IR magnification factor (represents the total value, with all subcomponents included). }
\tnt{a}{Background component assumes a gaussian point source.}
\enddata
\label{tab:sourceprop}
\end{deluxetable*}

The center of the foreground mass profile is significantly offset from the stellar light profile ($\sim 0.20''$ or $1.2$ kpc), but this separation could be due to the dust-lane partially obscuring the true center of the stellar emission or the foreground galaxy not being perfectly edge-on. The near-IR model also predicts a smaller Einstein radius ($0.56''\pm0.01$ vs. $0.77\pm0.03$) and  magnification factor than the sub-mm lens model ($\mu_{{\rm NIR}}=5.3^{+1.4}_{-0.4}$ vs. $\mu_{880} = 10.3\pm1.7$). We note that as it currently stands, it is difficult for both lens models to account for the different observed lensing morphologies in the near-IR and sub-mm. In order to constrain the lens model, data in which the extended dusty star-forming regions and the point-source AGN component are detected at high significance is needed.\\

\textbf{HFLS08 (Grade A1):} The {\it HST} image shows an arc-like morphology east of the foreground lens. A counter-image located south-west from the foreground lens centroid is also detected at $>5\sigma$ after surface brightness profile subtraction. Since there are multiple regions of emission that could all potentially be associated with the arc, we use an initial fitting region that encloses all the suspected features for our preliminary models. We also tried models in which the background galaxy is described  by multiple components, or a two component mass profile. None of these solutions successfully account for the compact emission $\sim3''$ south of the foreground lens. We are unable to produce a configuration that accounts for the faint regions northeast and southeast of the foreground lens shown in the residual image. Therefore, we consider it unlikely that these features are from the lensed galaxy. Spectroscopy is required to confirm whether all the emission is associated with the background SMG. Since a single background component provides the best fit to the lensed arc, that is the model that we retain, and that is presented in Fig.~\ref{fig:lm}. \\

\textbf{NB.v1.43 (Grade A1):} This object was presented in \citet{Bussmann13} and \citet{George13} and will be further analyzed in Fu et al. (in prep.). This object could potentially be lensed by a cluster, as discussed in \citet{Bussmann13}. The $K_{\rm s}$-band and $H$-band images (Fig~\ref{fig:mw}) show a much more elongated morphology than the sub-mm data, but there is little curvature. The lack of additional counter-images and a central position for the lensing mass places very weak constraints on the configuration, so we do not provide a lens model for this source.  \\

\textbf{G09v1.40 (Grade A2):} The lens model for the $880\,\mu$m emission for this source was presented in \citet{Bussmann13}. The near-IR model for the background galaxy is a highly elongated, extended object with $a_{\rm{eff}} = 0.18$, which is roughly three times the size of the sub-mm model. In the near-IR, the background galaxy is nearly in perfect alignment with the foreground lens, producing the observed Einstein ring. This configuration shows a slight contrast with the sub-mm data, which show two peaks in the emission which could represent a double configuration, as supported by their lens model. However, the near-IR magnification $\mu_{\rm{NIR}} = 11.4^{+0.9}_{-1.0}$ is consistent with the SMA data, $\mu_{880} = 15.3\pm3.5$), which suggests that the lensing configurations are similar and the two peaks seen in the SMA map are likely a result of having poor spatial resolution compared to Keck AO. \\

\textbf{HCOSMOS01 (Grade A3):} The $K_{\rm s}$-band image shows an incomplete Einstein ring in which three well-separated arcs are visible. The F110W image (Fig.~\ref{fig:mw}) shows a consistent configuration but appears to be fainter. Only one component is required to reproduce the observations and using multiple components results in only a marginal improvement in the fit. The wide range of magnifications ($\mu_{\rm{NIR}} = 9^{+5}_{-2}$), is due to the compact size of the background galaxy ($a_{\rm{eff}}\sim0.04''$) and its location relative to the caustics. The residual image shows areas of under and over subtraction, also reflected by a relatively poor fit $\chi^{2}_{\nu}=1.86$,  indicating that the {\sersic} profile could be an over-simplified model to describe the background SMG\textbf{ or be due to systematic effects in the data}. \\
\textbf{HLock04 (Grade A3):} The double arc lensing morphology of HLock04 is detected in both the near-IR and sub-mm, which makes it ideal for multi-wavelength studies. This morphology is consistent in the $J$, $H$, and $K_{\rm s}$, but is brightest at the $K_{\rm s}$-band, shown in Fig.~\ref{fig:mw}. We calculate a slightly higher magnification factor of $\mu_{\rm{NIR}} = 8.1^{+0.2}_{-0.3}$ compared to $\mu_{\rm{NIR}} = 6.17\pm0.03$ from \citet{Wardlow13}, but is consistent in the sub-mm \citep[$\mu_{880}=7.1\pm1.5$][]{Bussmann13}. This is likely due to the background galaxy being located outside, near the central caustic, which is a region with a steep magnification gradient \citep{Hezaveh12}. A slight positional offset between the two lens models could then cause a significant change in magnification value.  
\\
\textbf{HFLS02 (Grade A3):} This object was included in the supplementary sample of \citet{Wardlow13}. The {\it HST} imaging shows an asymmetric Einstein ring lens morphology that suffers blending with the foreground lens. The residual image shows areas of under-subtraction, which could be either due to the presence of substructure in the source plane or left-over emission from the foreground lens. This is also a rare case in which the background source has a larger angular size than the foreground lens. 
\\
\textbf{HECDFS05 (Grade A4): }Subtracting the foreground lens emission reveals a counter-image ($> 7\sigma$) east of the foreground lens, exhibiting a double configuration. The residual image shows an under-subtracted region to the south of the foreground lens, which could be an arc. However, the low signal to noise feature is not reproduced in the lens modeling and may not be part of the lensed SMG. The source plane reconstruction shows a strongly magnified ($\mu_{\rm{NIR}} = 4.0^{+0.8}_{-0.7}$), compact ($a_{\rm{eff}} = 0.11\pm0.01$), spherical ($\epsilon_{\rm{s}} \sim 0$) galaxy. 
\\
\textbf{HECDFS02 (Grade A4): } This source was discussed in \citet{Wardlow13} and we present an updated lens model in this paper. The {\it HST} image shows an arc with two knots north-east of the foreground lens. We detect a counter-image at $> 10\sigma$ after subtracting the foreground lens. the best-fit lens model contains two background sources of similar size ($\sim0.15''$), with their centroids separated by $\sim0.4''$. The SPIRE colors suggest a redshift of 2.4, which corresponds to two $\sim1$ kpc objects separated by $\sim3$ kpc. Both background sources are distorted by the lensing galaxy to produce a double configuration in the image plane, where the fainter counter-image of both sources are in the same region and blended in our data. Leaving the ellipticity as a free parameter in the two-component model consistently caused it to converge to zero ($\epsilon=0$ corresponds to circular symmetry), which is the lower limit, so we fix this parameter to this value in our best-fit model. The background source is reminiscent of merger-like systems presented in figure 2 of \citet{Chapman03}. A single-component model gives a slightly worse fit ($\chi^{2}_{\nu}=1.2$), which yields a mass profile that is significantly elongated ($\epsilon\sim0.6$) in contrast to the rounder light profile ($\epsilon\sim0.1$) and a cusp configuration similar to HFLS08.  

\section{Results and Discussion}
\label{sec:res}
\subsection{Differential Lensing and Source Sizes}
\label{sec:dl}

\begin{figure}
\begin{minipage}[b]{0.5\textwidth}
\includegraphics[width=\textwidth,keepaspectratio]{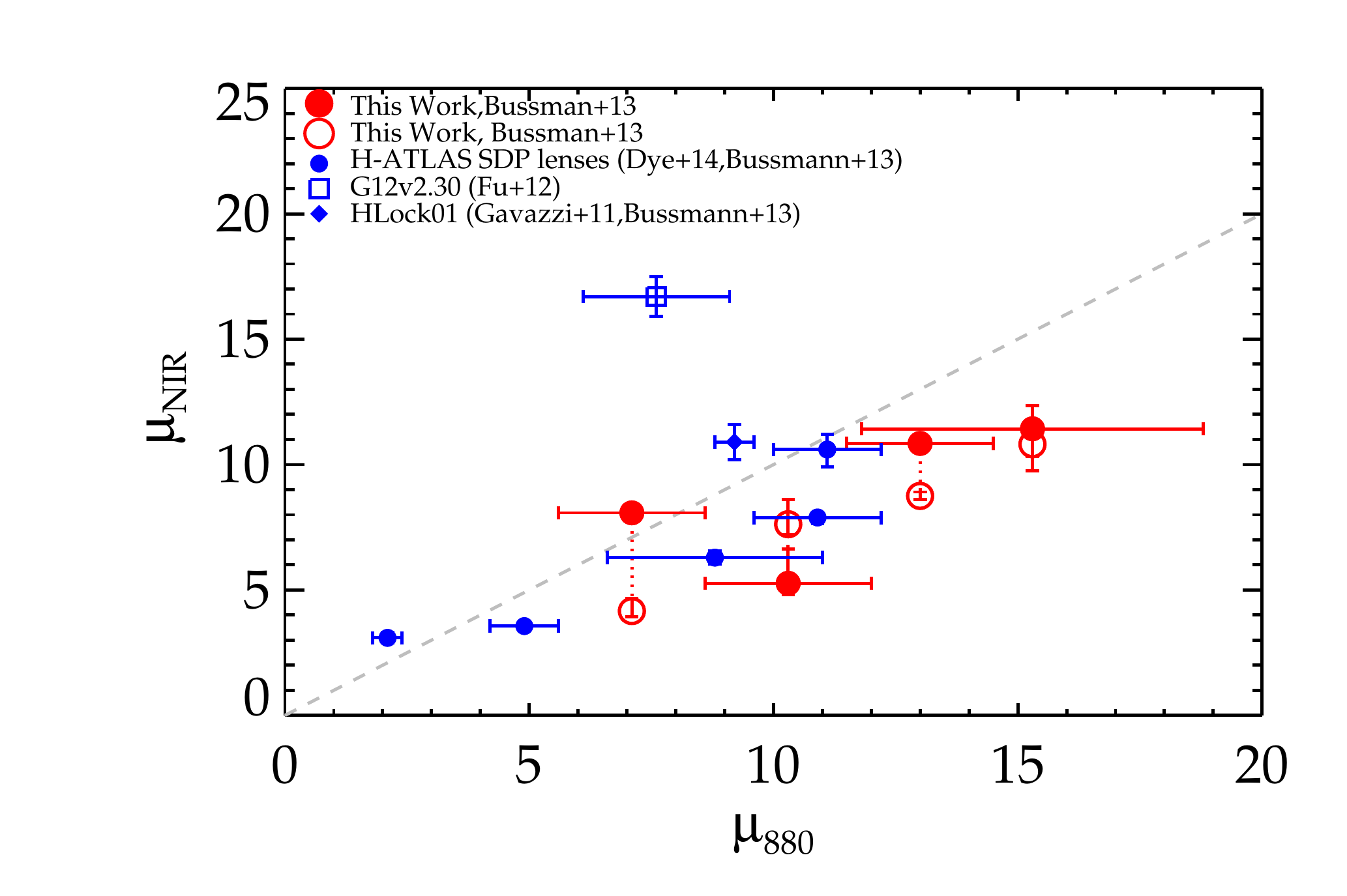} 
\caption{$\mu_{\rm{NIR}}$ vs $\mu_{880}$. Filled symbols are magnification values from independent near-IR and sub-mm lensing analyses. Open symbols denote best-fit lens models using consistent foreground lens parameters in the near-IR and sub-mm. For our work, we fix sub-mm lens parameters from \citet{Bussmann13} to our near-IR data. The blue circles, diamonds and square are near-IR data points from \citet{Dye13,Gavazzi11}, and \citet{Fu12}, respectively,  with the corresponding sub-mm magnifications from \citet{Bussmann13}, if available. The dashed line shows one-to-one correspondence between {\munir}and {\musubmm}.  Most sources lie below this line, with {\munir} $<$ {\musubmm}. Differential magnification is observed and is likely due to spatial variations or a morphological difference between the near-IR (stellar) and sub-mm (dust) emission.}
\label{fig:dl}
\end{minipage}
\end{figure}

Differential lensing is caused by spatial variations within the background galaxy, which, if they have different colors or SEDs, effectively corresponds to different wavelength regimes. This effect is more pronounced in galaxy-galaxy lensing than cluster lenses because of the steeper gradients of the magnification factors mapped onto the source plane. Recent simulations predict the effect of  differential lensing in galaxy-galaxy SMG systems~\citep{Hezaveh12,Serjeant12}, but few observations studies have successfully measured it \citep{Gavazzi11,Fu12, Dye13}. In order to measure the effects of differential lensing, a consistent mass profile to describe the foreground galaxy must be applied on lens modeling multi-wavelength data sets of the same background source. Here, we search for evidence of differential lensing by comparing the sub-millimeter lens models (from Bussmann et al. 2013) with our near-IR lens models. Figure~\ref{fig:dl} compares {\munir} with {\musubmm} for the systems in our sample that are also in \citet{Bussmann13}, where we show both our best-fit near-IR magnifications, and the values calculated using the same foreground lens parameters from sub-mm data. To verify that the difference in lens modeling methods between the near-IR and the sub-mm is not a dominant source of error, we also model sub-mm data from \citet{Bussmann13} and are able to recover consistent magnifications values. The results of applying sub-mm foreground lens parameters on near-IR data are summarized in Fig.~\ref{fig:dl} and Table~\ref{tab:dl}. For comparison, we also show the lensed SMGs with both near-IR and sub-mm magnification measurements from \citet{Dye13}, \citet{Fu12}, \citet{Gavazzi11}, and \citet{Bussmann13} \footnote{Differential magnification for G12v2.30 was measured in \citet{Fu12} by applying the near-IR foreground lens parameters in the sub-mm. However, we note that an updated model for this source was discussed \citep{Bussmann13}, due to additional SMA EXT data. The studies of SDP lenses featured in \citet{Dye13}, HLock01 in \citet{Gavazzi11}, and \citet{Bussmann13} use independent foreground lens parameters.}. Our overlapping sample has $\mu_{\rm{NIR}} < \mu_{880}$, in most cases, with {\musubmm}/{\munir}$ ~\sim1.5$ on average, providing observational evidence of differential lensing 500-$\mu$m selected galaxies. This result is likely due to the fact that the selection preferentially identifies sources that have boosted sub-mm fluxes and this bias is weakened in the near-IR. Therefore, in cases where magnification factors can only be measured in one regime, caution should be used when interpreting physical quantities at other wavelengths. However, it is also important to note that the measurement uncertainties are often greater than the average effect of differential magnification \citep[e.g. stellar masses have systematic uncertainties from 2-5][]{ Michalowski10, Wardlow11, Michalowski12,Targett13, Simpson13}.

\begin{figure}
\begin{minipage}[b]{0.5\textwidth}
\includegraphics[width=\textwidth,keepaspectratio]{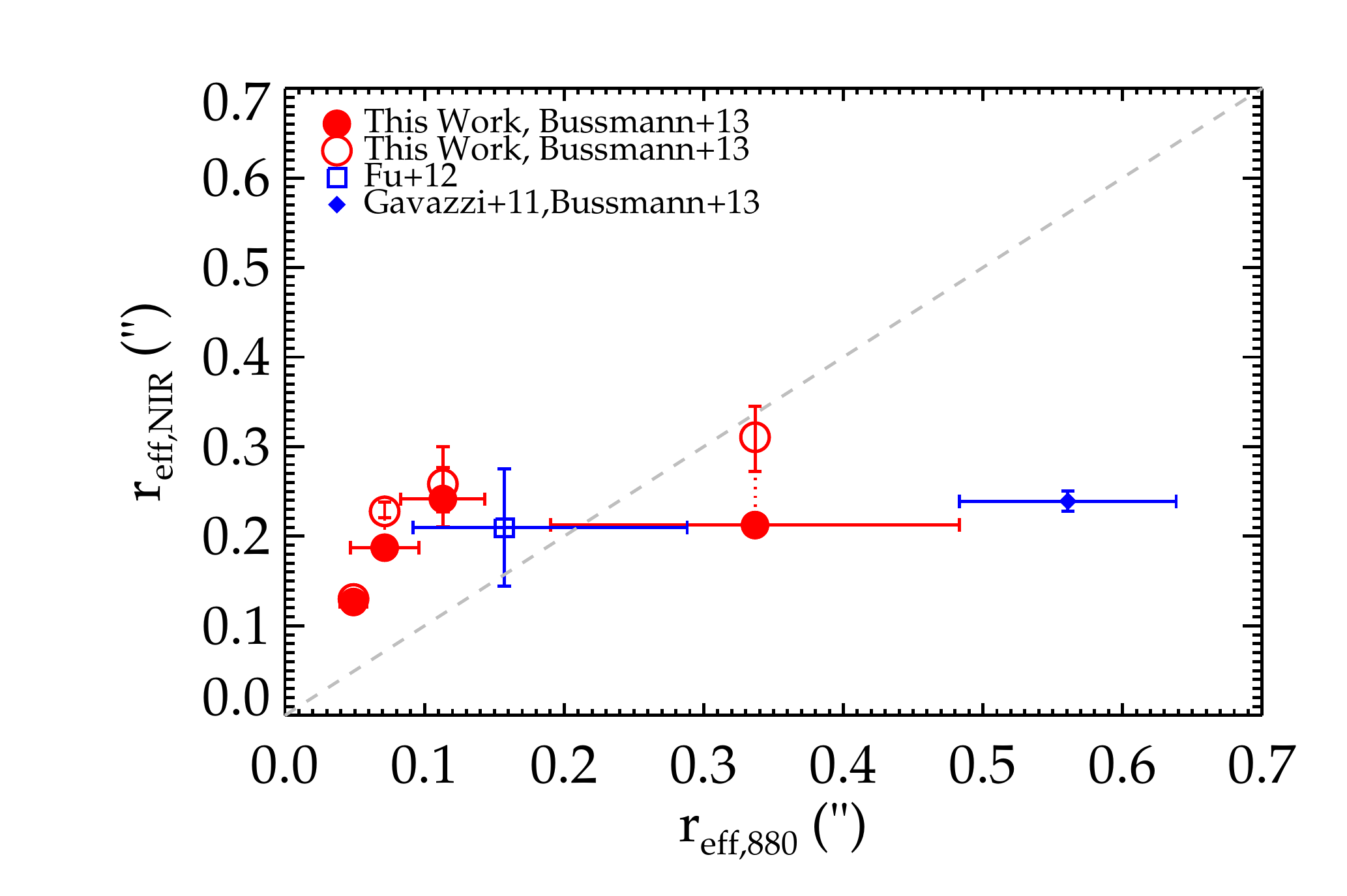} 
\caption{Intrinsic effective radii of lensed SMGs in the near-IR compared with $880\,\mu$m. Filled symbols are from independent analyses in the near-IR and sub-mm. Open symbols denote consistent foreground lens parameters between the near-IR and sub-mm. Here, the foreground lens parameters are fixed to those derived from the sub-mm \citep{Bussmann13}. Most of the SMGs lie above the line of one-to-one correspondence (dashed line), showing that their dust emission is typically less extended than the rest-frame optical (likely stellar) emission. This is consistent with the observed differential magnification (Fig.~\ref{fig:dl}), and suggests that smaller emission regions are generally more highly magnified.}
\label{fig:sizes}
\end{minipage}
\end{figure}

Lensing magnification values are generally negatively correlated to intrinsic sizes of the lensed background source. Therefore, Fig.~\ref{fig:dl} could suggest that the near-IR emission regions in lensed SMGs are larger than sub-mm emission regions in the source plane. Physically, this could imply that the lensed dusty star-forming regions have clumpier morphologies than the older stellar distribution. We further explore this, by showing in Fig.~\ref{fig:sizes}  the circularized effective radius ($r_{\rm{eff}}=\sqrt{a_{\rm{eff}}b_{\rm{eff}}}$) of the most extended background component in our near-IR models compared with the sub-mm emission. Indeed, in most cases the dust emission does appear to originate from a smaller region than the stellar light (as proxied by the observed frame near-IR data).

It is difficult to assess whether the disagreement at larger values of $r_{\rm{eff},880}$ is generally true for lensed SMGs. Lensed sources that are intrinsically extended in the sub-mm are also less magnified, which means a lower probability for detection in near-IR observations. HLock04 is the only source from our analysis with a smaller measured intrinsic size in the near-IR relative to the sub-mm, which could be due to the uncertainty in the observed sub-mm lensing configuration as discussed in the Appendix. The results of Fig.~\ref{fig:dl} and~\ref{fig:sizes} could be a direct consequence of the bias that exists in selecting lensing events in the sub-mm. Simulations predict that detections of sub-mm selected gravitationally lensed galaxies are subject to an angular size bias towards the most compact emission regions that are both comparable to the size of, and near the source-plane caustics \citep{Hezaveh12,Serjeant12,Lapi12}. The bias towards compact sub-mm sources translates to larger values of $\mu_{880}$. However, this effect is reduced in the near-IR and hence contributes to the deviation from the one-to-one correspondence line in Fig.~\ref{fig:dl}. If this bias has the same effect on sources that are less amplified, more extended sources in the sub-mm \citep{Bussmann13}, then its possible that our result in Fig.~\ref{fig:sizes} could also hold true for larger values of $r_{\rm{eff,880}}$.

Spatially resolved radio and gas/dust continuum observations \citep{Chapman04,Biggs08, Ivison08, Tacconi08, Engel10} of SMGs have measured the emission due to star-formation to be as extended as $\sim10$ kpc. This is also in agreement with high-resolution sub-mm observations \citep{Younger08,Younger09,Hodge13}. While in the near-IR regime, SMGs have a typical size range of $2-4$ kpc \citep{Swinbank10,Targett11,Targett13,Aguirre13}. For our sample of lensed SMGs that overlap in the near-IR and sub-mm, we calculate a median intrinsic physical size of $\sim2$ kpc in the near-IR, compared to $\sim1$ kpc in the sub-mm \citep{Bussmann13}. These results are in contrast to the larger values of the previous findings but could also be demonstrating one of the main drawbacks of galaxy-scale lenses. The area of high magnification in galaxy-scale lenses is smaller compared to cluster-scale lenses, so it is entirely possible that only a sub-region of the total emission in both near-IR and the sub-mm is being amplified and detected. Future high-resolution sub-mm observations using the full capabilities of the Atacama Large Millimeter Array (ALMA) with sub-arcsecond spatial resolutions (0.10-0.4$''$) will be able to confirm this by measuring the sizes of star-forming clumps in unlensed SMGs.

Figures~\ref{fig:dl} and~\ref{fig:sizes} also give a measure of the variation of {\munir} and $a_{\rm{eff}}$ from performing lens models independently (i.e., without using $880\,\mu$m parameters). On average, using $880\,\mu$m foreground lens parameters to derive magnification factors and intrinsic sizes are in agreement relative to our independent analysis to within $\sim30\%$. Less deviation is observed in the magnification measurements when the lensing morphology provide strong constraints and show similar configurations in both the sub-mm and near-IR. 


\begin{figure}
\begin{minipage}[b]{0.5\textwidth}
\includegraphics[width=\linewidth,keepaspectratio]{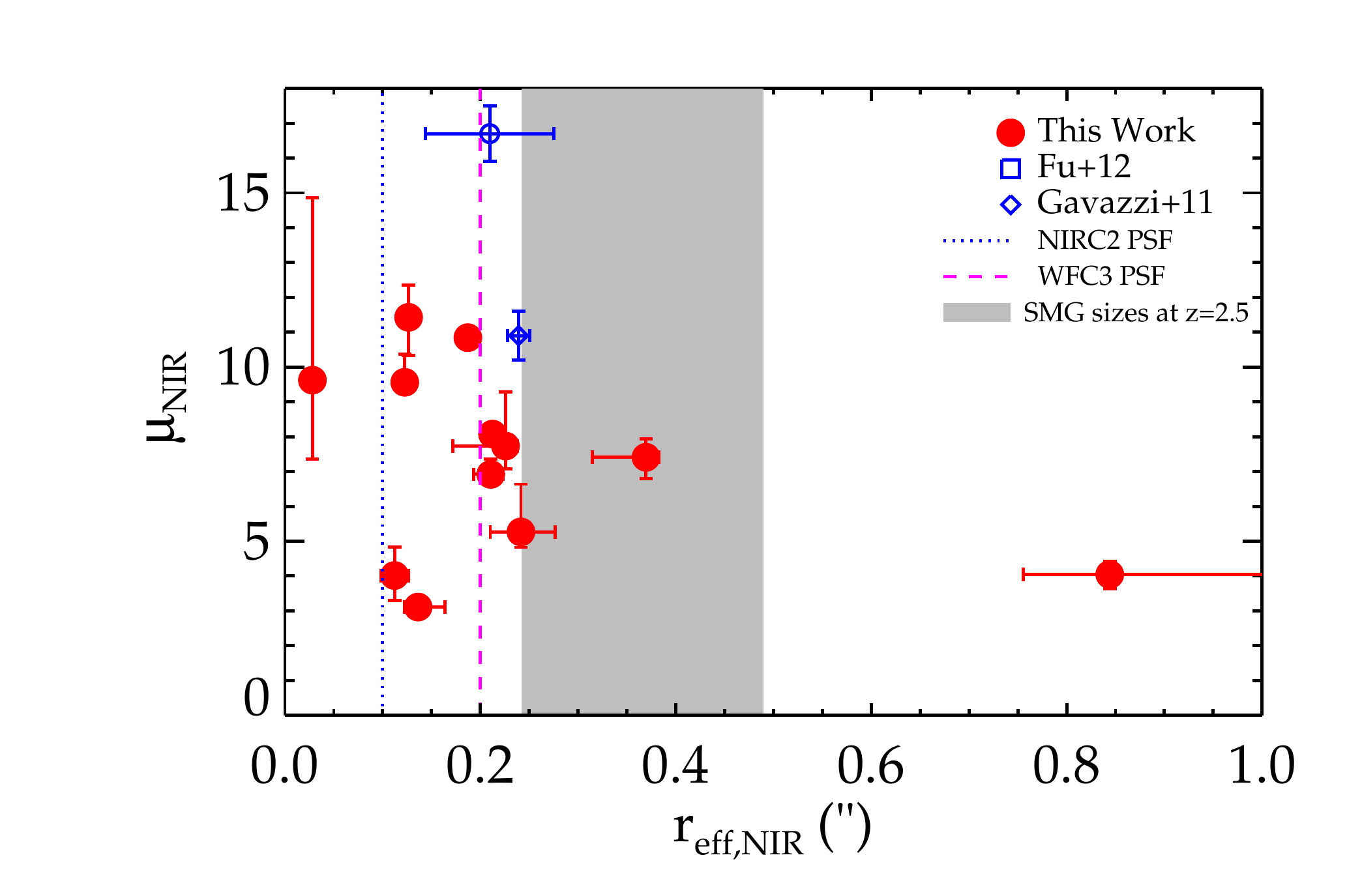} 
\caption{Magnification and intrinsic effective radius in the near-IR for lensed SMGs. For sources with multiple components, we plot the most extended component. Vertical dashed lines show typical spatial resolutions of our NIRC2-LGS/AO and {\it HST} F110W WFC3 data. The grey shaded region covers the range of $2-4$ kpc for unlensed $880\,\mu$m-selected SMGs at $z=2.5$, based on high resolution near-IR analyses of \citet{Swinbank10,Targett11,Targett13}, and \citet{Aguirre13}. A size bias for sub-mm selected lensing systems is observed in the near-IR, in which compact sources typically have larger magnifications. The near-IR emission for \herschel-selected lensed SMGs is generally more compact than previous size measurements of unlensed classical SMGs.}
\label{fig:muvssize}
\end{minipage}
\end{figure}

The analysis of \herschel-selected SMGs in \citet{Bussmann13} confirmed the angular size bias present in sub-mm selected lensing systems. We investigate whether this bias also affects near-IR observations of lensed SMGs in Fig.~\ref{fig:muvssize}, where we show the observed near-IR magnification factors against the intrinsic size of the lensed galaxy. For objects with multiple components, we use the one with the largest angular size. We find a hint of negative correlation between magnification factors and size, albeit with large scatter, but consistent with simulations and sub-mm observations.

In Fig.~\ref{fig:muvssize} we also highlight sizes of $0.24''-0.48''$, which corresponds to 2-4 kpc at $z=2.5$, the range measured for the observed-frame near-IR median sizes of 850 $\mu$m selected unlensed SMGs \citep{Chapman03, Swinbank10, Aguirre13, Targett13}. Few of our targets are more extended than this, and most are smaller than $0.24''$. If 500 $\mu$m selected lensed SMGs are evolutionarily similar to unlensed 850  or $880\,\mu$m-selected galaxies (as is likely, since the sample from~\citet{Bussmann13} have $S_{880}\ge4$ mJy, when corrected for magnification, comparable to the classical SMG selection. Also, see Section~\ref{sec:mb} for a discussion), then it appears that the lensed galaxies are preferentially those with the smallest near-IR emission regions. Thus, it appears that the sub-mm selection method, which is biased towards the highest sub-mm fluxes, and therefore highest sub-mm magnifications and smallest intrinsic sub-mm emission region \citep{Bussmann13} also selects the galaxies with the most intrinsically compact near-IR emission regions. This follows from Fig.~\ref{fig:dl}, which shows a correlation between {\munir} and {\musubmm}. 

\begin{figure}
\begin{minipage}[b]{0.5\textwidth}
\includegraphics[width=\textwidth,keepaspectratio]{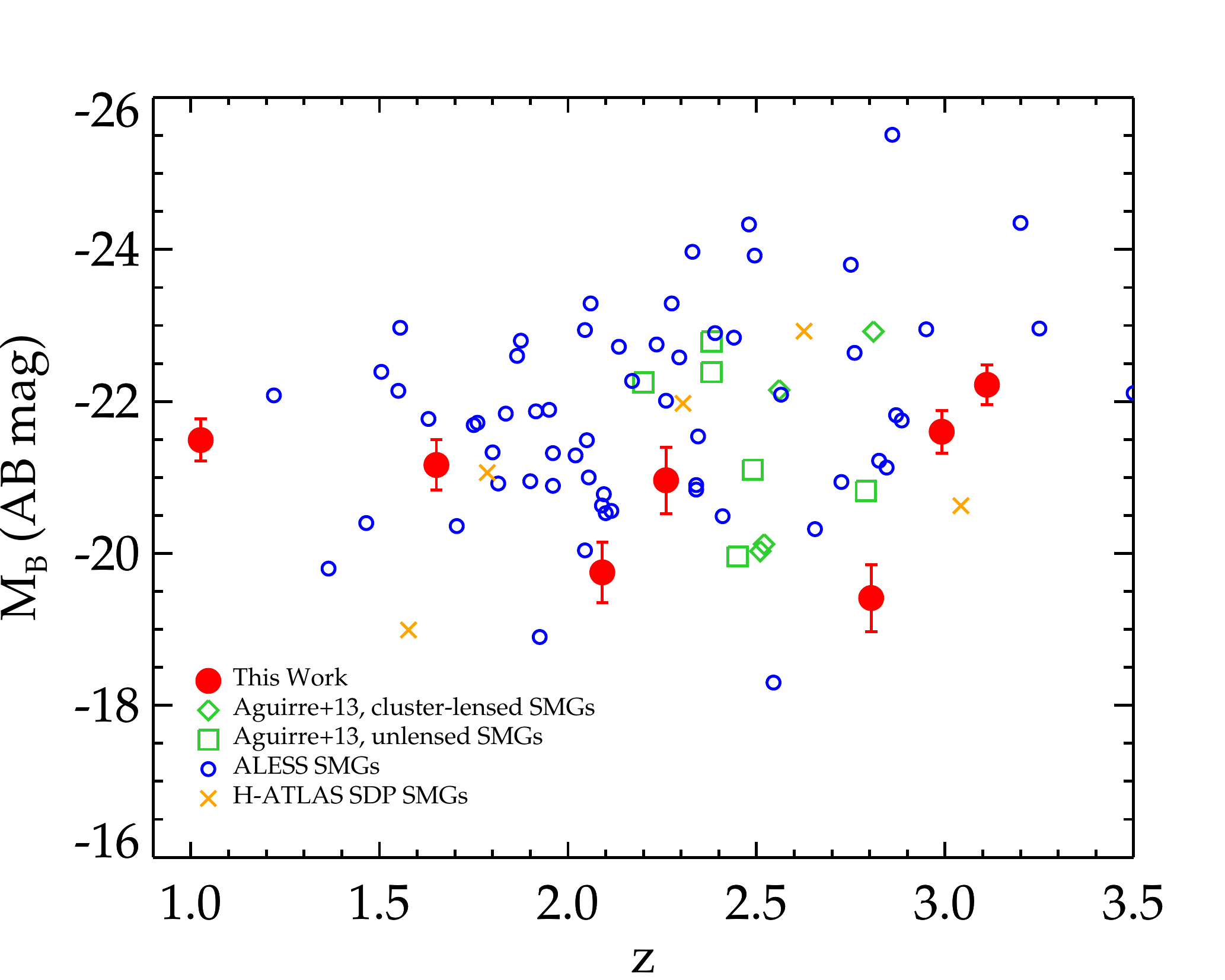}
\caption{Rest-frame magnification-corrected absolute $B$-band magnitudes ($M_{B}$) for Grade A1 and A2 candidates versus redshift.  Open diamonds and squares represent cluster-lensed and unlensed SMGs from \citet{Aguirre13}, respectively. Open circles are unlensed ALESS SMGs from \citet{Simpson13}. The $M_{B}$ values for lensed SMGs are consistent with unlensed SMGs at $z>1$, but tend to lie towards the fainter end of the distribution.}
\label{fig:mbvsz} 
\end{minipage}
\end{figure}
\begin{figure}
\begin{minipage}[b]{0.5\textwidth}
\includegraphics[width=\textwidth,keepaspectratio]{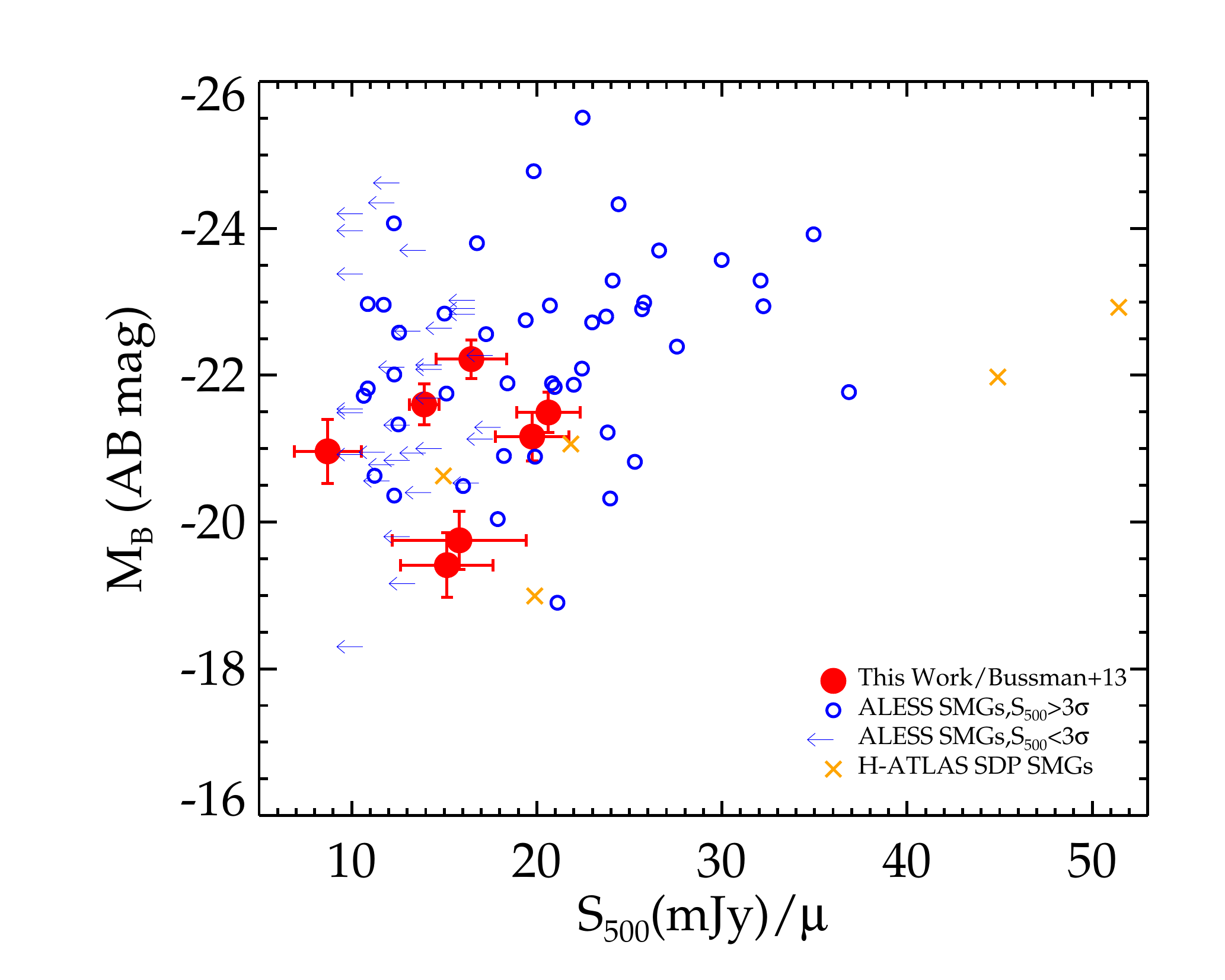}
\caption{Rest-frame magnification-corrected absolute $B$-band magnitudes ($M_{B}$) for Grade A1 and A2 candidates versus magnification corrected SPIRE $S_{500}$. We use the sub-mm magnification from \citet{Bussmann13} when available to correct for the observed $S_{500}$.  Open circles are unlensed ALESS SMGs from \citet{Swinbank14} and \citet{Simpson13}. Our sample of lensed SMGs have consistent $S_{500}$ values for a given $M_{B}$ relative to the unlensed population, suggesting that \herschel-selected lensed SMGs are similar to classical unlensed $850\,\mu$m-bright SMGs.}
\label{fig:mbvs500} 
\end{minipage}
\end{figure}

In our sample of lensed SMGs, we calculate a median intrinsic size of $2.3$ kpc for sources with secure redshifts and if we include sources with photometric redshifts derived from SPIRE colors (Grade A3 and A4 sources), this number is reduced to $1.9$ kpc. If we also assume that the photometric redshift subset have a redshift range of $z=1-4$ \citep{Chapman05, Chapin09,Wardlow11, Michalowski12b, Yun12, Wardlow13, Simpson13, Umehata14}, the maximum angular size scale variation is $\sim1.5$ kpc arcsec$^{-1}$, which we use to constrain a minimum and a maximum median intrinsic size of $1.66$ and $2.03$ kpc for our whole sample. This difference is not significant, given the large uncertainties associated with photometric redshifts. These values are smaller than the median sizes previously found for $850\,\mu$m selected SMGs in the near-IR ($r_{\rm{eff}}=2.5-2.7$ kpc, \citealt{Aguirre13}; $r_{\rm{eff}}=4.0$ kpc,\,\citealt{Targett13}, $r_{\rm{eff}}=3.1$ kpc,\,\citealt{Targett11}; $r_{\rm{eff}}=2.3-2.8$ kpc,\,\citealt{Swinbank10}). Although the smaller measured intrinsic size could be due to the sub-mm size bias, our study of lensed SMGs is performed at spatial resolutions well above the observational limitations of the current near-IR facilities. Therefore, if we are observing the total near-IR emission from the background source, it could represent the typical size scales for this galaxy population.

\subsection{Morphological Comparison With Previous Near-IR Studies of SMGs}

Previous studies of SMGs in the rest-frame optical have revealed a variety of morphologies.~\citet{Aguirre13}  identified that five out of 10 SMGs from their sample observed with {\it HST}/WFC3's F110W and F160W band have multiple components. The stellar mass ratio calculations of these multi-component SMGs showed that they could be associated with major and minor mergers. In contrast, the same study found that some of their most massive SMGs are single-component systems and have morphologies that resemble hydrodynamic simulations of rapidly star-forming galaxies \citep{Dave10}. In agreement with this picture are the near-IR observations of unlensed SMGs in \citet{Targett11} and \citet{Targett13}, in which SMGs appeared to be compact star-forming disks and are simply extreme examples of normal star-forming galaxies at $z\sim2.5$. Many factors can provide an explanation for this discrepancy: varying levels of dust-obscuration in the rest-frame optical that gives rise to distinct observed morphologies \citep{Swinbank10}; SMGs or their substructure having intrinsic sizes that are either comparable or smaller than the measured seeing could cause them to appear smoothed; or SMGs could simply be a heterogeneous sample with different galaxy formation mechanisms. 

The boost in both flux and spatial resolution from gravitational lensing should reduce some of the limiting factors present in previous studies, provided that there are no significant morphological differences between $500\,\mu$m-selected and $850\,\mu$m-selected SMGs. Indeed, this assumption is justified since \citet{Bussmann13} shows that the majority of the $500\,\mu$m selected lensed SMG sample observed in $880\,\mu$m have magnification corrected flux densities consistent with $S_{880}\ge 4$~mJy. Of the 12 systems with lens models featured here, four are best fit with multiple components in the source plane. In three of these systems (NB.v1.78, H{\bootes}02, G15v2.19) the rest-frame optical SMG consists of a smaller component embedded in a larger one. We note that these multiple component systems also place a lower limit on the size of substructure ($0.02'' \sim 0.2$ kpc at $z\ge1$) found in lensed SMGs, which would otherwise not be readily detected with current instrumentation. 

These compact components could be interpreted as SMGs hosting an optically-bright AGN, small regions of star-formation embedded in a larger galaxy, or the remnants of a merger. Our findings suggest that near-IR studies of unlensed SMGs described as single components could have complicated morphologies that are unresolved even when using instruments that offer the highest spatial resolution. The morphologies of the SMGs in \citet{Aguirre13} could support this claim, given that all their single component SMGs are unlensed and four out of five with multiple components are lensed by a nearby cluster.  We also note that HECDFS02 is similar to the SMGs shown in \citet{Chapman03}; however a more accurate redshift and velocity information for each individual component is needed to confirm if this source is indeed in the process of a major-merger.

The remaining eight gravitationally lensed galaxies in our Grade A sample are composed of a single component that dominates the surface brightness profile of the background source, consistent with the axisymmetric models in \citet{Targett11,Targett13,Aguirre13} and simulated SMGs in \citet{Dave10}. We note that five systems have excess flux in the residual images, which could be due to some substructure in the background galaxy, although our data cannot robustly determine whether this, or substructure in the foreground lens is responsible. The median {\sersic} index for the subset that are best fit with a single component is $n\sim2.5$, a significant deviation from the disk-like morphologies in \citet{Targett13} ($n\sim1.5$) but comparable with the measured values from \citet{Swinbank10} ($n\sim2.0$). However, we note that the statistical uncertainties associated with the best-fit \sersic~indices, which are on the level of $10-30\%$ is likely underestimated since it does not account for the assumptions used in the lens modeling that can affect the morphology of the background source, such as the shape of the PSF or the assumed mass profile.

\begin{deluxetable*}{ l c c c c c c c c c }[h!]
\tablecolumns{10}
\tablewidth{8.0in}
\tablecaption{Near-IR Photometry of Lensed SMGs}
\tablehead{
\colhead{Name} &
\colhead{$F$\tnm{a}$_{\rm{F110W}}$} &
\colhead{$\sigma$\tnm{b}$_{\rm{F110W,stat.}}$} &
\colhead{$\sigma$\tnm{c}$_{\rm{F110W,tot.}}$} &
\colhead{$F$\tnm{a}$_{H}$} &
\colhead{$\sigma$\tnm{b}$_{H,\rm{stat.}}$} &
\colhead{$\sigma$\tnm{c}$_{H,\rm{tot.}}$} &
\colhead{$F$\tnm{a}$_{K_{s}}$} &
\colhead{$\sigma$\tnm{b}$_{K_{s},\rm{stat.}}$} &
\colhead{$\sigma$\tnm{c}$_{K_{s},\rm{tot.}}$}
\\
\colhead{} &
\colhead{($\mu$Jy)} &
\colhead{($\mu$Jy)} &
\colhead{($\mu$Jy)} &
\colhead{($\mu$Jy)} &
\colhead{($\mu$Jy)} &
\colhead{($\mu$Jy)} &
\colhead{($\mu$Jy)} &
\colhead{($\mu$Jy)} &
\colhead{($\mu$Jy)}
}
\startdata
NB.v1.78    &       \nodata &       \nodata &       \nodata &       2.5 &      0.1 & 0.2 &        3.9    &        0.1    &        0.2     \\     
HLock12      &        3.5    &        0.4    &        0.7    &       \nodata &       \nodata &       \nodata &       \nodata &       \nodata &       \nodata  \\     
HLock06      &       \nodata &       \nodata &       \nodata &       \nodata &       \nodata &\nodata&        2.4    &        0.2    &        0.2     \\     
G15v2.19    &       \nodata &       \nodata &       \nodata &       14.17 &       2.5 & 2.5 &       12.4    &        1.0    &        1.7     \\     
H{\bootes}02\tnm{b}    &       $<0.12$\tnm{a}&       \nodata &       \nodata &       $<0.36$\tnm{a} &       \nodata & \nodata &        2.5    &        0.7    &        1.4     \\     
HFLS08     &        0.7    &        0.2    &        0.2    &       \nodata &       \nodata &       \nodata &       \nodata &       \nodata &       \nodata  \\     
HCOSMOS01\tnm{b}    &       0.49 &       0.3 &       0.3 &       \nodata &       \nodata &       \nodata &        2.8    &        1.5    &        1.5    \\      
HLock04      &       0.5 &       0.1 &       0.1 &       3.0 &       0.5 &       0.5 &        6.1    &        0.2    &        0.5    \\      
HFLS02      &        1.0    &        0.1    &        0.2    &       \nodata &       \nodata &       \nodata &       \nodata &       \nodata &       \nodata \\      
HECDFS05    &        0.5    &        0.1    &        0.2    &       \nodata &       \nodata &       \nodata &       \nodata &       \nodata &       \nodata \\      
HECDFS02    &        0.9    &        0.1    &        0.2    &       \nodata &       \nodata &       \nodata &       \nodata &       \nodata &       \nodata \\      
\enddata
\tablecomments{The following columns describe the near-IR photometry: $F=$ measured flux density, corrected for magnification, $\sigma_{\rm{stat}}=$ 1$\sigma$ error due to statistical noise, which accounts for the error in the background and magnification, $\sigma_{\rm{tot}}=$  Total noise, which accounts for both systematic and statisical errors. Systematic errors are dominated by the zero-point derivations from UKIDSS flux calibrations. }
\tnt{a}{These values represent $3\sigma_{\rm{tot}}$ limits.}
\tnt{b}{The measured errors for these sources are dominated by the error in their magnification values.}
\label{tab:nirphot}
\end{deluxetable*}

\subsection{Rest-Frame Optical Photometry}
\label{sec:mb}
Given the average redshift of our sample ($z\sim2.5$) and the fact that half of the Grade A sources we present are only observed in a single near-IR band, it is impossible to derive  well-constrained physical quantities (e.g., stellar masses) without making sweeping assumptions about the effects of dust extinction, different star-formation histories, and inferred mass-to-light ratios of the near-IR SED. Instead, we opt to report observable quantities to minimize sources of systematic uncertainty and aim to use this paper as a starting point for future studies once sufficient multi-wavelength data have been acquired. The rest-frame wavelength range in the observed $J$ and $K$ band of our Grade A candidate lensing systems with secure redshifts (Grade A1 and A2) corresponds to $\sim0.3-0.6\,\mu$m. We use SMG SED templates from \citet{Michalowski10} and our measured magnification corrected photometry, listed in Table~\ref{tab:nirphot}, to extrapolate the rest-frame $B$ band ($\lambda=0.450\,\mu$m) flux density. To measure the uncertainty of our extrapolated $B$-band magnitudes, we perform the same calculation using the near-IR data from the H-ATLAS SDP sample in \citet{Negrello14} and calculate the scatter between the values using our fitting method and from their best-fit SED. On average, we find that the extrapolated $B$-band values are in agreement within $0.2$ mag and show this as part of the errors shown in Fig.~\ref{fig:mbvsz} and ~\ref{fig:mbvs500}. For sources with one near-IR band, we simply normalize the SEDs to the observed datapoint and quote the average redshifted $B$ band flux density and the standard deviation as an additional source of error. 

Figure~\ref{fig:mbvsz} shows that the magnification-corrected $B$-band absolute magnitudes ($M_{B}$) for our lens Grade A1 and A2 sources are consistent with both $880\,\mu$m and $500\,\mu$m selected unlensed and lensed SMG samples \citep{Simpson13, Negrello14, Aguirre13}, with our sample typically on the fainter end of the distribution.  We obtain a similar result in Fig.~\ref{fig:mbvs500} if we compare magnification corrected $500\,\mu$m flux densities. For sources that have a lens model from \citet{Bussmann13}, we use the sub-mm magnification factors to correct for the observed $S_{500}$, otherwise we use the values from the near-IR lens modeling. Our sample typically has intrinsic $S_{500}\le20$ mJy, which corresponds to the $\sim3\sigma$ limit (confusion and instrumental noise) for unlensed SMGs \citep{Swinbank14}. 
This result is likely due to the benefits of flux amplification from lensing, which allows fainter objects to be detected at a higher significance. Although we find that lensed SMGs are on average intrinsically fainter in the rest-frame optical and far-IR compared to the unlensed populations, they are consistent with the observed flux distribution. This adds further evidence that the lensed SMGs in this paper are lensed analogs of the unlensed population, consistent with the findings of \citet{Harris12} and \citet{Bussmann13}.

\section{Conclusions}
We have obtained deep, high-resolution near-IR imaging that traces the rest-frame optical emission of 87 $500\,\mu$m$-$bright candidate lensing systems. The main results from our studies are:

\begin{enumerate}
\item Out of the current sample of 87 candidate lensing systems, 15 have definitive features of lensing and are highly prioritized for analysis, with nine, one, three, and two having existing redshifts for both foreground lens and background source (Grade A1), the background source (Grade A2), the foreground lens (Grade A3), and neither (Grade A4), respectively. We find that the Grade A sources typically have larger $500\,\mu$m flux densities (median $S_{500}\sim120$ mJy) than their lower priority counterparts, with median $S_{500}\sim90$ and 80 mJy for Grade B and C sources, respectively. This is expected from the selection method, since galaxies with larger sub-mm flux densities have a higher probability of being lensed. We find that 32\% of the sources with $S_{500}\ge100$ mJy are classified as Grade A, demonstrating a lower success rate in identifying strong lensing events than spatially resolved sub-mm studies of {\it Herschel} SMGs \citep[$\sim80\%$]{Bussmann13}. This is likely due to the rest-frame optical emission suffering heavy dust-obscuration, as well as the varying depth in our observations, being significantly spatially offset from regions of high-magnification in the source plane, or because \herschel-selected SMGs are typically at high redshift. 

\item We generate lens models for 12 Grade A systems to derive near-IR magnification factors and reconstruct the morphologies of SMGs. Our lensed SMGs have an average magnification factor of {\munir} = $7\pm3$ and typically have rest-frame emission that extends out to angular sizes of $0.3''$, which is $\sim2$ kpc at $z\ge1$. For sources with multiple components, we calculate an upper limit of $0.02''$ (0.2 kpc at $z\ge 1$) for the size of substructures within the background galaxy. These angular sizes have been measured as lower limits from previous studies of the unlensed SMGs. While these smaller angular sizes could represent the typical size scales for this galaxy population, it could also be due to the lensing of a subregion that is located near areas of high magnification in the source plane. Future simulations using lens models of mock data with known sizes should resolve this degeneracy.

\item For the subset of sources that overlap with \citet{Bussmann13}, we derive near-IR magnification factors using foreground lens parameters derived in the sub-mm. Differential lensing is observed in all cases, with $\mu_{{\rm NIR}} = \mu_{880}/1.5$, typically. A size comparison reveals that the near-IR background source models are generally $2\times$ more extended than their sub-mm counterparts in the same galaxies. This indicates that the lensed stellar emission regions in SMGs are typically more extended than the lensed dust emission regions, in the same galaxies. 

\item The rest-frame absolute $B$-band magnitude values and $500\,\mu$m flux densities, both corrected for magnification, show that the lensed SMGs are intrinsically similar to unlensed SMGs from previous studies, but with our sources typically at the fainter end of the distribution. 
\end{enumerate}

\begin{acknowledgements}
We would like to thank the anonymous referee and Ian Smail for their thoughtful feedback and insightful comments to improve the paper.

The data presented herein were obtained at the W.M. Keck Observatory, which is operated as a scientific partnership among the California Institute of Technology, the University of California and the National Aeronautics and Space Administration. The Observatory was made possible by the generous financial support of the W.M. Keck Foundation. The authors wish to recognize and acknowledge the very significant cultural role and reverence that the summit of Mauna Kea has always had within the indigenous Hawaiian community.  We are most fortunate to have the opportunity to conduct observations from this mountain.

Support for program $\#$GO-12194 and $\#$GO-12488  was provided by NASA through a grant from the Space Telescope Science Institute, which is operated by the Association of Universities for Research in Astronomy, Inc., under NASA contract NAS 5-26555.

The {\it Herschel}-ATLAS is a project with {\it Herschel}, which is an ESA space observatory with science instruments provided by European-led Principal Investigator consortia and with important participation from NASA. The H-ATLAS website is \textcolor{blue}{http://www.h-atlas.org/}.

This research has made use of data from the HerMES project (\textcolor{blue}{http://hermes.sussex.ac.uk/}). HerMES is a {\it Herschel} Key Programme utilizing Guaranteed Time from the SPIRE instrument team, ESAC scientists and a mission scientist. The data presented in this paper will be released through the HerMES Database in Marseille, HeDaM (\textcolor{blue}{http://hedam.oamp.fr/HerMES/}).

SPIRE has been developed by a consortium of institutes led by Cardiff Univ.\,(UK) and including: Univ. Lethbridge (Canada); NAOC (China); CEA, LAM (France); IFSI, Univ. Padua (Italy); IAC (Spain); Stockholm Observatory (Sweden); Imperial College London, RAL, UCL-MSSL, UKATC, Univ. Sussex (UK); and
Caltech, JPL, NHSC, Univ. Colorado (USA). This development has been supported by national funding agencies: CSA (Canada); NAOC (China); CEA, CNES, CNRS (France); ASI (Italy); MCINN (Spain); SNSB (Sweden); STFC, UKSA (UK); and NASA (USA). 

JAC, AC, BM, CMC, JMO, NT, and CT acknowledge support from NSF AST-1313319.

M.N. acknowledges financial support from ASI/INAF agreement I/072/09/0 and from PRIN-INAF 2012 project ÔLooking into the dust-obscured phase of galaxy formation through cosmic zoom lenses in the {\it Herschel} Astro- physical Large Area SurveyÕ

LD, SJM and RJI acknowledge support from the European Research Council (ERC) in the form of Advanced Investigator programme,
{\sc cosmicism}

This work was supported in part by the National Science Foundation under Grant No. PHYS-1066293 and the hospitality of the Aspen Center for Physics.

Support for CARMA construction was derived from the Gordon and Betty Moore Foundation, the Kenneth T. and Eileen L. Norris Foundation, the James S. McDonnell Foundation, the Associates of the California Institute of Technology, the University of Chicago, the states of California, Illinois, and Maryland, and the National Science Foundation. Ongoing CARMA development and operations are supported by the National Science Foundation under a cooperative agreement, and by the CARMA partner universities.

The Sub-millimeter Array is a joint project between the Smithsonian Astrophysical Observatory and the Academia Sinica Institute of Astronomy and Astrophysics and is funded by the Smithsonian Institution and the Academia Sinica.

The Dark Cosmology Centre is funded by the Danish National Research Foundation (DNRF).

\end{acknowledgements}
\label{sec:conc}


\appendix
\section{Lens Models using Sub-mm Parameters}
\begin{figure*}[h]
\begin{minipage}[b]{0.5\textwidth}
\includegraphics[width=\linewidth,keepaspectratio]{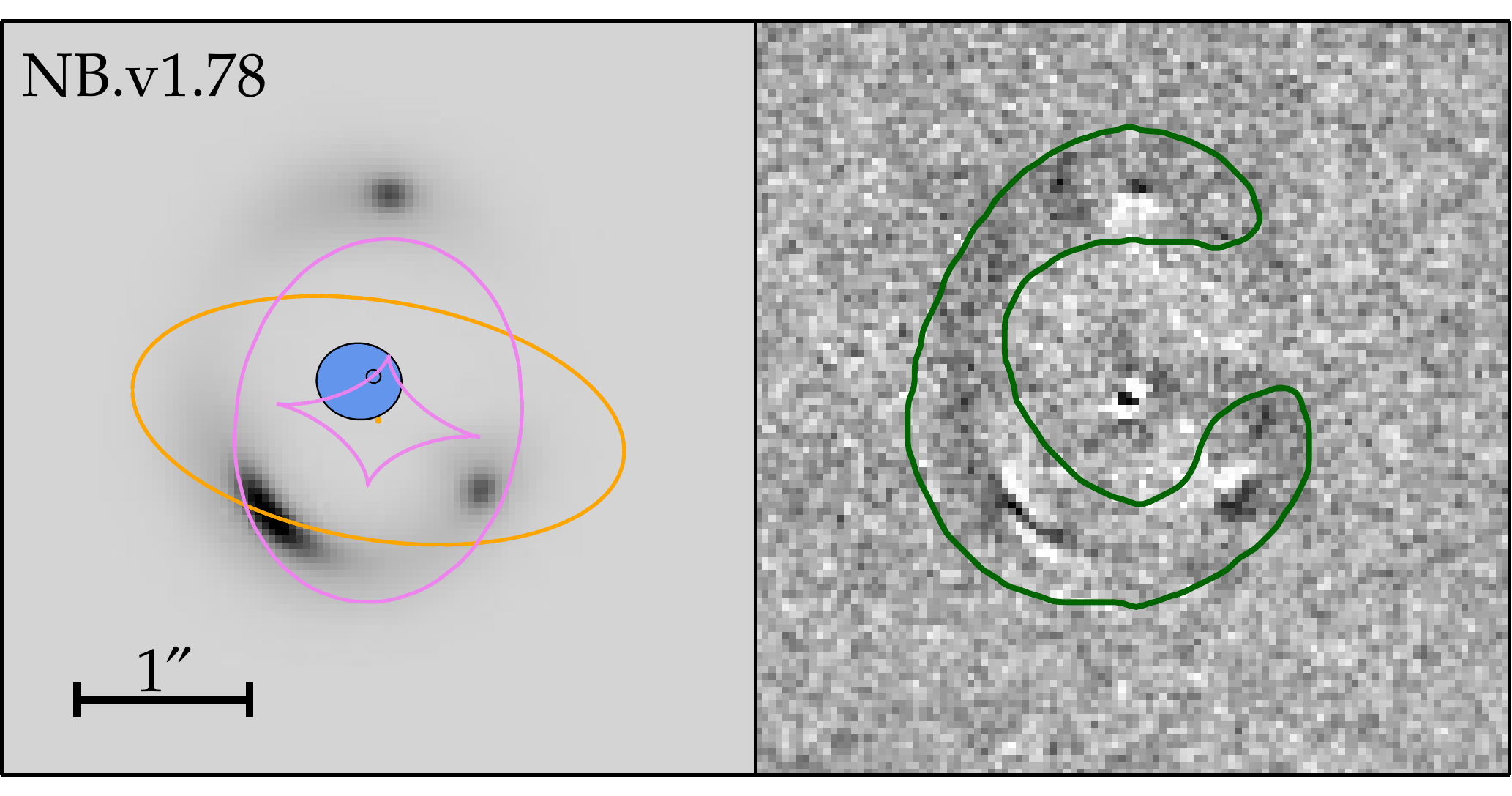} 
\includegraphics[width=\linewidth,keepaspectratio]{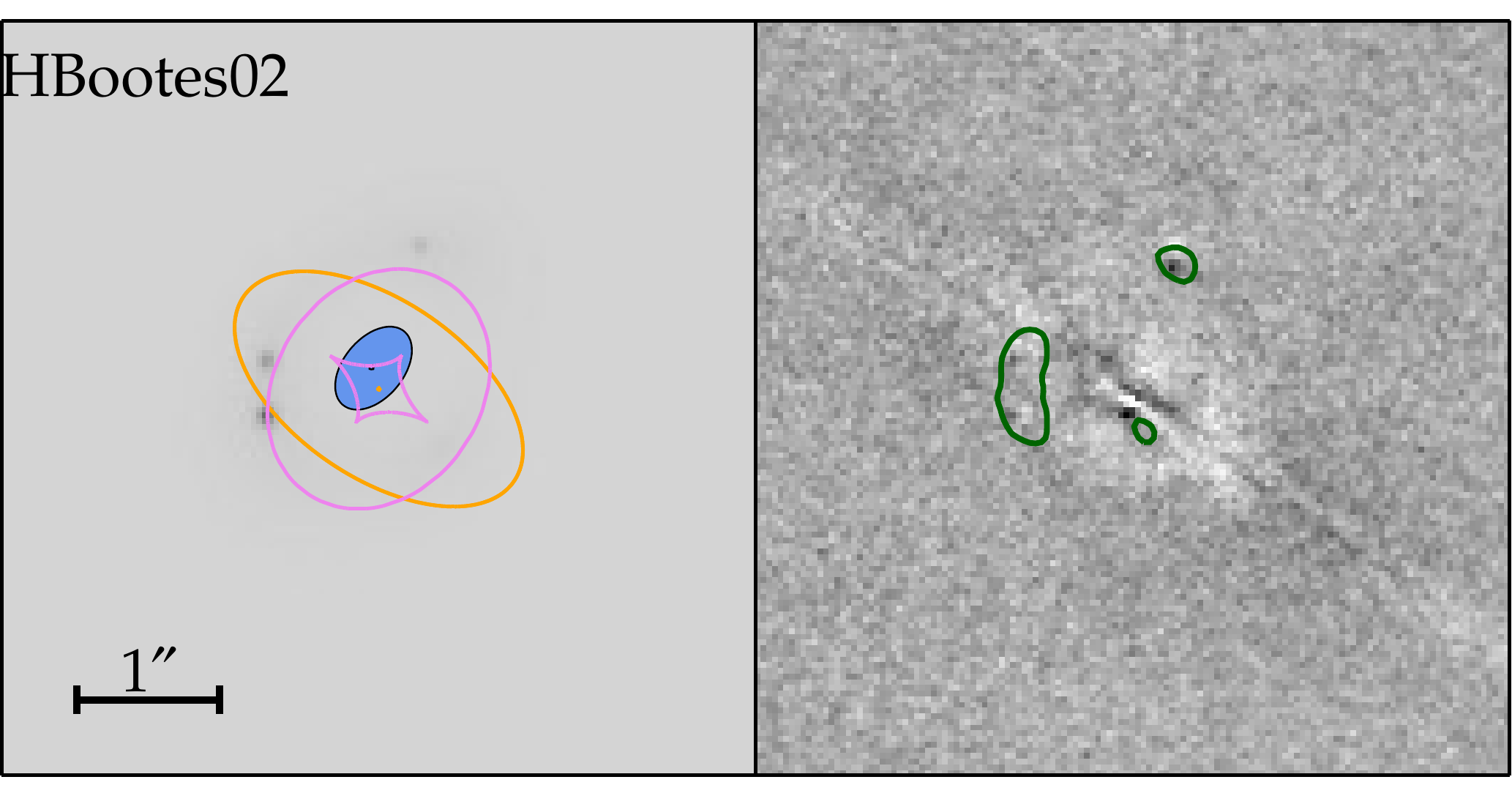} 
\end{minipage}
\begin{minipage}[b]{0.5\linewidth}
\includegraphics[width=\linewidth,keepaspectratio]{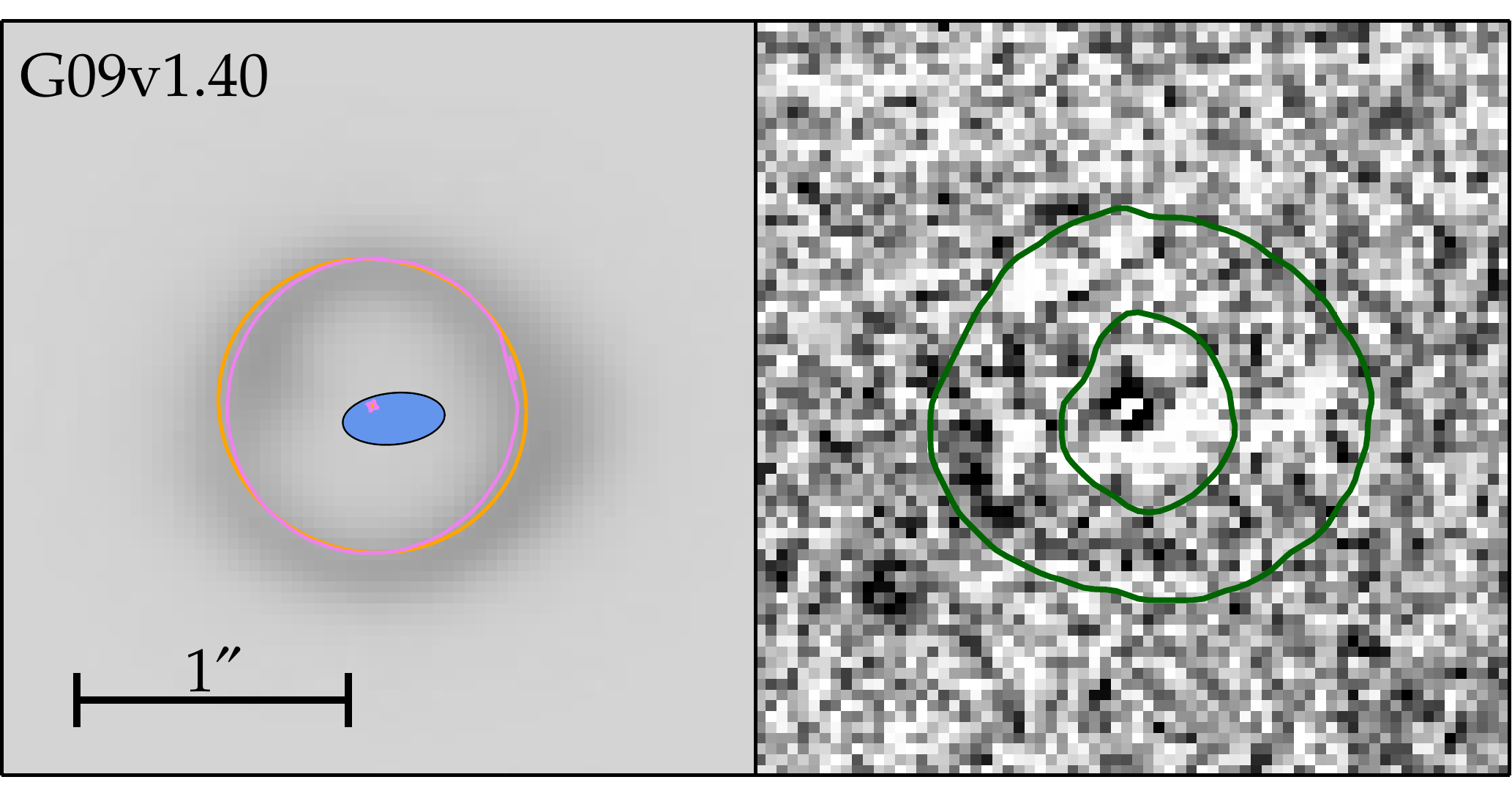} 
\includegraphics[width=\linewidth,keepaspectratio]{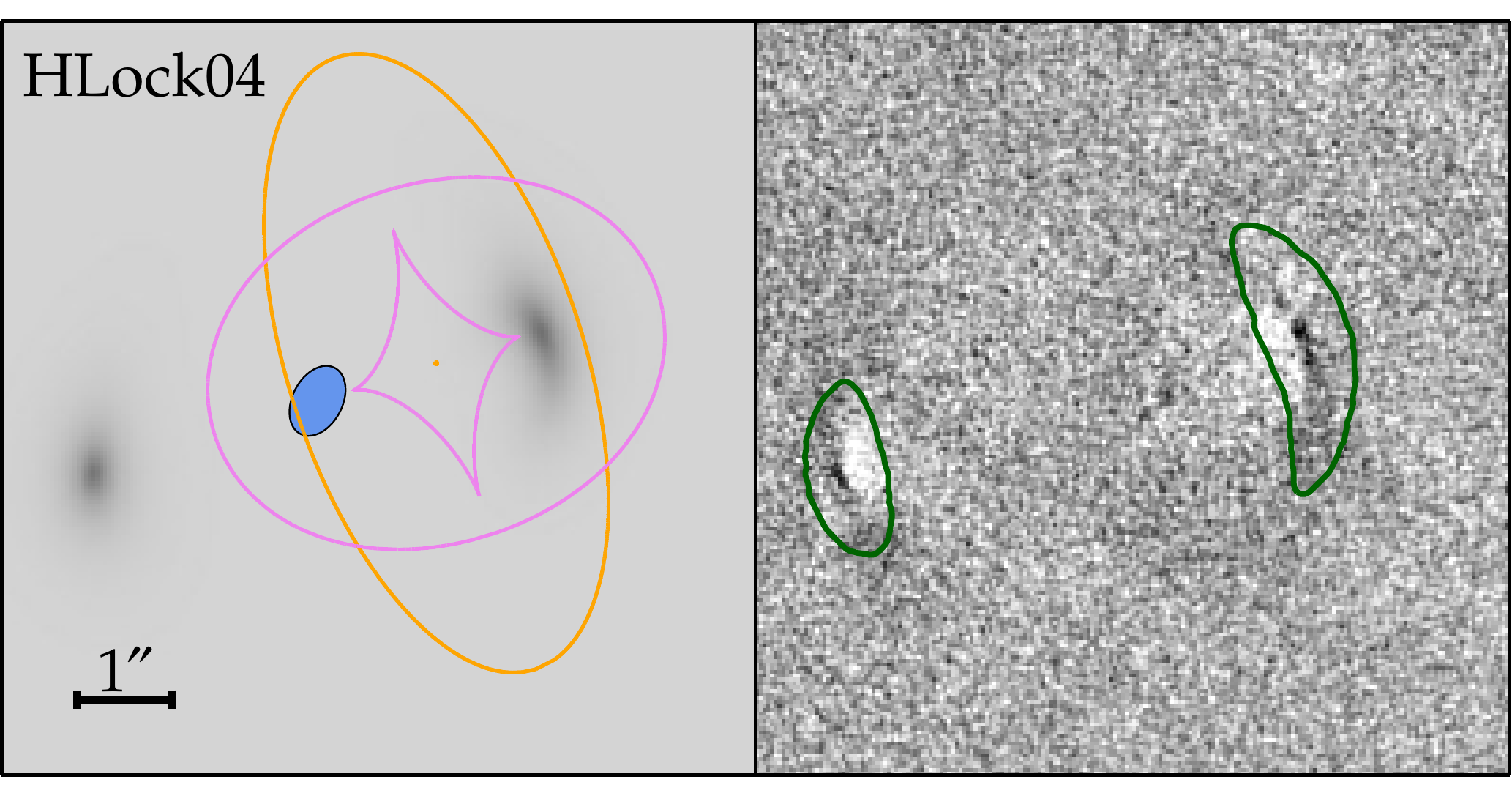} 
\end{minipage}
\caption{Lens models of the lens Grade A subsample that overlaps with \citet{Bussmann13}, with the foreground lens parameters fixed to sub-mm derived values. The images are displayed at the same scale as their Fig.~\ref{fig:lm} counterparts. North is up and east is left for all the panels.}
\label{fig:dl_lm}
\end{figure*}

In this section we describe the lens models shown in Fig.~\ref{fig:dl_lm} and summarized in Table~\ref{tab:dl} for four sources that also have sub-mm data. We fix their foreground lens parameters to sub-mm derived values \citep{Bussmann13} as a test for differential magnification as discussed in Section~\ref{sec:dl}.

\textbf{NB.v1.78:} The near-IR data is more poorly fit, with $\chi_{\nu}^{2} = 1.08$ compared to $\chi_{\nu}^{2} = 0.77$  for our original solution. The lens model is able to reproduce the configuration demonstrated by the brightest knots, similar to the sub-mm emission. However, it fails to fully account for the extended emission producing the fainter Einstein ring.

\textbf{H{\bootes}02:} A similar configuration with an incomplete quad can be reproduced using sub-mm foreground lens parameters. However, the position of the northern counter-image is offset by $\sim0.1''$, which is a significant offset, since it is comparable to the size of the NIRC2 PSF. The orientation of the extended component in the source plane compared to the original model is significantly different, offset by $\sim90^{o}$. This could indicate that the observed configuration of the fainter extended emission in the image plane causes the lens model to be poorly constrained.

\textbf{G09v1.40:} A consistent result compared to our original near-IR model is obtained if we instead model the system using sub-mm foreground lens parameters. We measure a marginally lower magnification ($\mu_{\rm{NIR}}=10.8^{+0.9}_{-1.1}$), although is comparable to the sub-mm magnification value ($\mu_{880}=15.3\pm3.5$). 

\textbf{HLock04:} The overall fit is significantly degraded ($\chi^{2}_{\nu}=1.27$ compared to the original $\chi^{2}_{\nu}=0.63$ ) when sub-mm foreground lens parameters are used. However, this is because the larger beam size of the $880\,\mu$m image shows a configuration that is less constrained. While the near-IR image shows a clear double arc configuration, the sub-mm image is more ambiguous and the model from  \citet{Bussmann13} statistically favors the cusp-configuration. 

\begin{deluxetable*}{l c c c c} 
\tablecolumns{5}
\tablewidth{\textwidth}
\tablecaption{Properties of Background Galaxies Using Sub-mm Foreground Lens Parameters} 
\tablehead{
\colhead{Name} & 
\colhead{$\epsilon_{\rm{s}}$} &
\colhead{$a_{\rm{eff}}$} &
\colhead{$\mu_{\rm{NIR}}$} &
\colhead{$\chi^{2}_{\nu}$} 
\\
\colhead{} & 
\colhead{} & 
\colhead{$''$} & 
\colhead{} &
\colhead{} 
}
\startdata
NB.v1.78 & $0.11^{+0.04}_{-0.02}$ & $0.24^{+  0.01}_{    -0.01}$  &  $8.8^{+     0.2}_{    -0.1}$ & 1.08 \\
\nodata & $     0.09^{+     0.03}_{    -0.02} $ & $     0.035^{+     0.001}_{    -0.001} $ & \nodata & \nodata \\
 H{\bootes}02 & [0.0] & $0.010^{+     0.001}_{    -0.001} $ & $  7.6^{+     1}_{    -0.4}$ & 1.75 \\
 \nodata &  $     0.4^{+     0.1}_{    -0.1} $ & $     0.33^{+     0.05}_{    -0.03} $ & \nodata & \nodata \\
 G09v1.40 &  $     0.51^{+     0.03}_{    -0.1} $ & $  0.18^{+     0.01}_{    -0.01} $ & $        10^{+         1}_{        -1}$ & 0.63 \\
 HLock04 & $     0.3^{+     0.1}_{    -0.1} $ & $     0.38^{+     0.04}_{    -0.04} $ & $     4.2^{+     0.5}_{    -0.2}$ & 1.27 \\
\enddata
\label{tab:dl} 
\end{deluxetable*}

\clearpage
\begin{figure*}[h]
\section{Supplementary Near-IR Images}
In Figures~\ref{fig:lg2} and~\ref{fig:lg3} we show high-resolution near-IR images of Grade B and C sources, respectively. Figure~\ref{fig:mw} shows the currently available high-resolution multi-wavelength near-IR data for Grade A sources, which we use to measure near-IR photometry.\\

\includegraphics[width=\textwidth,keepaspectratio]{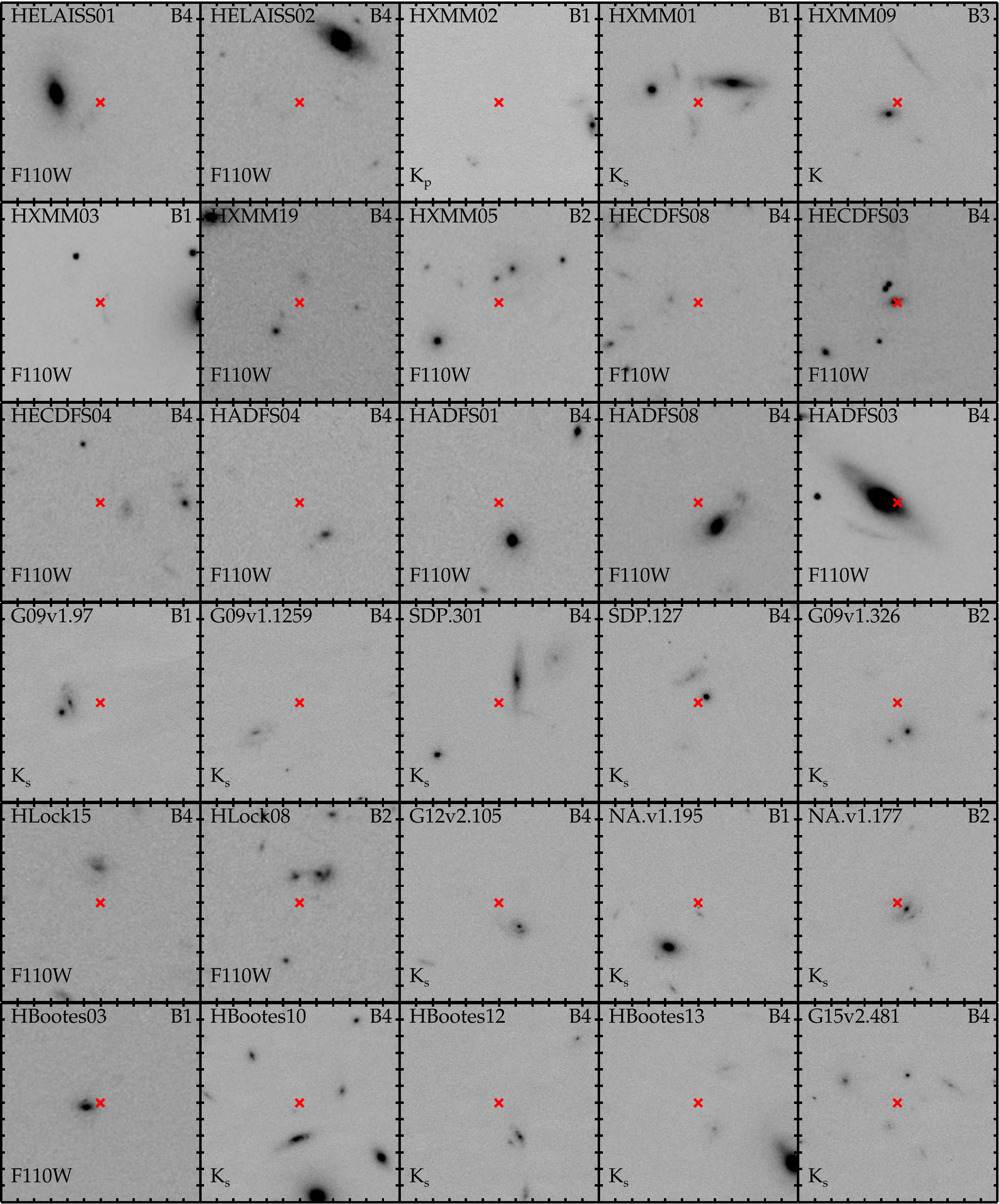}
\caption{Near-IR images of Grade B sources, oriented north is up and east is left for all images. Each tick mark is $1''$ and the size of each panel is $12''$. The near-IR band and the complete lens grade are shown in the lower left and upper right corners, respectively. The red crosses represent the measured {\herschel} position.}
\label{fig:lg2} 
\end{figure*}

\begin{figure*}[]h
\includegraphics[width=\textwidth,keepaspectratio]{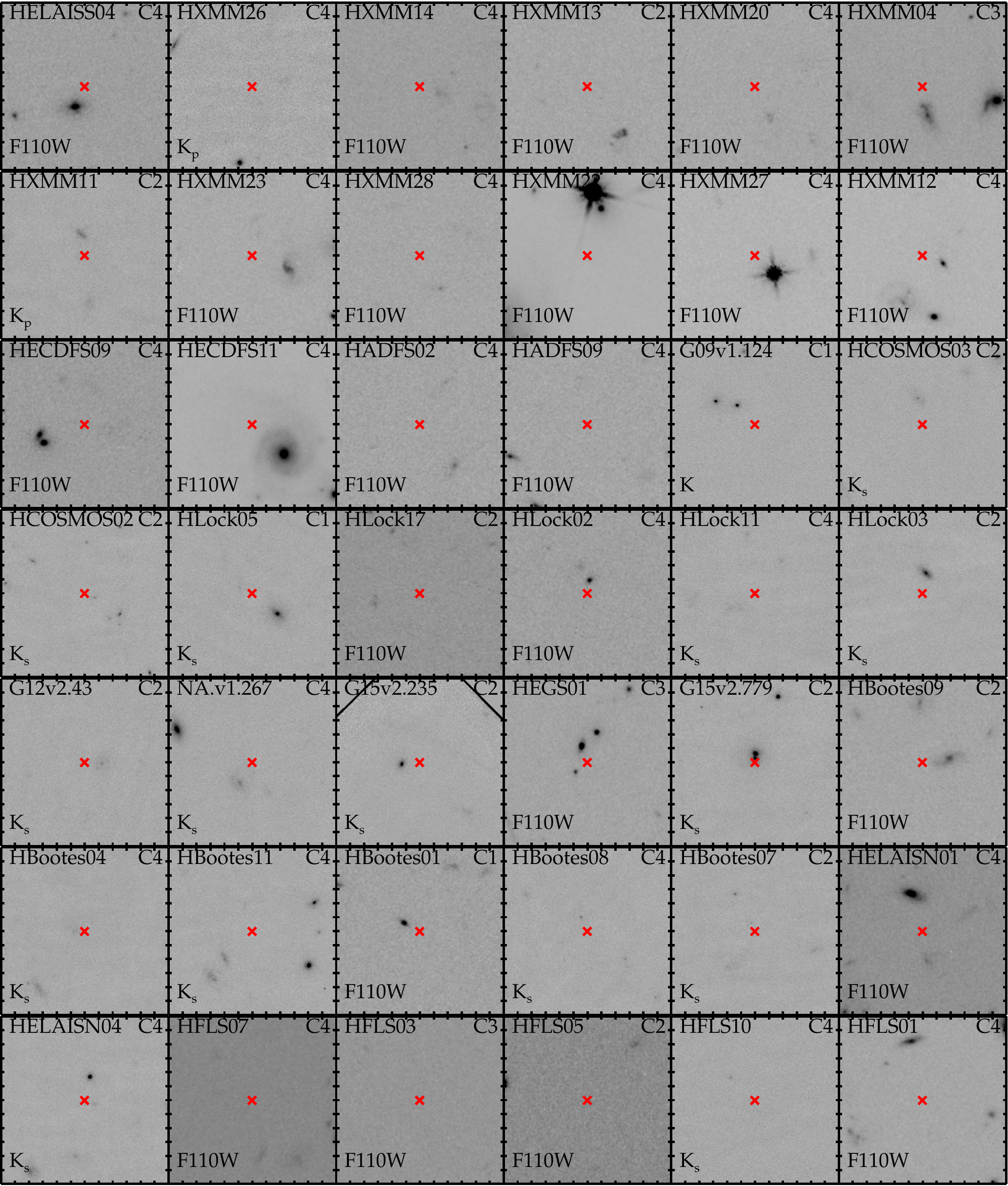}
\caption{Near-IR images of Grade C sources, oriented north is up and east is left for all images. Each tick mark is $1''$ and the size of each panel is $12''$. The near-IR band and the complete lens grade are shown in the lower left and upper right corners, respectively. The red crosses represent the measured {\herschel} position.}
\label{fig:lg3} 
\end{figure*}

\begin{figure*}
\includegraphics[width=0.5\textwidth,keepaspectratio]{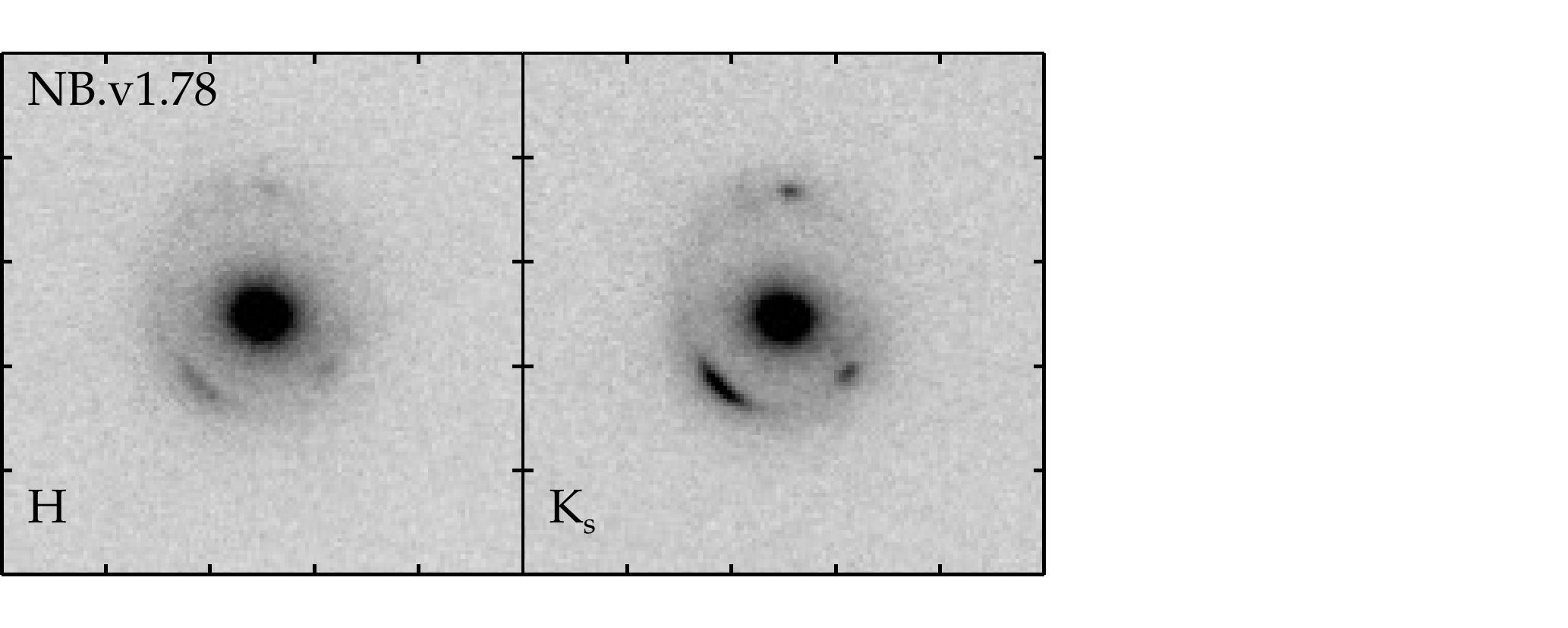} \hspace{-27.5mm}
\includegraphics[width=0.5\textwidth,keepaspectratio]{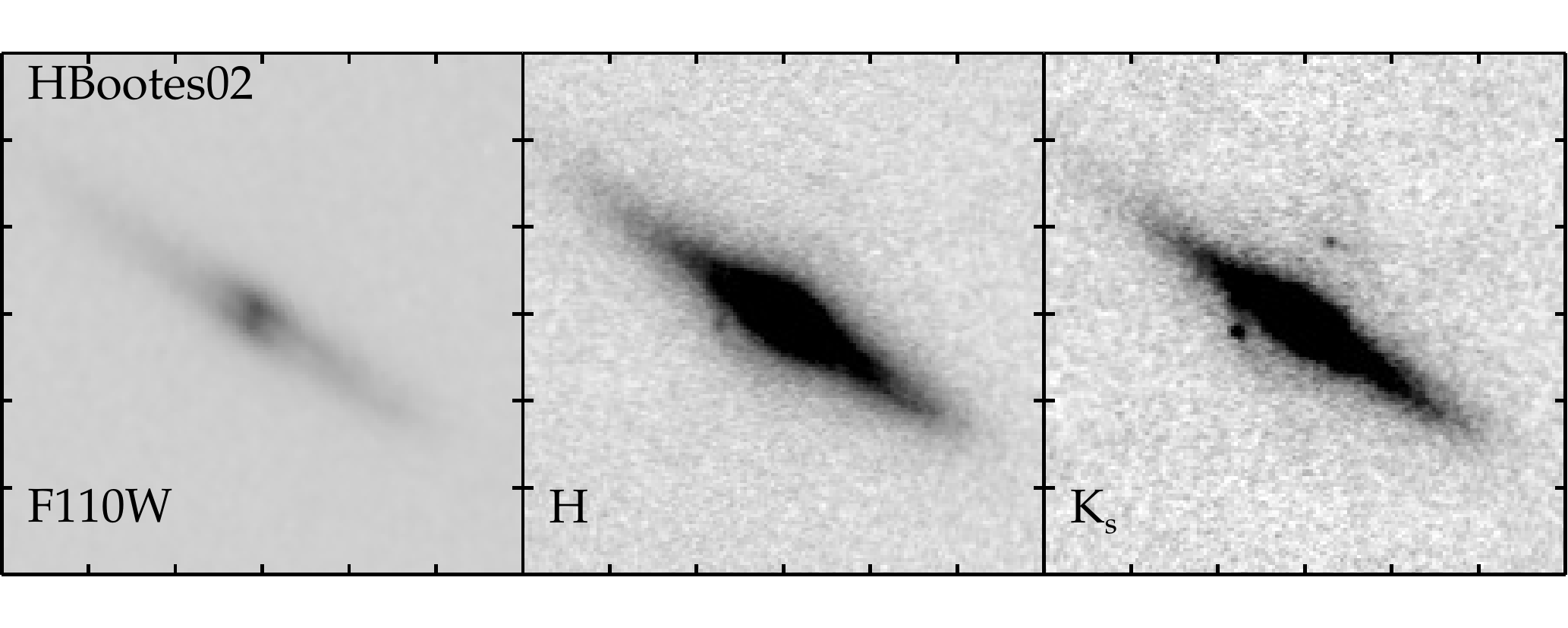}\\
\vspace{-10.0mm}

\includegraphics[width=0.5\textwidth,keepaspectratio]{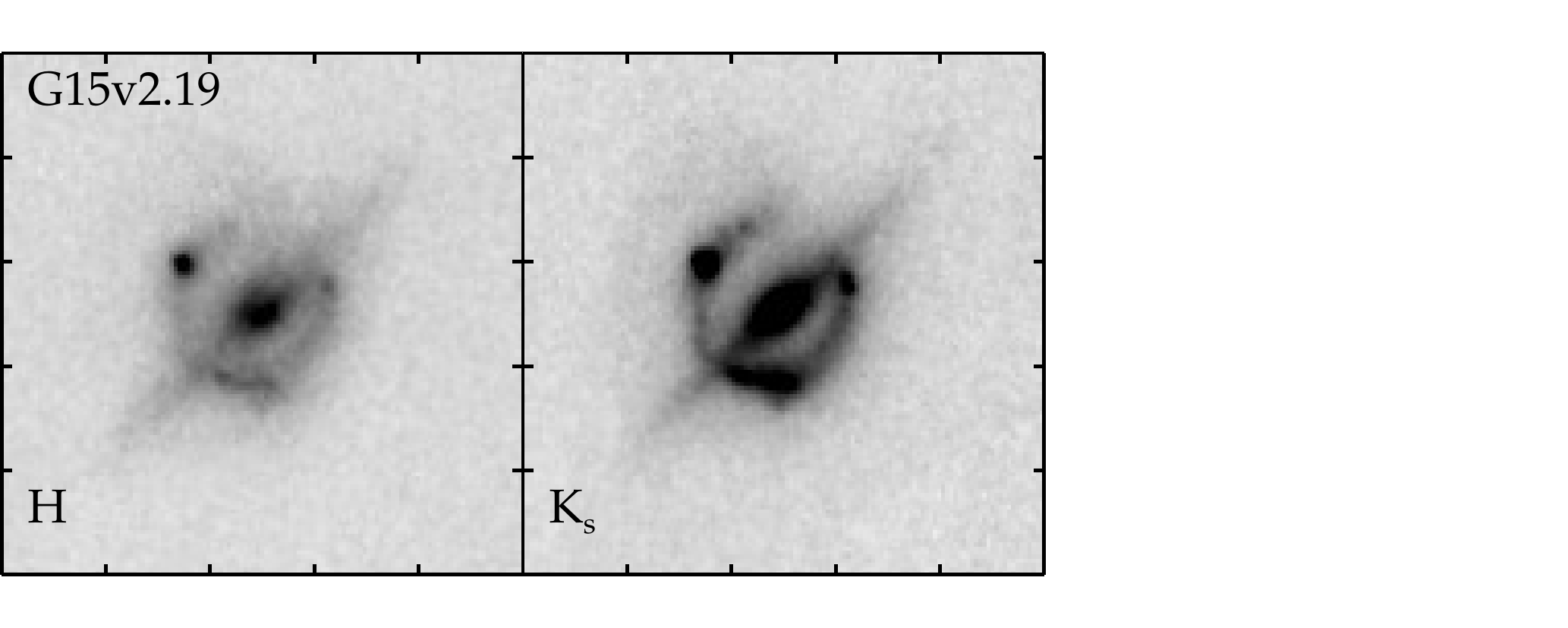} \hspace{-27.5mm} 
\includegraphics[width=0.5\textwidth,keepaspectratio]{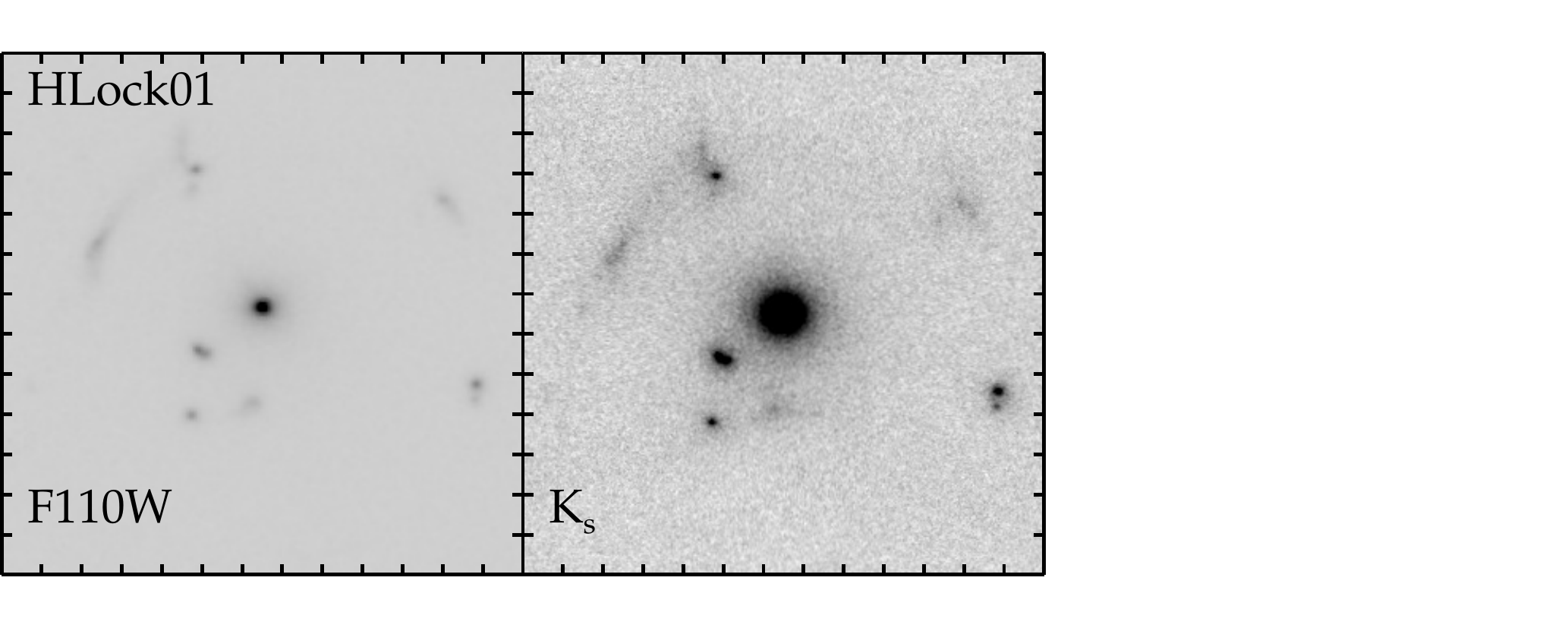}\\
\vspace{-10.0mm}

\includegraphics[width=0.5\textwidth,keepaspectratio]{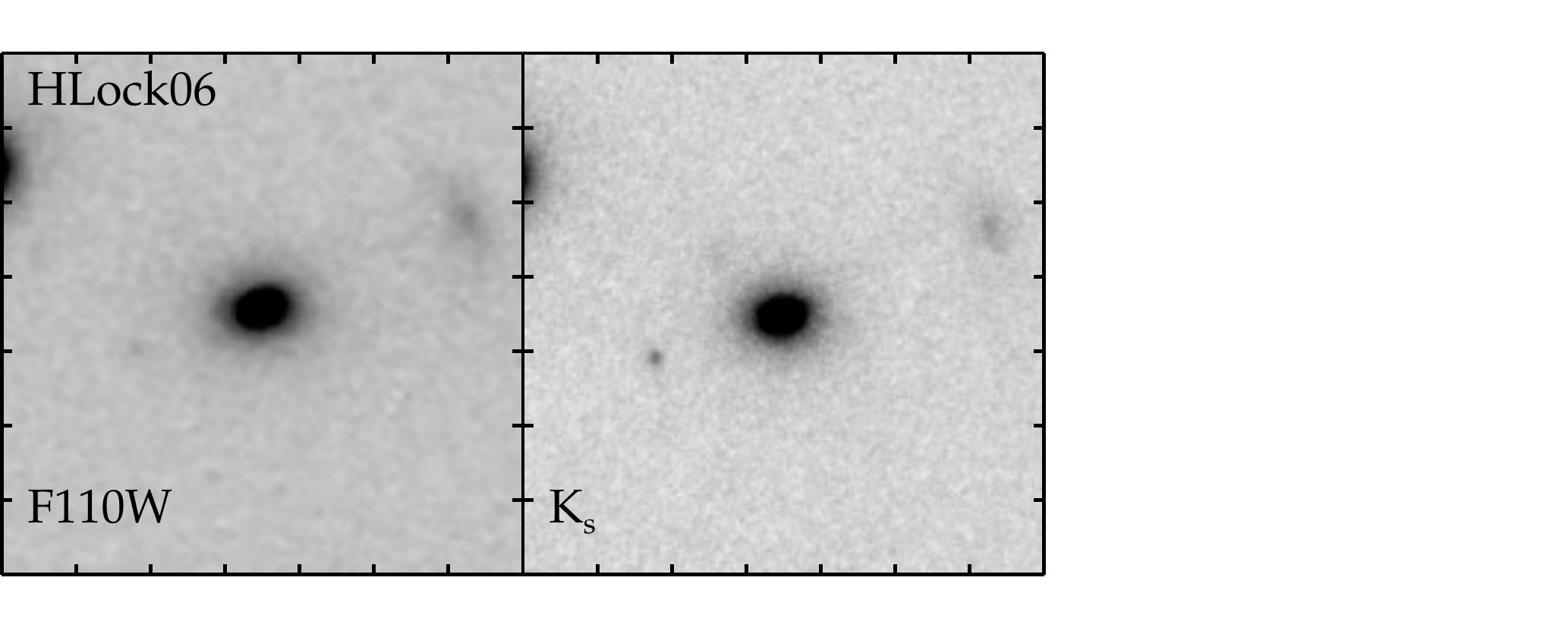} \hspace{-27.5mm}
\includegraphics[width=0.5\textwidth,keepaspectratio]{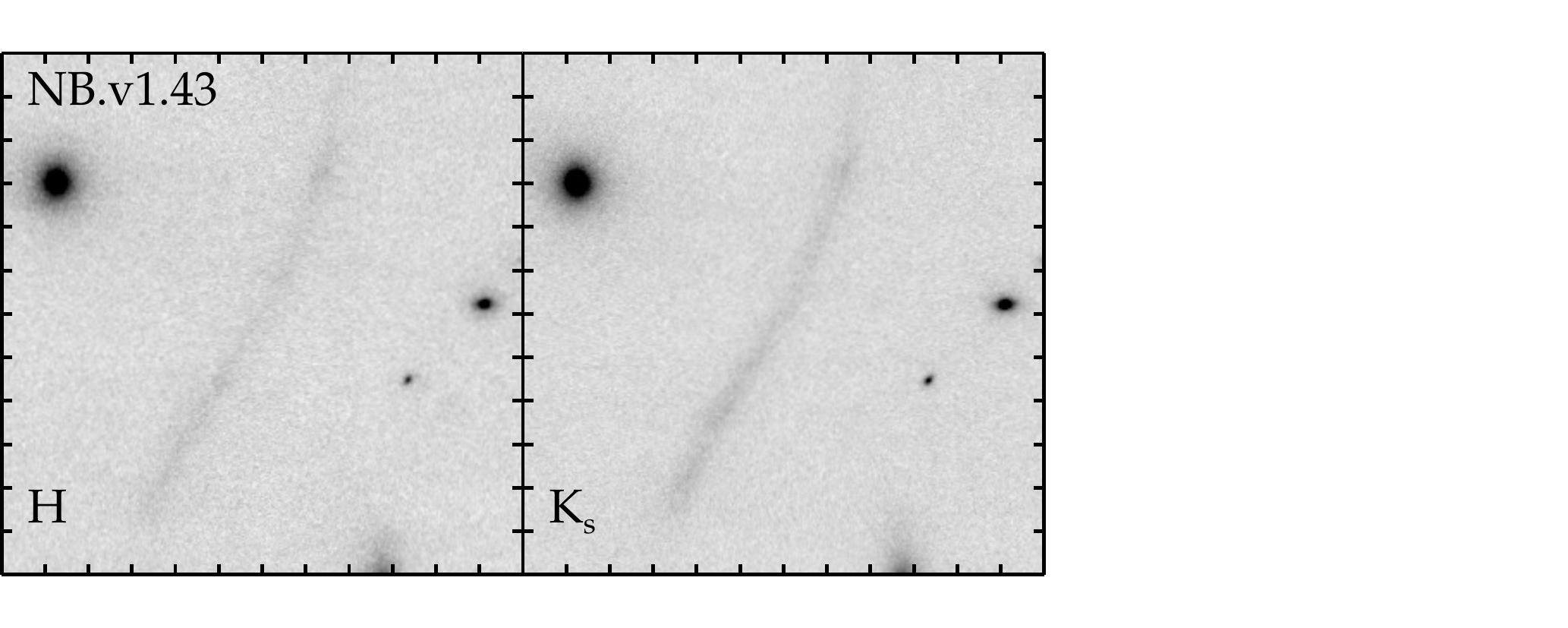}\\ 
\vspace{-10.0mm}

\includegraphics[width=0.5\textwidth,keepaspectratio]{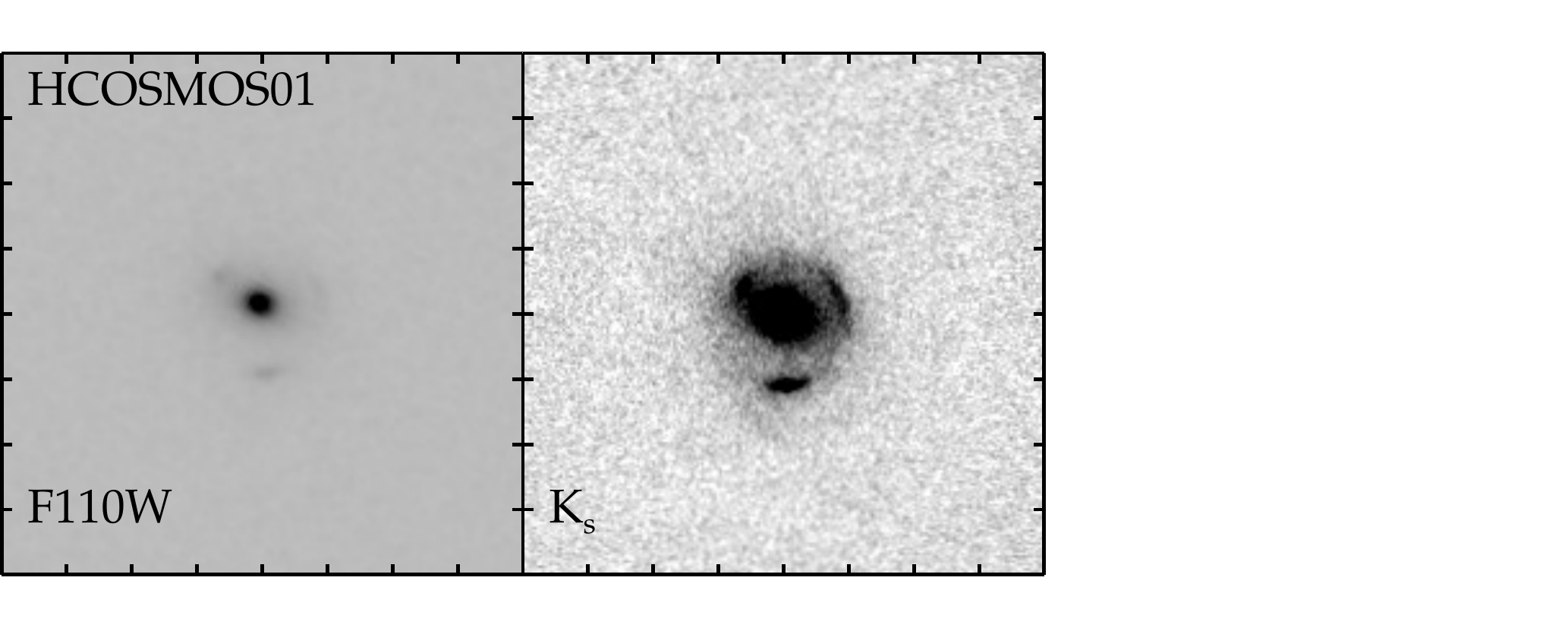} \hspace{-27.5mm}
\includegraphics[width=0.5\textwidth,keepaspectratio]{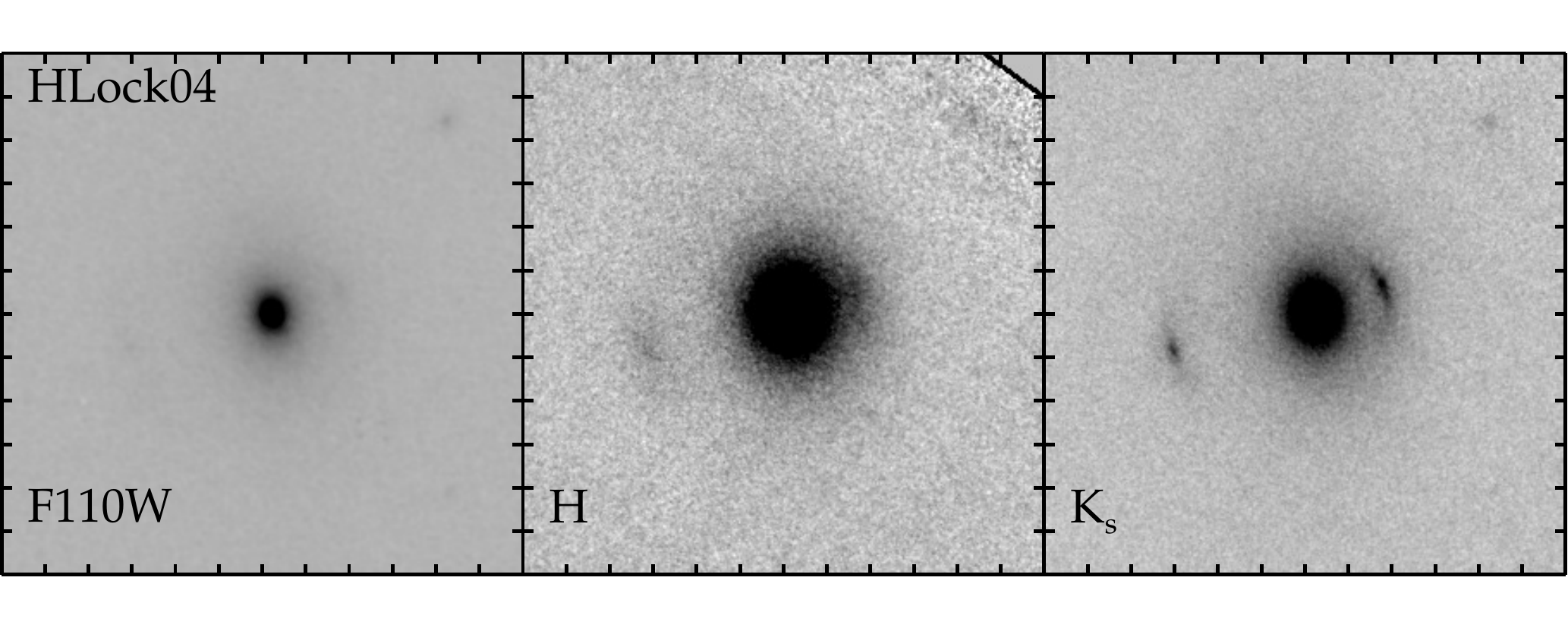}\\

\caption{Multi-wavelength high-resolution near-IR for Grade A lensed SMGs, oriented north is up, east is left for all images. The near-IR band is labeled on the lower left corner. Each tick mark represents $1''$. All images are scaled to have consistent brightness units.}
\label{fig:mw}
\end{figure*}

\end{document}